\documentclass[twocolumn]{aastex62}

\hypersetup{linkcolor=red,citecolor=blue,filecolor=cyan,urlcolor=blue}
\usepackage{color}
\usepackage{captcont}
\usepackage{amsmath}

\newcommand\wise{\textit{WISE}}

\graphicspath{{./}{figures/}}

\submitjournal{ApJ}

\shorttitle{Obscured Quasars}
\shortauthors{Patil et al.}
\begin{document}

%\title{A VLA Study of %\textit{WISE}-NVSS Selected Obscured Quasars}
%\title{High-resolution VLA Imaging of Obscured Quasars: Young Radio Jets Caught in a UniqueEvolutionary Stage}
\title{High-resolution VLA Imaging of  Obscured Quasars: Young Radio Jets Caught in a Dense ISM}

\correspondingauthor{Pallavi Patil}
\email{pp3uq@virginia.edu}

\author[0000-0002-9471-8499]{Pallavi Patil} \altaffiliation{Grote Reber Pre-Doctoral Fellow}
\affil{Department of Astronomy, University of Virginia, 530 McCormick Road, Charlottesville, VA 22903, USA}
\affiliation{National Radio Astronomy Observatory, 520 Edgemont Road, Charlottesville, VA 22903, USA}

\author{Kristina Nyland}
%\affiliation{National Radio Astronomy Observatory, 520 Edgemont Road, Charlottesville, VA 22903, USA}
\affiliation{National Research Council, resident at the Naval Research Laboratory, Washington, DC 20375, USA}

\author{Mark Whittle}
\affil{Department of Astronomy, University of Virginia, 530 McCormick Road, Charlottesville, VA 22903, USA}

\author{Carol Lonsdale}
\affiliation{National Radio Astronomy Observatory, 520 Edgemont Road, Charlottesville, VA 22903, USA}

\author{Mark Lacy}
\affiliation{National Radio Astronomy Observatory, 520 Edgemont Road, Charlottesville, VA 22903, USA}

\author{Colin Lonsdale}
\affiliation{Massachusetts Institute of Technology, Haystack Observatory, Westford, MA 01886, USA }

\author{Dipanjan Mukherjee}
\affiliation{Inter-University Centre for Astronomy and Astrophysics, Post Bag 4, Ganeshkhind, Pune - 411007, India.}
\affiliation{Dipartimento di Fisica Generale, Universita degli Studi di Torino , Via Pietro Giuria 1, 10125 Torino, Italy}
\affiliation{INAF, Osservatorio Astrofisico di Torino, Strada Osservatorio 20, 10025 Pino Torinese, Italy
}

\author{A.C. Trapp}
\affiliation{Department of Physics and Astronomy, University of California Los Angeles, CA, 90095-1562, USA}

\author{Amy E Kimball}
\affiliation{National Radio Astronomy Observatory, 1003 Lopezville Rd, Socorro, NM 87801, USA}

\author[0000-0002-3249-8224]{Lauranne Lanz}
\affiliation{Department of Physics and Astronomy, Dartmouth College, 6127 Wilder Laboratory, Hanover, NH 03755, USA}
\affiliation{Department of Physics, The College of New Jersey, 2000 Pennington Road, Ewing, NJ 08628, USA}

\author{Belinda J. Wilkes}
\affiliation{Harvard-Smithsonian Center for Astrophysics, Cambridge,
MA 02138, USA}

\author{Andrew Blain}
\affiliation{Department of Physics \& Astronomy, University of Leicester, University Road, Leicester LE1 7RH, UK}

\author{Jeremy J. Harwood}
\affiliation{Centre for Astrophysics Research, School of Physics, Astronomy and Mathematics, University of Hertfordshire, College Lane, Hatfield AL10 9AB, UK}

\author{Andreas Efstathiou}
\affiliation{School of Sciences, European University Cyprus, Diogenis Street, Engomi, 1516, Nicosia, Cyprus}

\author{Catherine Vlahakis}
\affiliation{National Radio Astronomy Observatory, 520 Edgemont Road, Charlottesville, VA 22903, USA}

%\author{Add Your Name Here}
%\affiliation{Your Institution}



\begin{abstract}
We present new sub-arcsecond-resolution Karl G. Jansky Very Large Array (VLA) imaging at 10~GHz of 155 ultra-luminous ($L_{\rm bol}\sim10^{11.7-14.2}$~L$_\odot$) and heavily obscured quasars with redshifts  $z \sim0.4-3$. The sample was selected to have extremely red mid-infrared (MIR)-optical color ratios based on  data from \textit{Wide-Field Infrared Survey Explorer} (\wise)~along with a detection of bright, unresolved radio emission from the NRAO VLA Sky Survey (NVSS) or Faint Images of the Radio Sky at Twenty-Centimeters (FIRST) Survey.  Our high-resolution VLA observations have revealed that the majority of the sources in our sample (93 out of 155) are  compact on angular scales $<0.2^{\prime \prime}$
($\leq$ 1.7~kpc at  $z \sim2$). The radio luminosities, linear extents, and lobe pressures of our sources are similar to young radio active galactic nuclei (AGN; e.g., Gigahertz Peaked Spectrum, GPS, and Compact Steep Spectrum, CSS, sources), but their space density is considerably lower.  
Application of a simple adiabatic lobe expansion model suggests relatively young dynamical ages  ($\sim10^{4-7}$~years),  relatively high ambient ISM densities ($\sim1-10^4$~cm$^{-3}$), and modest lobe expansion speeds ($\sim30-10,000$~km s$^{-1}$).  Thus, we find our sources to be consistent with a population of newly triggered, young jets caught in a unique evolutionary stage in which they still reside within the dense gas reservoirs of their hosts. Based on their radio luminosity function and dynamical ages, we estimate only $\sim20\%$ of classical large scale FRI/II radio galaxies could have evolved directly from these objects. We speculate that the \wise-NVSS sources might first become GPS or CSS sources, of which some might ultimately evolve into larger radio galaxies.
%We speculate that following blowout of the dense ISM, subsequent episodes of jet production might create GPS and CSS sources, some of which would then evolve into FRI/FRII radio galaxies. 
\end{abstract}

\keywords{galaxies: active - galaxies: evolution - galaxies: jets - radio continuum: galaxies - quasars: general}


\section{Introduction} \label{sec:intro}
The active galactic nucleus (AGN) phenomenon, driven by accretion onto supermassive black holes (SMBHs), %is of fundamental importance to the 
is believed to play an important role in the 
evolution of 
%the host galaxies 
galaxies over cosmic time. There is now compelling evidence interlinking SMBH growth with host galaxy star formation and mass buildup.  The primary evidence supporting %the 
SMBH-galaxy co-evolution includes the empirical relation found between SMBH mass and the stellar velocity dispersion 
%in bulge of the galaxy 
in galactic bulges (\citealt{kormendy+13} and references therein) and the similarities in the cosmological evolution of AGN space densities and the star formation rate densities  (\citealt{heckman+14, madau+14}, and references therein).

The energy released by AGN can have an impact on
the surrounding interstellar (ISM) or circumgalactic medium (CGM) via a variety of radiative and mechanical processes. Such interactions, often termed AGN feedback, can shock and/or expel the gas causing suppression or triggering of star formation in the host galaxy. 
Improving our understanding of SMBH-galaxy co-evolution requires direct observations of AGN feedback in action during the peak epoch of stellar mass assembly and SMBH growth at $1<z<3$.  However, this phase of galaxy evolution is believed to take place in the presence of thick columns of gas and dust, leading to heavily obscured systems that are challenging to observe at optical and X-ray wavelengths \citep{hickox+18}.

In dust-obscured systems, emission at optical, UV and X-ray wavelengths  from the AGN and/or nuclear starburst is absorbed by dust and re-radiated in the infrared.  
 Mid-infrared (MIR) color diagnostics using 
infrared satellites such as the \textit{Spitzer Space Telescope} \cite[e.g.,][]{lacy+04, stern+05, hatz+05, lacy+07, lacy+13, donley+12}, AKARI \cite[e.g.,][]{oyabu+11}, and \textit{Wide-Field Infrared Survey Explorer} (\wise;~e.g.,  \citealt{stern+12, mateos+12, wu+12b, assef+13, lonsdale+15}) have 
provided an effective means of identifying both obscured and unobscured AGN populations. 
Recent studies have suggested that the heavily reddened AGN population 
%is suggested to be in 
represents a transient phase of peak black-hole fueling and stellar mass assembly \cite[e.g.,][]{eisenhardt+12,wu+12,jones+14, assef+15, tsai+15, diaz+16}. The most extreme population of these galaxies, identified  based on very red \textit{WISE} colors, are called Hot Dust Obscured Galaxies (Hot DOGs) 
due to the presence of hot dust and high luminosity MIR emission \citep{eisenhardt+12, wu+12, bridge+13}. 

One way to favor obscured AGN emission over obscured star formation is to additionally require a significant radio source. If the radio flux is greater than the MIR flux, the source is likely to be an AGN \citep[e.g.,][]{ibar+08}. Thus, surveys that combine MIR and radio can identify obscured powerful jetted AGN (e.g., \citealt{condon+02}). Ideally, these sources will be similar to the Hot DOGs discussed above -- they are AGN caught at an early stage in their evolution -- but with the additional possibility of showcasing jet-driven feedback. 

\citet{lonsdale+15} define such a sample, with an additional requirement that the optical counterparts are faint which favors sources at intermediate redshift, $z \sim 1-3$. This sample forms the basis of the present study. As it stands, however, the \citet{lonsdale+15} sample only made use of relatively low-resolution radio observations.  In the current paper, we present high-resolution X-band (8---12~GHz) Karl G. Jansky Very Large Array (VLA) images of this sample, which allow us to place much stronger constraints on the radio source properties. In particular, we wish to establish whether the sources are young, reside in a dense ISM, and may be caught in a state of expansion. In a companion paper (Patil et al. in prep.) we will use multi-frequency observations to explore the radio spectral shapes, using these to further investigate the nature of the radio sources and the nature of the near-nuclear environments.

Section~\ref{sec:sample} summarizes the sample selection and the MIR properties of the sample. %explained in detail in our first paper \citep{lonsdale+15}. 
The VLA observations and data reduction are described in the Section~\ref{sec:datared}. We present source measurements and properties in Sections~\ref{sec:smeasure} and~\ref{sec:properties}, respectively. We analyze our sample's radio luminosity function in Section~\ref{sec:lumfunc}.  Section~\ref{sec:discussion} discusses how our sample might fit into an evolutionary framework with the other known classes of compact and extended radio sources. We also use an adiabatic expanding lobe model to derive some important source properties. Section~\ref{sec:conclusion} summarizes our conclusions. 
%Subsequent papers will address VLBA imaging of 90 sources (Colin Lonsdale et al., in preparation) 
We adopt a $\Lambda$CDM cosmology with $H_{0}$ = 67.7  km s$^{-1}$ Mpc$^{-1}$, $\Omega_{\Lambda}$ = 0.691 and $\Omega_{\textrm{\small{M}}}$ = 0.307 \citep{planck+15}.

\section{Sample Selection}\label{sec:sample}
A detailed description of our sample selection is given in \citet{lonsdale+15}. Briefly, point sources from the \wise~AllSky catalog \citep{wright+10} with S/N $> 7$ in the 12 or 22~$\mu$m bands were cross-matched with sources from the National Radio
Astronomy Observatory Very Large Array Sky Survey (NVSS; \citealt{condon+98})
%with the matching 
or, when available, the Faint Images of the Radio Sky at Twenty-centimeters (FIRST; \citealt{becker+95}) catalog. An important requirement was that the source be unresolved in NVSS ($\theta_{\rm FWHM} < 45^{\prime\prime}$) and FIRST ($\theta_{\rm FWHM} < 5^{\prime\prime}$) catalogs in order to exclude sources dominated by large scale, evolved radio emission.   
%This step could potentially bias the sample against sources that also possess a more recently triggered very small and young powerful jet, however such systems would be significantly harder to interpret as we would not be able to easily distinguish which radio jet episode might have made significant feedback impacts on the host galaxy.  
We also required the candidates to have relatively large radio---to---MIR flux ratios, ($q_{22}$ = log$(f_{22\,\mu \textrm{m}} /f_{20\,\textrm{cm}}) < 0$) to favor AGN emission as opposed to star formation \citep{appleton+04, ibar+08}.
%Other selection criteria help favor the inclusion of radio AGN \textbf{with possibly small angular extents}: (a) the sources were unresolved in  NVSS ($\theta_{\rm FWHM} < 45^{\prime\prime}$) and/or FIRST ($\theta_{\rm FWHM} < 5^{\prime\prime}$); and (b) they have relatively large radio-to-MIR flux ratios: $q_{22}$ = log$(f_{22\,\mu \textrm{m}} /f_{20\,\textrm{cm}}) < 0$.

%Similar to the studies mentioned above, we 
The selection also includes only objects with very red MIR colors,
with a color cut defined by $(W1-W2) + 1.25(W2-W3) >7$ \footnote{We note that this infrared color selection criterion contained an error in Section 2 of \citet{lonsdale+15}.  The error was a typo only and did not impact the analysis or any of the figures in \citet{lonsdale+15}.  The color cut defined here is the correct version.} and a flux density cut of 7~mJy at 22~$\mu$m. Coupled with the limit on the $q_{22}$ parameter from above, this introduces a 1.4~GHz flux limit of about 7 mJy.

To minimize contamination by the non-AGN population, the sample excludes sources within $10^o$ of the Galactic plane.  

Each source was inspected using the Sloan Digital Sky Survey (SDSS; \citep{york+00}) or Digitized Sky Survey (DSS; if not within the SDSS footprint), and only objects that were relative optically faint or undetected were kept. We have not defined any specific optical selection criteria to favor sources within the required redshift interval and to not create a bias against large amounts of scattered optical light.
We also relied upon follow-up spectroscopy to refine our sample by redshift.
This ensures that the objects are likely to be at intermediate or high redshift, and given the extreme MIR-to-optical color, they are also  likely to be heavily obscured.  Given the intermediate or high redshift, the bright MIR fluxes then suggest high bolometric luminosity.  A total of 167 sources met these selection criteria.
We will discuss the completeness of the sample in Section~\ref{sec:lumfunc}.

\begin{figure}
    \centering
    \includegraphics[clip=true, trim=0.3cm 0.3cm 12.8cm 0cm, width=0.5\textwidth]{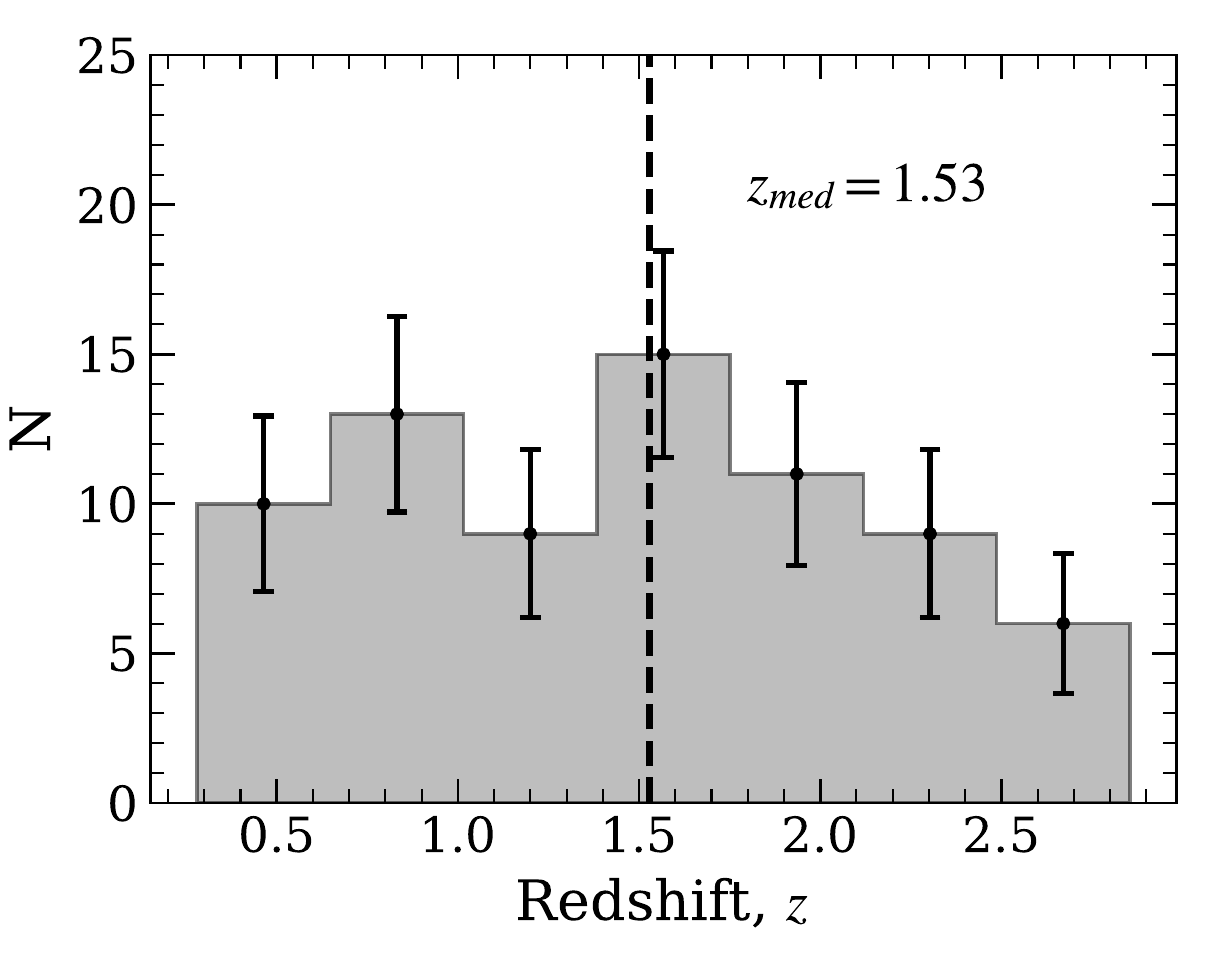}
    \caption{ Redshift distribution of our sample. We have spectroscopic redshifts available for 71 sources. 
The black dashed line 
 denotes the median value. The error bar in each redshift bin is the respective binomial uncertainty.}
    \label{fig:red}
\end{figure}

\subsection{Spectroscopic Redshifts}
We obtained spectroscopic redshifts for 71 out of 80 attempted sources using several telescopes \citep[see][for details]{lonsdale+15}. 
The remaining 9 sources were too faint to provide a reliable redshift.
Figure~\ref{fig:red} shows the redshift distribution which is seen to be approximately flat from $0.5<z<2$ with a possible decline from $2<z<2.8$.  The median value is  $z_{\textrm{\small{med}}} \sim1.53$. While the subset of sources targeted for redshift is likely biased to the optically brighter sources, it is unclear whether or not this translates to a bias in redshift -- while optically brighter galaxies might be at lower redshift, optically brighter quasars might be at higher resdhift. Taken at face value, our redshift distribution indicates that many of our sources lie in the epoch of peak star-formation and black hole fueling, some are nearer ($z\lesssim1$) and may be  suitable for detailed follow-up observations. 
%We discuss the redshit distribution further in Section~\ref{sec:linsize}.}

\subsection{MIR and Submm Properties}

870~$\mu$m Atacama Large Millimeter/Submillimeter Array (ALMA) imaging of 49 sources \citep{lonsdale+15}  and  850~$\mu$m James Clerk Maxwell Telescope (JCMT)-Submillimetre Common-User Bolometer Array imaging of 30 sources \citep{jones+15}  yielded 26/49 ALMA and 4/30 JCMT detections. Overall the MIR-submm SEDs of our sample is likely to be dominated in the MIR by AGN heated thermal dust emission. %from AGN with high-accretion rates. 
The extremely red optical-WISE colors  and bright 22~$\mu$m emission revealed that these sources have high IR and bolometric luminosities ($L_{bol}\sim 10^{11.7-14.2} L_\odot$) with a few reaching the Hyper-Luminous Infrared Galaxy (HyLIRG) regime. AGN populations identified using ultra-red WISE color diagnostics are now known to belong to a class of  IR-luminous obscured quasars such as Hot DOGs  \citep[e.g.,][]{eisenhardt+12, wu+12, assef+15}. The MIR signatures and high-ionization lines in the spectra of our sample (\citealt{kim+13}, Ferris et al. submitted) are consistent with a population of radiative-mode obscured quasars. We refer our readers to \citet{lonsdale+15} for more details on the MIR and submm properties of our sample.

\section{New VLA Data}
\label{sec:datared}

\subsection{Observing Strategy}
We observed 
167 sources from \citet{lonsdale+15} at X-band (8--12 GHz) with the VLA in the A- and B-arrays through projects 12B-127 and 12A-064, respectively. Due to the complexity of dynamic scheduling for such a large sample, 12 sources were not observed in any array, and 32 were observed in only one array. 
Therefore, the sample discussed in this paper consists of 155 sources, 26 of which lack imaging with the A-array and 6 of which lack imaging with the B-array.
The A-array observations were divided into 13 separate scheduling blocks (SBs), and a total of 129 sources were observed between October and December 2012. The B-array observations were divided into 7 different SBs, and 149 sources were observed from June to August 2012. 

Sources closer to each other on the sky were scheduled in groups, with phase calibrators interleaved.  However, to maximize observing efficiency, the same calibrator was not always re-observed after each target. This strategy was worth the inherent risk of failing to obtain phase closure for a few targets, because most of the sources were expected to be bright enough for self-calibration. 

%The WIDAR correlator was set in Open Shared Risk Observing full polarization mode and two basebands were tuned to utilize the complete 2048 MHz bandwidth. Both basebands had eight 128~MHz wide sub-bands with central frequencies 8.6~GHz and 11.4~GHz, respectively.
The observations took place during the Open Shared Risk Observing period when maximum bandwidths were limited to $\sim$2~GHz.  Our WIDAR correlator set-up consisted of two basebands with central frequencies 8.6~GHz and 11.4~GHz, respectively.  The bandwidth of each baseband was 1024~MHz divided among eight 128~MHz wide spectral windows.  The total bandwidth of our observations was 2~GHz.  The correlator setup was kept identical for both of the arrays. Our observing strategy aimed to obtain snapshot-imaging of the full sample with about 5 minutes of integration time per source with a theoretical rms noise level of about $\sim$ 13~$\mu$Jy~beam$^{-1}$.

\subsection{Calibration and Imaging}
We used the Common Astronomy Software Applications package (CASA; \citealt{mcmullin+07}) version 4.7.0 for data editing, calibration, and imaging. 
The initial step was to remove bad data with the help of the VLA operators log\footnote{\url{www.vla.nrao.edu/cgi-bin/oplogs.cgi}} %and 
followed by visual inspection of the data in the $uv$-plane using the task PLOTMS. 
Hanning smoothing was performed prior to calibration to remove the rigging effect from the Gibbs phenomenon caused by strong Radio Frequency Interference (RFI). 
The data were calibrated using the CASA VLA calibration pipeline\footnote{\url{www.science.nrao.edu/facilities/vla/data-processing/pipeline}} (version 1.3.9).

We then used the pipeline weblog and test images of the targets and phase calibrators to examine the quality of the calibration. If necessary, additional flagging was done, followed by a re-run of the calibration pipeline. We then used the CASA task 
%\textit{split} 
SPLIT to separate the $uv$-data for each target into individual datasets %to perform 
for self-calibration and final imaging. 

We ran a few rounds of phase-only self-calibration and one round of amplitude and phase calibration to correct artifacts due to residual calibration errors.
We used the CASA task 
%\textit{clean} 
CLEAN to produce the final continuum image. Because of the wide bandwidths made available by the new correlator, we formed images using the multi-frequency synthesis mode with two Taylor coefficients (by setting the CLEAN parameter nterms=2) to more accurately model the spectral dependence of the sky. Also, to mitigate the effects of non-coplanar baselines during imaging, we used the W-projection algorithm with 128 $w$-planes. The full-width half maximum (FWHM) of the synthesized beam of the final images in the A- and B-arrays are typically $\theta_{b} \sim0.2^{\prime\prime}$ and $\sim0.6^{\prime\prime}$, respectively.

Despite our careful calibration and imaging strategy, a total of 13 targets (11 in A-array and 2 in B-array) suffered from severe phase closure issues. As a result, 110 sources have imaging in both arrays, 8 sources have only A array imaging, and 37 sources have only B array imaging. %and as a result five out of 154 sources are excluded from further analysis. 
%A quality flag based on visual inspection is provided in Table~\ref{tab:obs_info}.
Thus, the analysis presented in the remainder of this paper is based on 155 sources.

\begin{figure}[hb]
    \centering
    \includegraphics[clip=true, trim=0.5cm 0.3cm 0.5cm 0.cm, width=\linewidth]{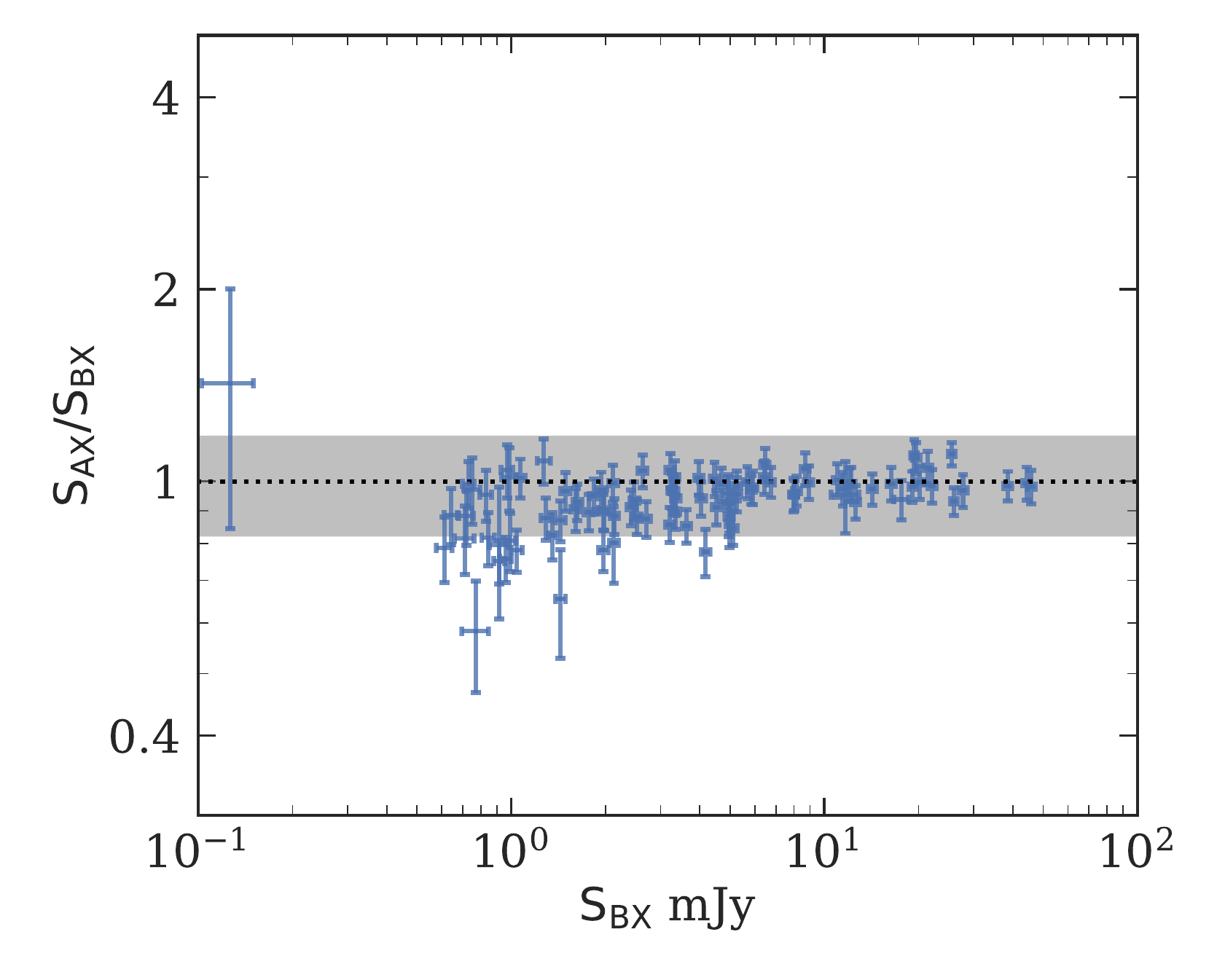}
    \caption{The ratio of the total flux measured in A- and B-arrays for 110 sources as a function of the flux measured from the B-array images. The black dotted line shows a ratio of unity. The normalized absolute median deviation of the flux ratios between the A- and B-array observations is 0.18 and is indicated by the gray shaded region.
    \label{fig:fluxd}}
\end{figure}

\section{Source Measurements}\label{sec:smeasure}

\subsection{Fluxes}\label{sec:flux}
To determine source parameters such as peak flux density, integrated flux, 
deconvolved shape parameters, and all corresponding uncertainties, we used the JMFIT task available in the 31DEC18 version of the Astronomical Image Processing Software (AIPS). 
In most cases, the radio sources have either single or multi-component Gaussian-like morphologies, and their 
flux and shape parameters may be estimated by fitting one or more two-dimensional elliptical Gaussian models.  For sources with extended, complex structures, we manually estimated the source parameters using the 
CASA Viewer \footnote{Following \citet{nyland+16}, we calculate   flux measurement uncertainties as $\sqrt{(N\times\sigma)^2 + (0.03\times S_{tot})^2}$, where N is total number of synthesized beam over 3$\sigma$ contour emission, $\sigma$ is the rms noise, and $S_{tot}$ is the integrated flux of the region.}
%in the situation of radio emission having a complex structure. 
The flux measurement uncertainties were calculated by adding the error provided by JMFIT and the 3\% VLA calibration error \citep{perley+13} in quadrature. We provide the clean beam dimensions, peak flux, and total flux from our A- and B-array observations in Table~\ref{tab:source_full}.

%A few examples of the basic source measurements are provided in Table~\ref{tab:sourcem}, with complete information for the entire sample available in Appendix Table~\ref{tab:source_full}. 
The total flux distributions in A- and B-array observations span the range $0.18-45$~mJy and $0.13-60$~mJy, respectively, with similar medians of $\sim3.3$~mJy.
Figure~\ref{fig:fluxd} compares the %total
integrated fluxes  of the 
110 sources with high-quality flux measurements from both  A- and B-arrays. 
The designation ``high-quality'' here simply indicates no hint of image artifacts. 
%There were 103/129 and 134/147 sources from the A- and B-configuration observations, respectively, with high-quality  final images that were suitable for flux and shape measurements. 
%We identified those high-quality images visually by removing any source that showed image artifacts.

%The range of 10~GHz integrated fluxes from our observations in the A and B configurations is $0.09-45$~mJy and $0.08-60$~mJy, respectively, with similar means of $\sim 2.8$~mJy.  %The normalized absolute median deviation of the flux ratios between our A and B configuration observations is 0.18 and is indicated by the shaded region in Figure~\ref{fig:fluxd}.
%Figure~\ref{fig:fluxd} shows the ratio between the total fluxes measured in the A and B configurations. 

We find that, for most of our sample, the total flux measurements from each array are in good agreement.  
%typically agrees within the margin of flux uncertainties.  %The typical resolution obtained at A-array is $\sim 0.2^{\prime\prime}$ and at B-array is $\sim 0.6^{\prime\prime}$. Additionally, the largest resolvable angular scale (LAS) for A and B configurations is $\sim 5.3^{\prime\prime}$ and $\sim 17^{\prime\prime}$, respectively. That means that  for a given source, A-array observation would be missing flux from any emission present on the intermediate scales between 5.3$^{\prime\prime}$ and 17$^{\prime\prime}$. However, the left plot in Figure~\ref{fig:fluxd} shows that our flux measurements are not affected by the issue of missing flux in the majority of the sources. 
%Majority of the sources in our sample are unresolved (See Section~\ref{sec:morph}) or do not 
%$\theta_{\rm FWHM}$
There are 4 sources that lie below the unity line in Figure~\ref{fig:fluxd} and have less flux recovered in the longer-baseline A-array observations.  These sources may have a diffuse emission component that has not been recovered in the A-array data\footnote{We note that the largest resolvable angular scale (LAS) for the 10~GHz images is $\sim5.3^{\prime\prime}$ and $\sim17^{\prime\prime}$ for the A- and B-array, respectively. That means that for a given source, the A-array image would be missing flux from any emission present on the intermediate scales between 5.3$^{\prime\prime}$ and 17$^{\prime\prime}$.}.  There is also one outlier in Figure~\ref{fig:fluxd} with significantly higher flux in the B-array data compared to the A-array, possibly as a result of intrinsic source variability or calibration error.  
%\textcolor{green}{[Should comment on the outliers.]}
%However, the left plot in Figure~\ref{fig:fluxd} shows that our flux measurements are not affected by the issue of missing flux in the majority of the sources.

\subsection{Source Angular Sizes}\label{sec:SAS}
  
We used the JMFIT task in AIPS to measure the angular sizes of our sources.  For resolved sources, JMFIT\footnote{We refer our reader to the online documentation of the JMFIT task for more details:  \url{http://www.aips.nrao.edu/cgi-bin/ZXHLP2.PL?JMFIT}} requires that 1) the integrated flux be larger than the peak flux density and 2) the deconvolved major axis is greater than zero (within the relevant uncertainties). If neither of these criteria were satisfied, the source was classified as unresolved.  The source fitting algorithm gives a cautionary message when only one of two criteria is satisfied. We discuss our morphological classification in the next section, including our approach to sources with ambiguous JMFIT results.

Deconvolved source sizes were taken directly from JMFIT. The uncertainties were calculated based on the formalism given by \citet{Murphy+2017}:
\begin{equation}
\frac{\sigma_\theta}{\sigma_\phi} = \Bigg[1 - \bigg(\frac{\theta_{b}}{\phi}\bigg)^2 \Bigg]^{-1/2}
\end{equation}
where $\sigma_\theta$ and $\sigma_\phi$ are the rms errors on the deconvolved ($\theta$) and measured ($\phi$) source sizes, respectively. The parameter $\theta_b$ is the FWHM of the synthesized beam. For the unresolved sources, we consider the maximum deconvolved angular size provided by JMFIT to be an upper limit on the source size. For extended sources with non-Gaussian morphologies, we measured the angular sizes using CASA viewer. 
%Given the uncertainty in the resolution of the sources designated as ``Caution'', we choose to flag these sources in our subsequent analyses involving the distribution of source sizes. 
%Tables~\ref{tab:sourcem} and~\ref{tab:source_full} provide clean beam sizes, deconvolved source sizes, respective uncertainties, and the morphological classification (see Section~\ref{sec:morph}). In the case of multi-component source structure, we list individual component sizes. 
Table~\ref{tab:ssize} provides the deconvolved source sizes and morphological classification. For sources with more than one component, separate measurements are given for each component.

%\subsection{Morphologies}\label{sec:morph}
\subsection{Morphological Classification}\label{sec:morph}
 
As described in the previous section, the JMFIT task in AIPS uses two basic criteria to determine if a source is formally resolved: the peak/total flux ratio and the deconvolved source size compared to the clean beam size. We use these criteria but modify the first to be more conservative by including a $3\%$ uncertainty in the flux calibration (see Section~\ref{sec:flux}).  

We classify as ``unresolved, U'' sources that satisfy both criteria, deconvolved sizes consistent with zero in both axes, and peak/total flux ratio of unity within the uncertainties. We classify as ``slightly resolved'' sources which show finite size along one of the two axes, and a peak/total flux ratio consistent with 1. We classify as ``resolved, R'' sources that show finite size along both axes and a peak/total flux ratio less than one (within 1 sigma, following \citet{owen+18}). 
%The cases where the source appears to be resolved only along the major axis are designated  
%``slightly resolved." We classify sources as resolved when they satisfy the following two conditions: $\theta_M, \theta_m > \theta_b$ and $f_{peak}/f_{total} < 1$, within the uncertainties (where subscripts $M$ and $m$ refer to the major and minor axes, respectively). 
Sources with more than a single distinct component are classified as double, triple, or multi-component morphologies. Figure~\ref{fig:morph_bar} shows the  distribution of morphologies in our sample. 
We note that the entire analysis is performed separately for the A- and B-array data, and when possible, A-array results are preferred for the morphological classification and further analysis.
%When available, we used the higher resolution A-array observations to determine the morphological classification.  
In summary, we categorize our sample sources into following morphological classes: 

\begin{figure}
\includegraphics[clip=true, trim = 0.3cm 0cm 0.3cm 0.3cm, width=\linewidth]{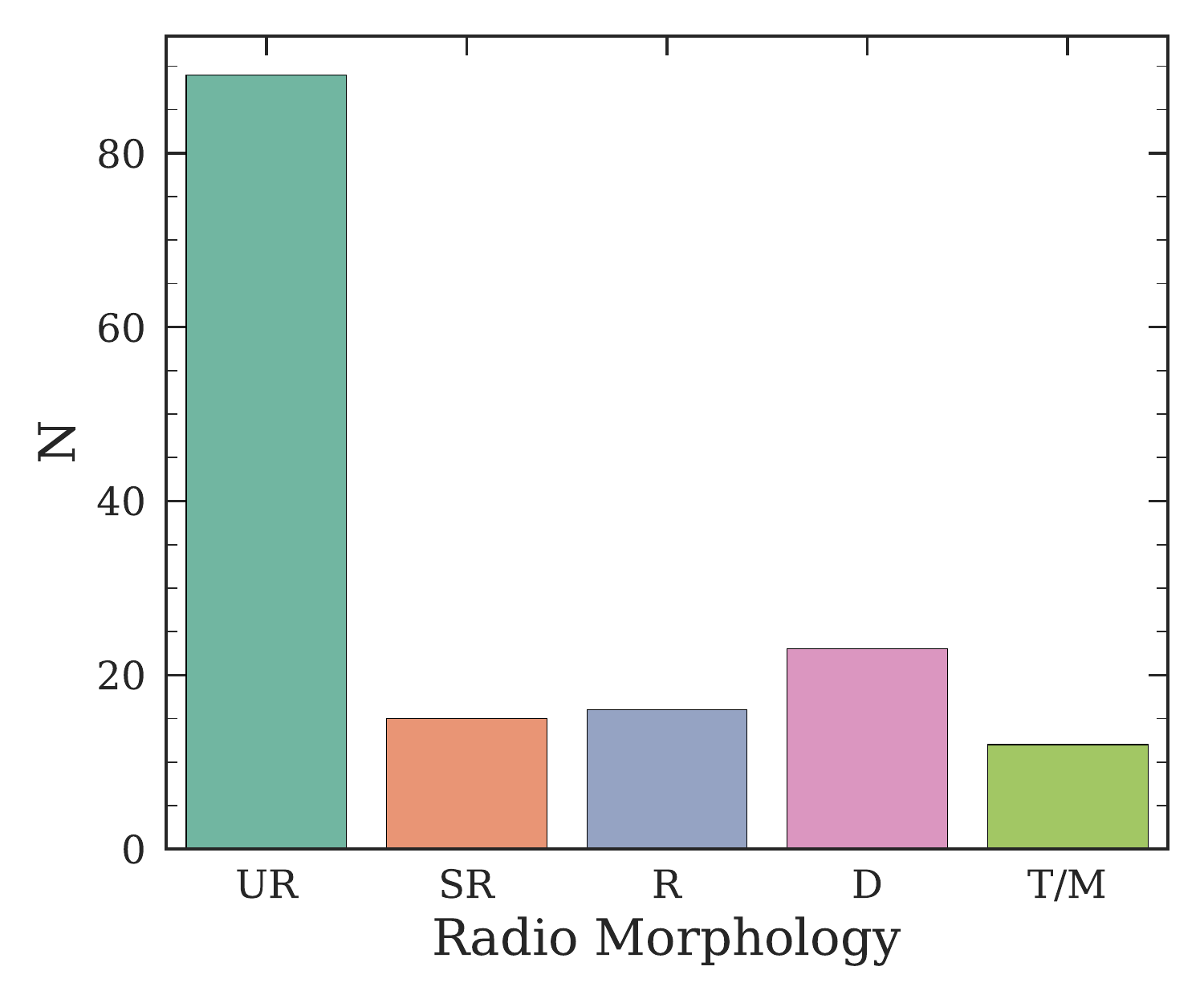}
\caption{The morphological distribution of the 155 sources from our sample. 
The six morphological classes are: UR: Unresolved, SR: Slightly resolved, R: Resolved, D: Double, T: Triple, and M: Multiple. 
Where available, A-array images are used, unless they were of poor quality.  $55.5\pm9.3$\% of the sources are unresolved, with linear extents $\leq 1.7$~kpc at $z\sim2$. 
\label{fig:morph_bar}}
\end{figure}

\begin{figure}[ht]
    \centering
    \includegraphics[clip=true, trim=0.5cm 0.4cm 0cm 0cm, width=\linewidth]{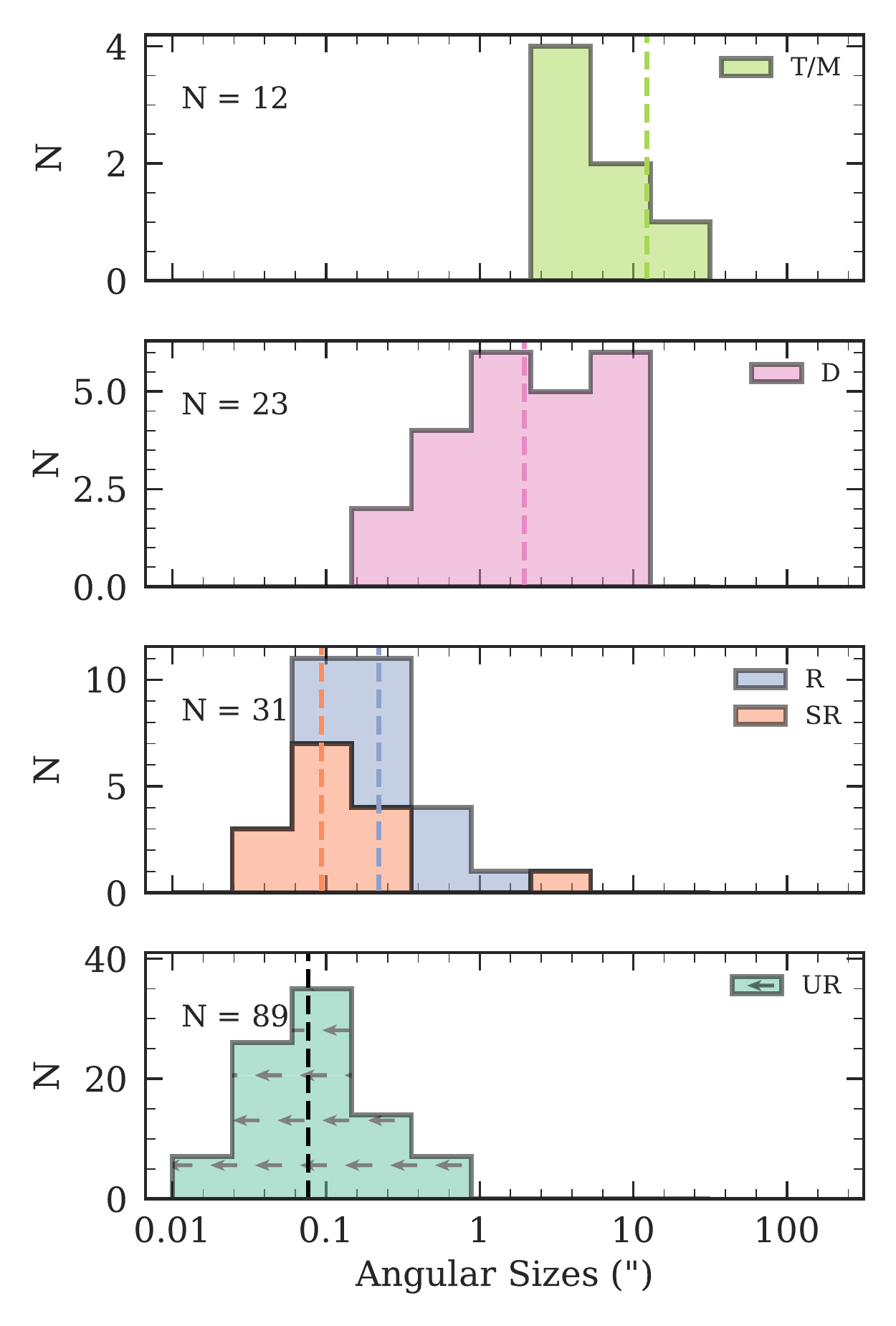}
    \caption{%The Figure shows the distribution of angular sizes broken down by morphological class. 
    The distribution of angular sizes from our new X-band observations broken down by morphological class.  The top two panels show the largest angular extents of the double (pink) and triple/multiple (light green) sources. The third panel from the top shows the angular sizes of slightly resolved (orange) and fully resolved (purple) sources. The bottom panel shows the upper limits on the source angular sizes of the unresolved sources (dark green). 
    The dashed line shown in each panel indicates the median angular size for each morphological class.}
    \label{fig:sas}
\end{figure}

\begin{enumerate}
\item {\bf Unresolved (UR):} 
The source is unresolved along both the major and minor axes and the peak/total flux ratio is unity within the 1$\sigma$ uncertainty.
\item {\bf Slightly resolved (SR):} 
The source is unresolved along one of the axes and the peak/total flux ratio is unity within the 1$\sigma$ uncertainty.
\item {\bf Fully resolved (R):} The radio source 
is resolved along both the axes, and the peak/total flux 
ratio is $<1$.

\item {\bf Double (D):} 
The source consists of two distinct components, each of which may be unresolved, slightly resolved, or fully resolved.

\item {\bf Triple (T):}  
The source consists of three distinct components, resembling the core-jet or core-lobe emission seen in large-scale radio galaxies. 

\item {\bf Multiple (M):} The source consists of more than three distinct components.
 \end{enumerate}

%\textbf{With the exception of two, we do not have any source with more than 3 components. Due to poor quality imaging, that multi-component source is excluded from our analysis.}

\begin{figure*}[!htb]

\includegraphics[clip=true,trim=0.4cm 0.3cm 0cm 0cm, width= \textwidth]{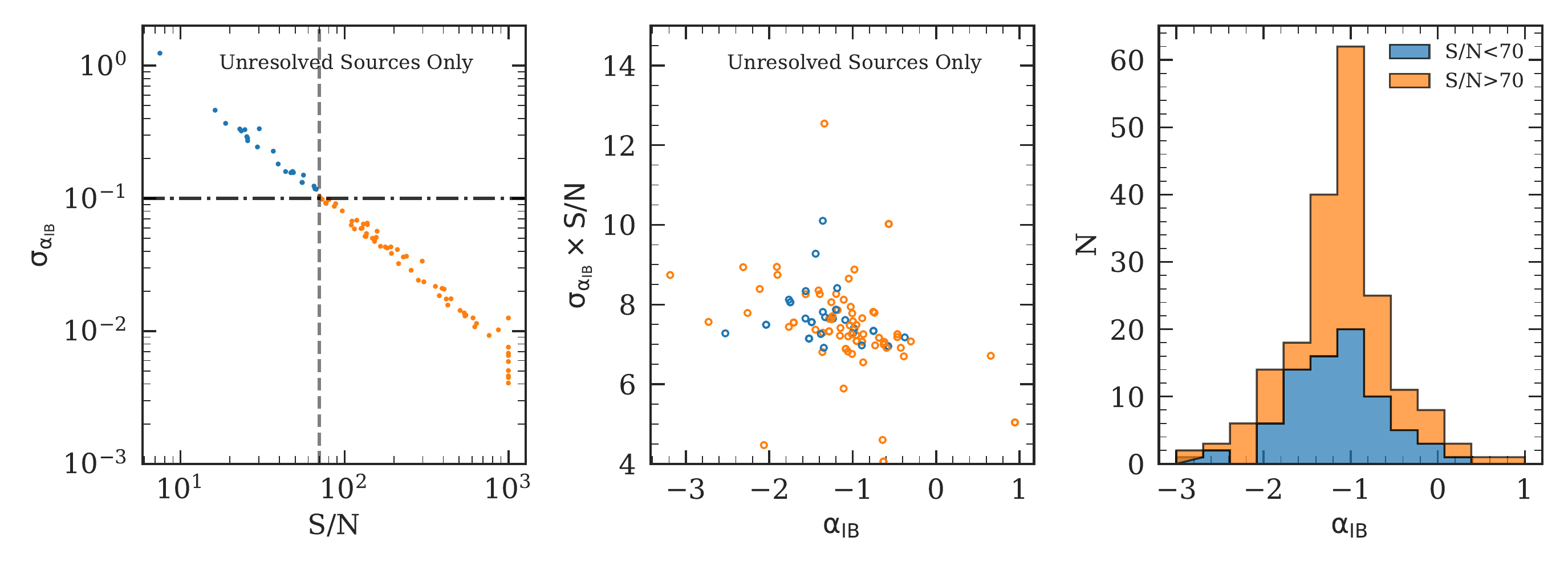}
\caption{Analysis of in-band spectral indices, $\alpha_{IB}$, and their errors, $\sigma_{\alpha_{IB}}$. \textbf{Left}: The relation between $\sigma_{\alpha_{IB}}$ and the average S/N of the 8.6 and 11.4~GHz images, evaluated by simple propagation of errors. A threshold S/N of $\sim$70 (vertical dashed line) ensures $\sigma_{\alpha_{IB}}< 0.1$ (horizontal dashed-dotted line). \textbf{Center}: The product $\sigma_{\alpha_{IB}} \times S/N$ from our simple analysis confirms a theoretical analysis by \citet{Condon+2015} that predicts a value of $\sim$8. \textbf{Right}: The distribution of measured $\alpha_{IB}$ colored according to high S/N ($>$ 70; orange) or low S/N ($<$ 70; blue).\\}
\label{fig:alphad}
\end{figure*}

Figure~\ref{fig:morph_bar} shows the 
morphological classifications of the 155 sources in our final sample. 
Expressed as percentages, $55.5\pm9.3\%$ are unresolved, $13.5\pm4.6\%$ are slightly resolved, $7.7\pm3.5$\% are fully resolved single sources, $14.8\pm4.8$\% are double,  $6.4\pm3.1$\% are triple, and $1.3\pm1.4$\% are multi-component sources.  Figure~\ref{fig:sas} shows the distribution of angular sizes for each morphological class. There is a wide range of upper limit sizes for the unresolved sources due to the large span of source declinations and the use of both A- and B-array data. Deconvolved sizes are plotted for the slightly resolved sources, and outermost peak separation sizes are given for double, triple, and multiple sources.

%Due to the high angular resolution of our new VLA X-band observations, we lack sensitivity to diffuse emission on scales larger than $5-17^{\prime\prime}$.  To determine whether diffuse radio emission is common in our sample, we compare fluxes from NVSS and FIRST for the 46 sources observed in both surveys in Figure~\ref{fig:fnvss}. These surveys were performed at 1.4~GHz with resolutions of 45$^{\prime\prime}$ and 5$^{\prime\prime}$ for NVSS and FIRST, respectively. %The median flux ratio in Figure~\ref{fig:fnvss} is 0.93 with a full range of 0.6-1.5.
%The FIRST/NVSS flux ratios shown in Figure~\ref{fig:fnvss} range from 0.6--1.5, with a median flux ratio of 0.93 and a normalized median absolute deviation of 0.1. There are a total of 9/2 sources that lie below/above the normalized absolute median deviation of flux ratios.

%As discussed in Section~\ref{sec:flux}, larger flux measurements based on higher-resolution observations may be caused by source variability or flux calibration issues. The largest outlier in Figure~\ref{fig:fnvss} has a FIRST/NVSS flux ratio of 0.33 caused by blending of two sources in the NVSS image.  Well-known issues with flux measurement reliability at low signal-to-noise ratios \citep{hopkins+15} likely contribute to the scatter at low fluxes for the remaining 8 sources with significantly less flux recovered in FIRST compared to NVSS.  However, we cannot rule-out the possibilities of a small component of diffuse emission for some sources or AGN variability. 

\subsection{In-band Spectral Indices}\label{sec:ibalph}

Our VLA X-band observations capture a wide range of frequencies, 8--12 GHz, offering the possibility of measuring ``in-band'' spectral indices, $\alpha$ (defined as $f_\nu \sim\nu^{\alpha}$). Although CASA generates a spectral index map with errors, we chose not to use it since its errors are calculated only as uncertainties to a polynomial fit and are less reliable at lower S/N \citep{cornwell+05, rau+11}. Instead, we have chosen a more classical approach to estimate the in-band spectral index and its uncertainty. 
By dividing our bandwidth into two halves (centered at $\nu_1 = 8.6$ and $\nu_2 = 11.4$~GHz), we imaged each %band 
half separately using identical 
%clean 
CLEAN parameters. 
We smoothed each 11.4~GHz image to match the resolution of the 8.6~GHz image using the task IMSMOOTH.  We then re-gridded the smoothed 11.4~GHz image using the corresponding 8.6~GHz image as a template (using the CASA task IMREGRID) to ensure matched coordinate systems in the two images.  Finally, we ran JMFIT to obtain source flux and shape measurements of all images. 

The in-band spectral index was estimated using the following equation:

\begin{equation}
%\alpha =  \frac{log_{10} (S_{\nu_1}/S_{\nu_2})}{log_{10} (\nu_1 / \nu_2)}
\alpha_{IB} =  \frac{log_{10} (S_{\nu_1}/S_{\nu_2})}{log_{10} (\nu_1 / \nu_2)}.
\label{eqn:alp}
\end{equation}
where $\nu_1$ and $\nu_2$ are 11.4 and 8.6~GHz.
%We used a simple error propagation formula to estimate the errors in the calculation as given by
Using standard propagation of errors, the uncertainty in the in-band spectral index is:

\begin{equation}
\sigma_{\alpha_{IB}}  = \frac{\big[(\sigma_{S_1}/ S_{\nu_1})^2 + (\sigma_{S_2}/ S_{\nu_2})^2 \big]^{1/2}}{log_{10}(\nu_1 / \nu_2)}.
\label{eqn:sigalp}
\end{equation}
%\textcolor{red}{We calculated parameter uncertainties using Monte-carlo simulations. The errors  }
The left panel in Figure~\ref{fig:alphad} shows the resulting uncertainty, $\sigma_{\alpha_{IB}}$, plotted against the average S/N of the 8.6 and 11.4~GHz images.  As expected, lower S/N yields larger uncertainties in $\alpha_{IB}$ with a threshold of S/N 
%$> \sim$70 for $\sigma_{\alpha_{IB}}< \sim0.1$,
$\gtrsim 70$ for $\sigma_{\alpha_{IB}} \lesssim 0.1$, which we take as a threshold of reliability for the calculated values of $\alpha_{IB}$.
%an error on $\alpha$. the uncertainty in $\alpha_{IB}$.

%The error propagation formula makes it clear that the uncertainties are dependent on the ratio of two-frequencies $\nu_1/\nu_2$ known as ``lever arm'' \citep{Condon+2015} and the S/N of the image ($S/\sigma_S$). %Therefore, we expect the errors to be more significant for faint sources. As a cross-check, we compare the in-band spectral indices derived using the classical method with the estimates provided by CASA in Figure~\ref{fig:alphad}.
%A study by \citet{Condon+2015} quantified  the accuracy of in-band spectral indices as a function of the S/N. The errors on the spectral index parameter are given by Equations 48 and 49 in \citet{Condon+2015}. The source sensitivity is dependent on the the intrinsic spectral index. For the case of pointed observations, when $\alpha = -0.5$ these equations simplify to 

\citet{Condon+2015} gives a theoretical analysis of in-band spectral indices and 
%its 
their uncertainties %which 
that broadly confirms our simple approach above. Combining %his Equations 48 and 49, for a spectral index near $\alpha_{IB}$ and spectral range 8--12~GHz, we find:
Equations~48 and 49 from \citet{Condon+2015} for an in-band spectral index $\alpha_{IB}$ over a bandwidth of 8--12~GHz, we find: 

\begin{equation}
%\sigma_\alpha \times S/N = \frac{\sqrt{12}}{\textrm{ln}(\nu_{max}/\nu_{min})}
\sigma_{\alpha_{IB}} \times S/N = \frac{\sqrt{12}}{\textrm{ln}(\nu_{max}/\nu_{min})}\,\, \sim\,\, 8 \end{equation}
where, $S/N$ is the signal-to-noise ratio of the %source image, 
source and %$nu_{max}$ 
$\nu_{max}$ and $\nu_{min}$ are the upper and lower %end 
ends of the observing bandwidth. The center panel in Figure~\ref{fig:alphad} shows the product %$\sigma_{\alpha_{IB}} \times S/N$ 
$\sigma_{\alpha_{IB}} \times S/N$ for our data, and broadly confirms this result, with values near 7--8 for a range of in-band %spectral index, $\alpha_{IB}$.  
spectral indices.

% \begin{figure}[t!]
%     \centering
%     \includegraphics[width=\linewidth]{Flux_NVSS_v4.pdf}
%     \caption{%The plot shows 
%     The ratio of fluxes measured 
%     %at 
%     in FIRST and NVSS as a function of NVSS flux.  The black dotted line indicates a ratio of unity. For the majority of our sample, FIRST is able to recover most of the flux measured by NVSS. The gray shaded region shows the normalized  median deviation ($\sigma_{\rm nmad} \sim0.1$) of the flux ratio. }
%     \label{fig:fnvss}
% \end{figure}

The %right panel in 
far-right panel of Figure~\ref{fig:alphad} shows the distribution of in-band %spectral index, $\alpha_{IB}$, 
spectral indices with values above/below our S/N threshold color coded as orange/blue.  The distribution is strongly peaked near the median value of $\alpha_{IB}$ = $-1.0$, with 80\% of the high-quality values within the range $-1.7$ to $-0.5$. We will discuss these spectral indices, together with the overall radio SEDs in a companion paper (Patil et al. in prep.). Briefly, the median spectral index is broadly consistent with optically thin synchrotron emission ($\alpha\sim-0.7$ near 1~GHz; e.g., \citealt{condon+16}), perhaps steepened somewhat via radiative losses as well as inverse Compton scattering from either the Cosmic Microwave Background or local infrared radiation fields. 
About $5\%$ of our sources might plausibly have a flat spectrum, consistent with an unresolved synchrotron core.
%be classified as flat spectrum, confirming that even our spatially unresolved sources are physically extended beyond the regions typically associated with self-absorbed cores.
This is also consistent with the absence of evidence for short timescale variability typical of beamed sources, indicated by the good overall agreement between the fluxes measured in our A and B configuration observations. We will address the spectral characteristics and the role of beamed core emission more thoroughly in the SED paper.

\section{Source Properties}\label{sec:properties}

\subsection{Diffuse Radio Emission?}\label{sec:extended}
%\subsection{Possible Diffuse Radio Emission?}\label{sec:extended}
Our sample was selected to have compact emission in the NVSS and FIRST catalogs. As discussed in Section~\ref{sec:smeasure}, the majority of our sources have compact morphologies in our new high-resolution X-band observations.  However, the presence of diffuse, extended emission on scales of a few arcseconds (which could be associated with earlier episodes of AGN activity) cannot be definitively ruled-out on the basis of the X-band data alone due to surface brightness sensitivity limitations. 

\subsubsection{Constraints from Radio Surveys}
To check on the incidence of such extended emission, we visually inspected images of all of our sources in NVSS and FIRST as well as two additional wide-field radio surveys: The GMRT Sky Survey (TGSS: \citealt{intema+17}) and the VLA Sky Survey (VLASS\footnote{We inspected the VLASS Epoch~1 ``quicklook" images available at \url{https://archive-new.nrao.edu/vlass/quicklook/}.  We caution readers that these images are preliminary only - higher quality survey products will be publicly available in the future, as discussed in \citet{lacy+19}.}; \citealt{lacy+19}).  The observing frequency, angular resolution, maximum resolvable scale, and 1$\sigma$ sensitivity for these surveys is summarized in  Table~\ref{tab:surveys}, along with similar information for our X-band observations.  The combination of our new X-band data with lower-resolution radio surveys provides a more complete picture of the radio morphologies of our sources, thus allowing us to constrain the presence of diffuse, extended emission. 

\begin{deluxetable}{cccccc}
\tablecaption{List of Radio Continuum Surveys. \label{tab:surveys}\\
Column 1: Name of the radio survey; Column 2: Frequency of the observation in GHz; Column 3: Typical angular resolution of the survey in arcseconds; Column 4: Largest resolvable angular scale  in arcseconds; Column 5: 1$\sigma$ rms noise in mJy/beam; Column 6: Number of our sources observed in each survey}
%\tablewidth{700pt}
\tablehead{
\colhead{Survey} & \colhead{$\nu$}  & \colhead{$\theta_{res}$} & \colhead{LAS} & \colhead{$\sigma_{rms}$}   & \colhead{$n_{sources}$}   \\
\colhead{} &  \colhead{GHz} &  \colhead{$^{\prime\prime}$} & \colhead{$^{\prime\prime}$} & \colhead{mJy/beam} &    \colhead{} 
} 
\colnumbers
\startdata
TGSS ADR1 & 0.15  & 25 & 4104 & 3.5   & 152 \\
NVSS & 1.4  & 45 & 970 & 0.45 &  155  \\
FIRST & 1.4 & 5 & 36 & 0.15 & 51 \\
VLASS & 3 & 2.5 & 58 & 0.12 & 153 \\
X-band-B & 10 & 0.6 & 17 & 0.03 & 149 \\
X-band-A & 10 & 0.2 & 5.3 & 0.03 & 129 \\
\enddata
\end{deluxetable}

 We re-confirmed that all of our sources are indeed compact in NVSS.  For the 51/155 sources included in the FIRST survey footprint, we inspected the FIRST images and found 6 sources that appear compact in NVSS but are either resolved into 2 distinct components or extended in FIRST.  In all six of these cases, the multiple components identified in FIRST appear to be associated with radio AGN jets/lobes. We provide a further comparison of the NVSS and FIRST properties of our sources in terms of their fluxes in Section~\ref{sec:NVSSFIRST}.

TGSS, which provides a factor of two higher angular resolution than NVSS and a much lower frequency of 150~MHz, is more sensitive to steep-spectrum emission from older radio sources.  We found a total of 15 sources with clearly resolved, extended emission and 3 sources with multiple components in TGSS.  %One source, J0010+16, is compact in NVSS but appears to consist of a double lobe + core morphology in TGSS. . . . discuss how its a candidate re-started AGN . . . mention limitations of sensitivity . . . 
Finally, we examined the 3~GHz VLASS images of our sources, which have two times higher resolution than FIRST.  We found 13 sources with extended morphologies and 8 sources with multiple components.  

Ultimately, the TGSS, FIRST, and/or VLASS images revealed extended or multi-component emission in a total of 25/155 unique sources.  Of these, 11 sources were not previously classified as being resolved in our X-band observations, thus leading to the re-classification of their morphologies.  A summary of the properties of all sources with resolved emission identified in radio survey images is provided in Table~\ref{tab:extended} and image cutouts are shown in Figure~\ref{fig:ext_source}. 
Thus, we conclude that the majority of our sources are indeed compact, even when observed at lower frequency and at lower resolution.
We emphasize that the discovery of extended emission only has an impact on our study by modifying our morphological classification and possibly indicating a prior episode of activity. However, the presence of more extended emission does not affect our primary analysis of the more compact central radio source. It is these sources that we are most interested in because they are likely to be associated with the denser gas responsible for the high dust column and high MIR emission.

\subsubsection{NVSS and FIRST Flux Ratios}
\label{sec:NVSSFIRST}
\begin{figure}
    \centering
    \includegraphics[width=\linewidth]{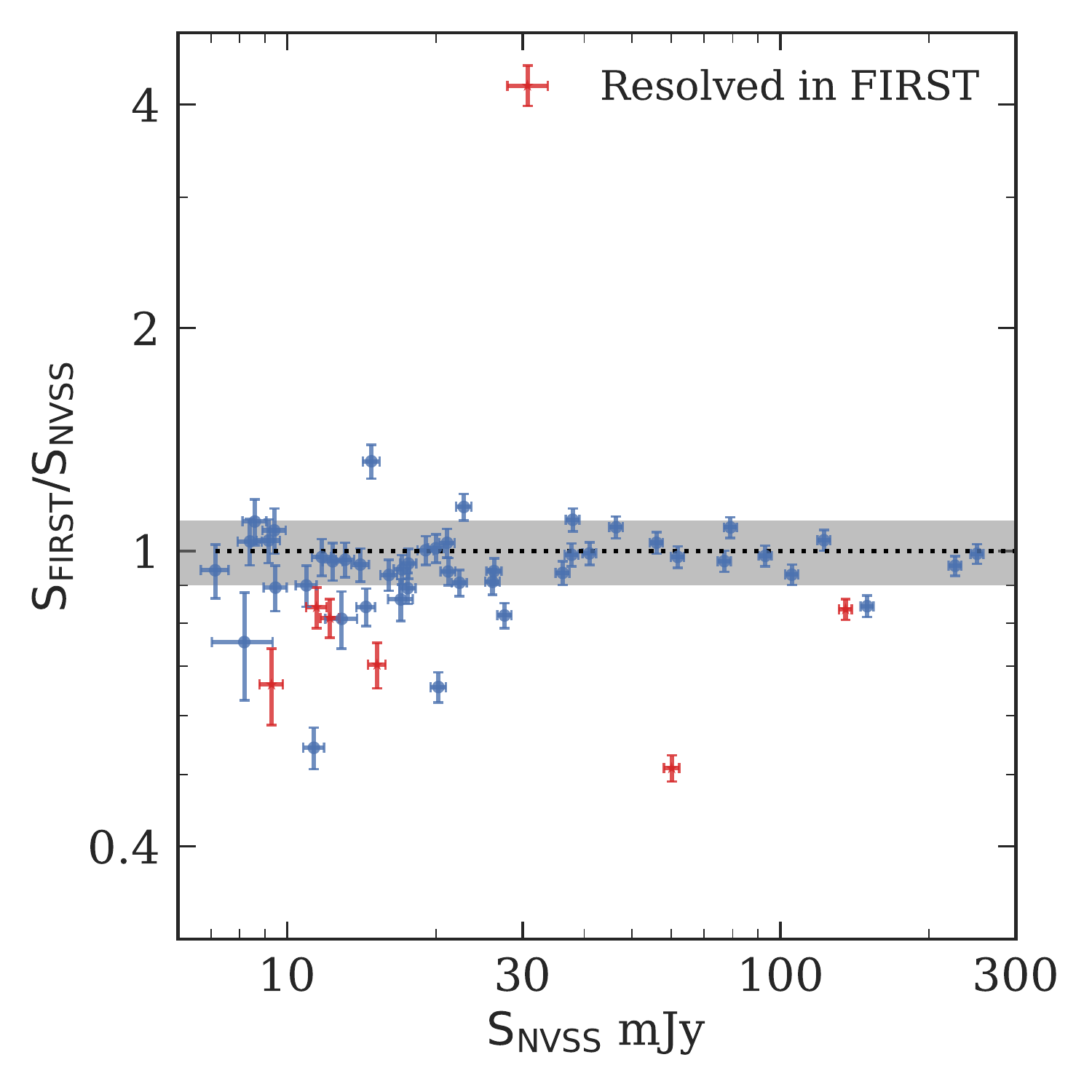}
    \caption{%The plot shows 
    The ratio of fluxes measured 
    in FIRST and NVSS as a function of NVSS flux.  The black dotted line indicates a ratio of unity. For the majority of our sample, FIRST is able to recover most of the flux measured by NVSS. The gray shaded region shows the normalized  median deviation ($\sigma_{\rm nmad} \sim0.1$) of the flux ratio. Six sources with  resolved morphologies in the FIRST are shown by the red symbols. 
    %We used total flux measurements from FIRST catalog for those six sources.
    \label{fig:fnvss}}
    
\end{figure}

As a further test for missed emission in our X-band observations, we compare in Figure~\ref{fig:fnvss} the 1.4~GHz NVSS and FIRST fluxes of our sources.  Excluding six sources that are resolved in FIRST but not in NVSS (J1025+61, J1138+20, J1428+11,  J1651+34,  J2145$-$06, J2328$-$2), the fluxes are in good agreement above 30~mJy
with slight ($\sim 5\%$) scatter to lower FIRST fluxes for weaker sources, 
with two outlier sources, J2322$-$00 and J1717+53, with flux ratios of 0.54 and 0.65 respectively. Neither of these sources shows any extended emission in TGSS, VLASS or FIRST, and since both NVSS and FIRST were corrected for ``CLEAN Bias'', it cannot explain the offsets. We note that other sources of bias exist for measurements at low S/N (e.g., \citealt{hopkins+15}). Variability might explain some of the outliers \citep{mooley+16}, though we emphasize that Figures~\ref{fig:fluxd} and ~\ref{fig:fnvss} indicate that the majority of our sources are not likely to be variable on the timescales sampled by our data.

\begin{figure}[ht!]
\centering
\includegraphics[clip=true, trim=13.1cm 0.3cm 0.3cm 0cm, width=0.45\textwidth]{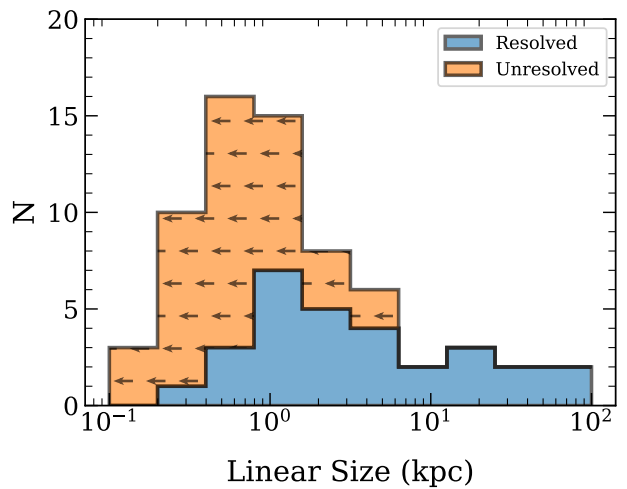}
\caption{  Linear sizes for %those 71 targets having 
the 71 sources with spectroscopic redshifts. We plot two separate histograms for the two broad morphological categories, 
resolved and unresolved. The blue histogram shows the largest linear extents for the resolved sources in our sample. The orange histogram with left arrows are the upper limits on the linear extents of unresolved sources and is stacked on top of the blue histogram.  \\}
\label{fig:linsiz}
\end{figure}

\subsection{Physical Sizes}\label{sec:linsize}
Figure~\ref{fig:linsiz} shows the distribution of physical source sizes, with the sample divided into resolved 
(including both slightly and fully resolved) and unresolved source morphologies. With the exception of 12 double or triple sources larger than 10~kpc, the rest are smaller than 5~kpc. Roughly 55\% of the sources are unresolved with median upper limit near 0.6~kpc. 
Given that our radio selection only requires sources to be compact on 40$\arcsec$ scales (NVSS, 100\% of the sample) or 5$\arcsec$ scales (FIRST, 30\% of the sample) we find essentially all our sources are significantly more compact  than these size limits, suggesting our joint selection with luminous and red \wise~ MIR emission is preferentially associated with compact radio sources. A further check of whether the MIR selection is associated with compact radio emission is to ask whether an MIR blind radio survey with similar flux threshold and redshift range yields many compact sources. 

Such a survey exists. The CENSORS sample of \citet{best+03} used NVSS to select sources brighter than 7.8 mJy and  cross-matched these with the ESO Imaging Survey (EIS). The resulting sample of 150 has similar median redshift and radio luminosity to our sample. However, the median radio source size for the CENSORS sample is 6$\arcsec$, which is significantly larger than our own median source size of $0.1-0.2\arcsec$.  Since the redshift distribution and flux cut for the two samples is similar, then we conclude that the smaller source size of our sample is tied to the additional selection criteria of extreme MIR colors and luminosities.

Having established that our radio sources are compact, are there any previously established classes of radio sources that closely resemble our sources? Clearly they are different from the classical Fanaroff-Riley (FR) type I and II \citep{fanaroff+74} radio sources which are much more extended. Similarly, our sources, with their steep spectral index (Section~\ref{sec:ibalph}), are also different from the compact flat spectrum sources. There are four known classes of steep spectrum radio sources that approximately match the angular and physical scales of our sample. These are the GPS (Gigahertz Peaked Spectrum; e.g., \citealt{fanti+90, odea+91, snellen+98,fanti+09, collier+18} ), CSS \citep[e.g., ][]{peacock+82, spencer+89, fanti+90, sanghera+95, fanti+01}, HFPs (High Frequency Peakers e.g., \citealt{dallacasa+00,stanghellini+09, orienti+14}), and FR0 classes (e.g., \citealt{baldi+18, sadler+14}). Of these, the FR0 class is significantly less luminous  ($<10^{24}$ W/Hz; \citealt{baldi+18}) and while the available GPS/CSS samples are somewhat more luminous than our sample (see next section), an SED analysis (Patil et al. in prep.) confirms that a significant fraction of our sources have curved or peaked spectra in the GHz range, similar to the GPS/CSS sources. Thus, since our sample seems to share a number of properties with the GPS/CSS sources, we will use these as a point of comparison in the following discussion.

\begin{figure*}[ht!]
\centering
\includegraphics[clip=true, trim=0cm 0cm 0cm 0cm, width=\textwidth, angle =0]{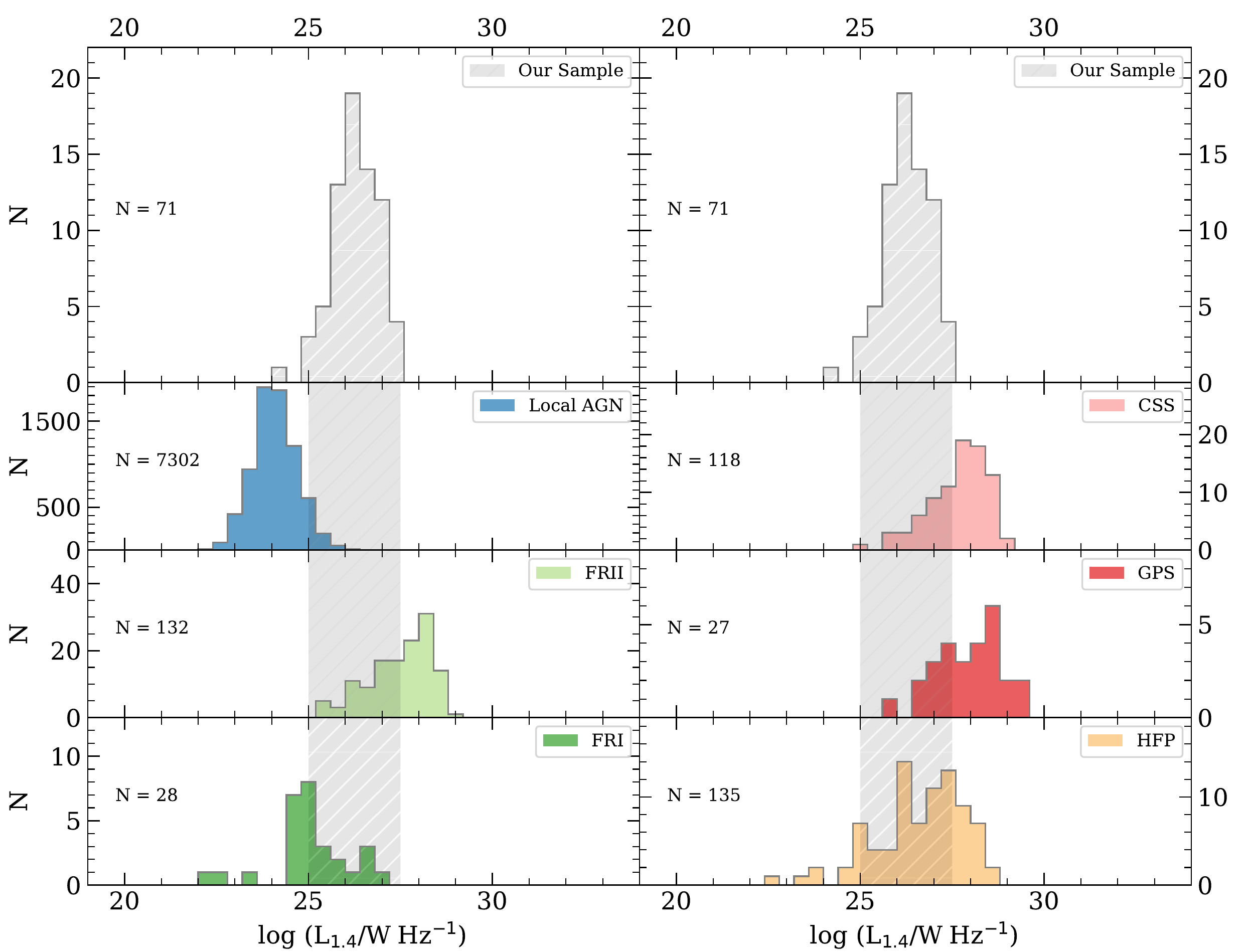}
\caption{Comparison of spectral radio luminosity at 1.4~GHz with other well-studied luminous radio source populations. The top panel shows the distribution of radio luminosities in our sample. The samples plotted in the left-hand panels are local radio AGN ($z<0.7$),  FRI, and FRII galaxies, respectively. The right-hand panels show compact radio AGN; CSS, GPS, and HFP, respectively. The total number of sources in each category 
is shown in the top left corner of the plot. The range of spectral luminosities for our sample  
is shown by the gray hatched area. 
The references for 
each source population are as follows: 
SDSS Local Radio-loud AGN: \citet{best+12}; 
FRI and FRII: \citet{laing+83}; %H$z$RGs: \citet{debreuck+10}; 
CSS and GPS:\citet{odea+98,sanghera+95,spencer+89,fanti+01}; HFP: \citet{dallacasa+00,stanghellini+09}. 
}
\label{fig:Lnvss}
\end{figure*}

%\subsection{ Comparing Radio Luminosities}
\subsection{Radio Luminosities}\label{sec:rlum}

%To help frame the nature of our galaxies, we compare the radio luminosities of our sample with other well-known luminous radio sources. In Figure~\ref{fig:Lnvss},  we present the distribution of radio luminosities at 1.4~GHz (with fluxes taken from NVSS) for the subset of our sample with measured redshifts. The distribution spans Log L$_{1.4}$ (W/Hz) $\sim25 - 27.5$ with median Log L$_{1.4} \sim 26.2$. 

Figure~\ref{fig:Lnvss} presents the 1.4 GHz radio luminosity of our sample, which spans the range $25 \lesssim \log(L_{1.4\,\rm GHz}/{\rm W~Hz}^{-1}) \lesssim 27.5$, with a median of $\log(L_{1.4\,\rm GHz}/{\rm W~Hz}^{-1}) \approx 26.3$. 
We also use Figure~\ref{fig:Lnvss} to compare with other well-known samples of radio AGN to help place our own sample within a wider ``zoo'' of radio sources. 

%This is not intended as a rigorous comparison because the selection criteria for the other samples are quite varied, but it does help frame our own sample in the ``zoo" of 
%various radio AGN populations.
A representative sample of local ($z<0.3$) radio AGN was presented by \citet{best+12} who cross-matched NVSS and FIRST sources with SDSS (radio luminosities calculated assuming a spectral index of $-0.7$). Clearly, our sample is roughly 2 dex more luminous than the local sample, confirming that our sample is much more luminous than the typical local radio AGN. 

Next we compare with the well-known low-frequency 3CRR survey, which is complete above S$_{178\,\rm{MHz}}$ = 10.9 Jy \citep{laing+83}. These span a wide range of redshift and luminosity, and broadly divide into large scale FRI and FRII radio sources \citep{fanaroff+74}. Our sample is, on average, 1.4 dex less luminous than the FRIIs and 1.3 dex more luminous than the FRIs, though there is considerable overlap with both these samples.

Turning to radio sources that are, perhaps, better matched to the redshifts and  physical scales of our own sources, the right side of Figure~\ref{fig:Lnvss} includes samples of CSS and GPS sources \citep{odea+98, sanghera+95, spencer+89, fanti+01} and HFP sources \citep{dallacasa+00, stanghellini+09}. These samples show considerable overlap, though the median luminosities of the CSS, GPS, and HFP samples are larger by $\sim$1, 1.8, and 0.5 dex, respectively.

Overall, then, while our sample is significantly more radio luminous than typical radio AGN, it has intermediate luminosity when compared to samples of powerful radio-loud AGN.

%\textbf{The overall radio luminosities of our sample probe the lower luminosity end of the compact radio AGN and are intermediate to that of FRI and FRII. 
%The radio luminosities of optically selected radio quasars also similar to our sample \citep{condon+13}. 
%}

%\begin{figure}
%    \centering
%    \plotone{Lum_func_v1.pdf}
%    \caption{We plot Radio Luminosity Function (RLF) for our sample sources along with three comparison samples.}
%    \label{fig:RLF}
%\end{figure}

\subsection{ Radio Lobe Pressures }\label{sec:pressure}

%An important aspect of any radio source is how it develops over time, and that in turn depends not only on the energy input via jets, but also the nature of the surrounding medium into which the jets deposit their energy and momentum. Both of these considerations affect how the lobes expand and on what timescale. In observational terms, one of the ways to access these issues is through the pressure in the radio source. To first order, the measured pressure likely reflects the pressure of the surrounding medium into which the radio source is expanding. If independent measures of this surrounding pressure are possible, then one can additionally ask whether the lobes are over- (or under-) pressured relative to their surrounding medium, and hence gain some insight into the nature of the lobe expansion. 

An important property of a radio source that affects how it develops is its internal pressure. To first order, the measured pressure likely reflects the pressure of the surrounding medium into which the radio source is expanding. If the radio source is over-pressured relative to the surrounding medium, perhaps being fed by a nuclear jet, then the radio source will expand.

%\textbf{Accurate measurements of energetics and dynamics of the radio sources  help us understand the impact of the expanding lobe on its surrounding medium (e.g.,), which ultimately can be used in a broader context to study the scaling relation between jet kinetic power and radio source luminosity \citep{godfrey+16} and jet-mode feedback \citep{}. Usually, for the calculations of lobe parameters the equipartition assumption is used where the mininum energy condition is satisfied between the relativistic particles and the magnetic field \citep{burbridge+56}. }

To estimate the internal lobe pressures in our sample sources, we use relations derived from synchrotron theory 
given in \citet{moffet+75} and \citet{miley+80}:
\begin{equation}
P_{l} \approx (7/9) (B^{2}_{min}/8\pi),
\end{equation}
where $P_{l}$ is the pressure in the lobe of a radio source in dyne cm$^{-2}$ and $B_{min}$ is the magnetic field in the magnetoionic plasma in Gauss, derived using the common ``minimum energy'' or ``equipartition'' assumption that energy is shared approximately equally between the particles and the magnetic field. The equation for this  magnetic field strength in Gauss can be written: 
\begin{equation}
\label{eqn:pres}
B_{min} \approx 2.93 \times 10^{-4}
  \Bigg[ \frac{a}{f_{rl}}\frac{(1+z)^{4-\alpha}}{\theta_{rx}\theta_{ry}} 
    \frac{S_\nu}{\nu^{\alpha}}    \frac{X_{0.5}(\alpha)}{\theta_{ry}r_{co}}\Bigg]^{2/7},
\
\end{equation}
% \begin{align}\label{eqn:pres}
%     B_{min} \approx & 2.93 \times 10^{-4} \\ 
%      & \Bigg[ \frac{S_\nu}{\theta_{rx}\theta_{ry}} \, \frac{a(1+z)^{4+\alpha_r}\nu^{\alpha_r}_{\textrm{GHz}} X_{0.5}(\alpha_r)}{f_{rl}\theta_{ry}r_{co}}\Bigg]^{2/7},
% \end{align}
where the radio source has flux $S_\nu$ in Jy with spectral form $S_\nu \propto \nu^{\alpha}$ and angular size $\theta_{rx} \times$  $\theta_{ry}$ arcsec, $z$ is the redshift of the source, and $r_{co}$ is the comoving distance in Mpc. We choose the filling factor for the relativistic plasma, $f_{rl}$, and the relative contribution of the ions to the energy, a, to be 
%1 and 2 respectively, the function 
1 and 2, respectively.  The function 
$X_{0.5}(\alpha)$ handles integration over the 
%spectrum between frequencies $\nu_1$ = 0.01~GHz and $nu_2$ = 100~GHz: 
frequency range from $\nu_l$ to $\nu_h$, where $\nu_l$ = 0.01~GHz and $\nu_h$ = 100~GHz, and is defined as: 

\begin{equation}
    X_{q}(\alpha) = (\nu^{q+\alpha}_{2} - \nu_{1}^{q+\alpha})/(q+\alpha),
\end{equation} 
where q is 0.5 in this case and represents the spectral shape function of the synchrotron emission.

\begin{figure}[t!]
\centering
\includegraphics[clip=true, trim=0.3cm 0.3cm 0.3cm 0.2cm, width=\linewidth]{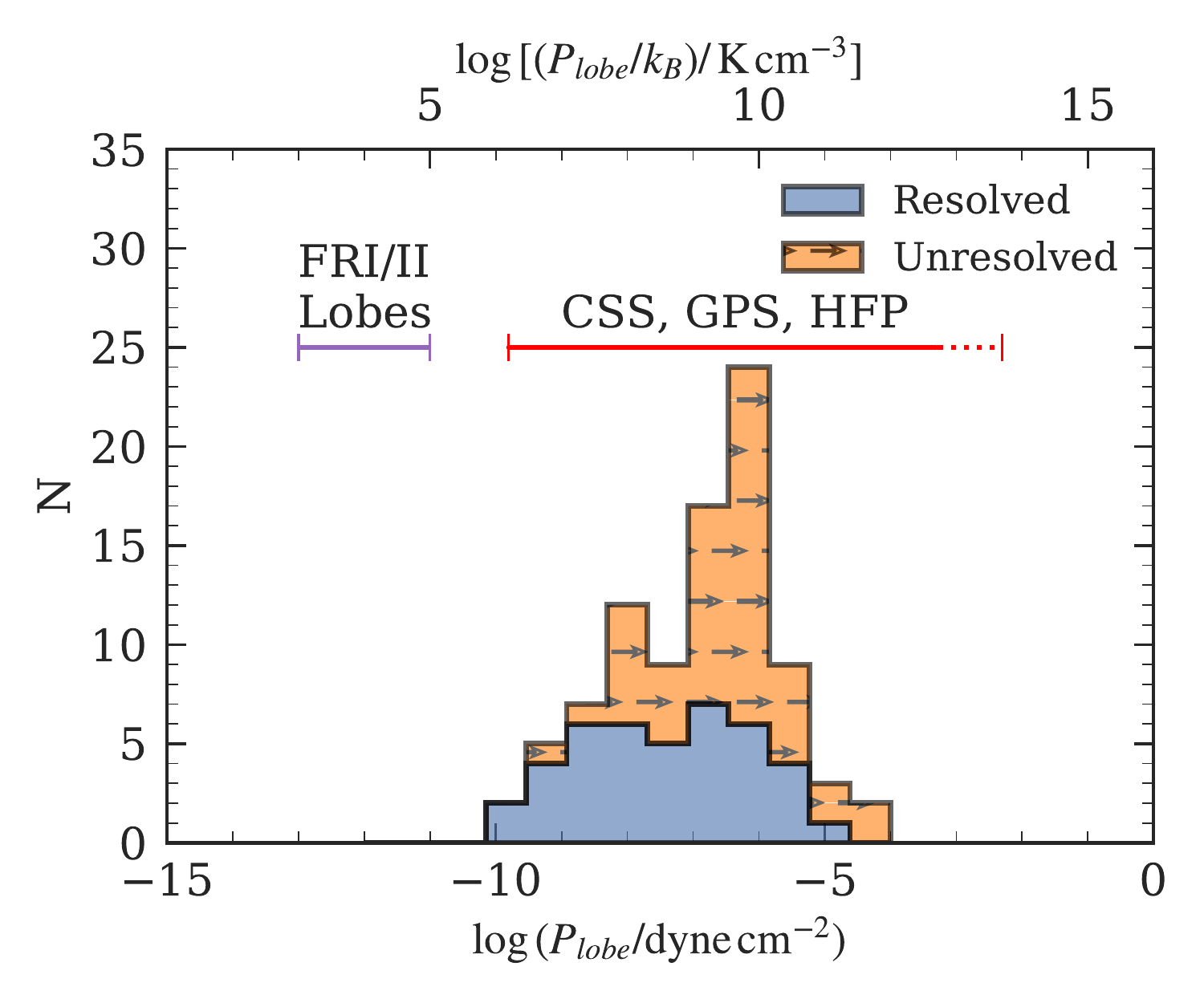}
\caption{The distribution of radio source pressures for our sample, with lower limits for spatially unresolved sources shown as arrows. The orange histogram is stacked on top of  the blue histogram. Also shown are typical ranges of source pressures for other classes of radio AGN (see text for references to the data that were used to generate these ranges). 
\label{fig:pressure}}
\end{figure}

Knowledge of the source size is required, since it feeds directly into the estimate of source pressure. For resolved single, double or triple sources we take the measured region sizes directly from JMFIT. For slightly resolved or unresolved sources we take a conservative approach and use the beam major axis as an upper limit to source size. This yields a conservative \textit{lower limit} for the source pressure. Higher resolution Very Long Baseline Array  (VLBA) images for a number of the unresolved sources (Patil et al., in prep.) usually reveal yet smaller scale double lobes with yet higher pressures. Thus, our current treatment of the VLA images yields useful, though conservative, lower limits to the radio source pressures in the unresolved sources. 

Figure~\ref{fig:pressure} shows the distribution of pressures for our sample, with lower limits for the unresolved sources. For the resolved sources, 
the median pressure is $\log(P_{l}/(\textrm{dyne~cm}^{-2}))=$ $-7.2$  or ( $\log[(P_{l}/k_B) / (\textrm{cm}^{-3} \textrm{K})]$ = +8.7). For the lower limits, these values are
$\log(P_{l}/(\textrm{dyne~cm}^{-2}))=$ $-6.3$  or ($\log[(P_{l}/k_B) / (\textrm{cm}^{-3} \textrm{K})]$ = +9.5).

To help  put our sample in context, 
we include the typical range of equipartition lobe pressures for a number of other classes of radio AGN.  On larger scales, the lobe pressures in FRI \citep[e.g., ][]{worrall+00,croston+08, croston+14} and FRII \citep[e.g., ][]{croston+05,  ineson+17, harwood+16, vaddi+19} radio galaxies are roughly 3 dex lower than our sample, almost certainly reflecting the much lower ambient pressures found on larger scales in the circumgalactic environment.
%\textbf{Range of pressures values are taken directly from the references given above.}

Figure~\ref{fig:pressure} also shows the range of equipartition lobe pressures for CSS, GPS, and HFP sources taken directly from various studies \citep{mutel+85, readhead+96b, orienti+14}. 
%For the CSS and GPS sources, the pressures were evaluated using their largest angular sizes and flux at 1.4~GHz from NVSS. \textbf{We assume the }\textbf{The pressures for the HFPs were calculated from the equiparition magnetic field estimates given in \citet{orienti+14} for four sources.} 
There is a considerable overlap between our source pressures and those of the CSS, GPS and HFP samples, possibly indicating a similarity in their properties and stage of development. However, a detailed comparison with these young radio AGN is not straightforward because most measurements for the CSS, GPS and HFP sources come from Very Long Baseline Interferometry (VLBI) observations with $\sim$milliarcsecond-scale angular resolution
capable of identifying much more compact radio structures.  Indeed, preliminary analysis of our own VLBA follow up survey shows that many of our unresolved sources also have more compact source components with significantly higher pressures ($\sim 1-3$ dex, Lonsdale et al. in prep.). In all these comparisons, we have verified that our approach to measuring source pressures reproduces the source pressures given in these other papers. 
%Indeed, many of our unresolved VLA sources are resolved by the Very Long Baseline Array (VLBA), extending the pressure distribution to significantly higher pressures. The results of our VLBA follow-up imaging will be presented in a forth-coming study (Lonsdale et al. in prep).

%\textbf{Observations have shown that the equiparition assumption doesn't always hold true (e.g.,) and the jet-power and source environment determine the radio lobe pressures \citep{croston+18}.    }
%In more detail, it has been observationally shown that lobes of FRIIs are over-pressured at their outer hotspots but otherwise roughly pressure-balanced with the ambient medium \citep{harwood+16, harwood+17b}, which was later supported by simulations \citep{turner+18}. Whereas, the lobes in FRIs  may  be  systematically  under-pressured,  possibly reflecting entrainment of non-radiating particles in this population \citep{croston+18}. 
%\textbf{Similar analysis would require estimation of ISM pressures via ALMA data  }

The compact nature of the radio sources, together with their implied high pressures seem to be a characteristic of the sample, and it is important to understand the origin of these high pressures. Unlike the lobes of extended FRI/FRII radio galaxies, the location of our radio sources deep within
the host galaxy means they are embedded within the relatively high-pressure environment of the central $\sim$1~kpc region. If the radio sources are in fact over-pressured relative to the ambient ISM, then that over-pressure may generate an expansion which, when coupled to the small size, may indicate a young source.  We will present a more quantitative analysis of the source pressures and ages in Section~\ref{sec:ages} when we use a simple model of jet-driven lobe expansion to fit the observed source sizes and pressures.

\begin{figure*}[t!]
\centering
\includegraphics[clip=true, trim=0.3cm 0cm 0cm 0cm, width=0.9\textwidth]{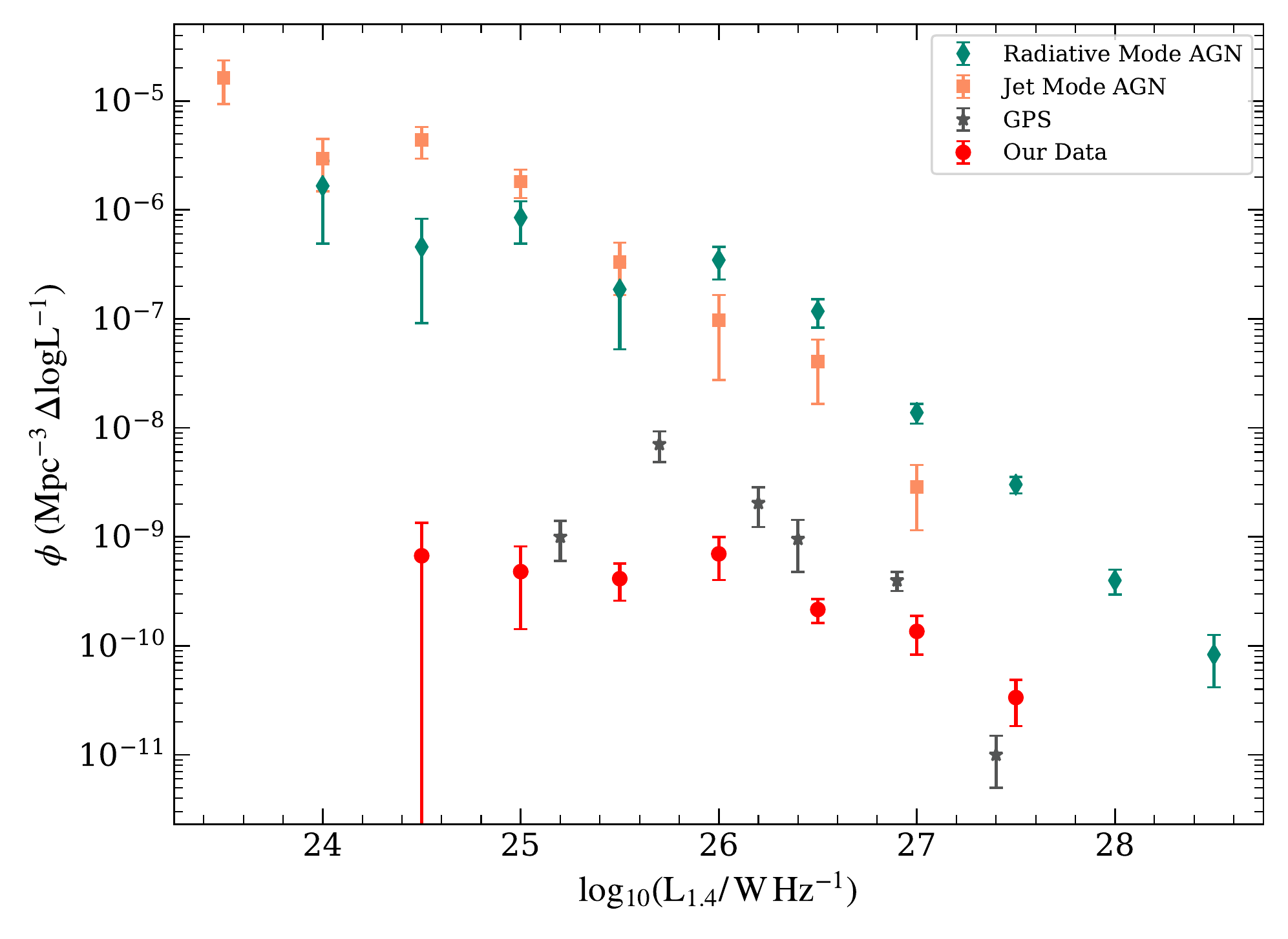}
\caption{Radio luminosity function (RLF) of our sample (red filled circles).  For comparison, we have also plotted the RLFs of the populations of radiative- and jet-mode AGN (green diamonds and orange squares, respectively) from \citet{best+14} as well as young radio AGN (gray asterisk) from the GPS samples presented in \citet{snellen+00} .\\}
\label{fig:RLF}
\end{figure*}

\section{Radio Luminosity Function}\label{sec:lumfunc}
%One way to quantify the relative abundance of a class of astronomical object is to calculate its luminosity function.
The radio luminosity function (RLF) measures the number of radio sources per dex of radio continuum luminosity per co-moving Mpc$^3$ \citep[e.g., ][]{condon+02}. To calculate the RLF for our sample, we use the standard 1/$V_{max}$ method \citep{schmidt+68}, which sums the space density for each source using a total volume within which that source could have been detected, given our sample selection criteria.

As described in Section~\ref{sec:sample}, our sample selection is somewhat complicated and involves a combination of cuts in  radio flux and source size as well as infrared fluxes and colors.  We therefore defined $V_{max,i}$ for the $i^{th}$ source as:

\begin{equation}
    V_{max,i} = \int_{z_{min,i}}^{z_{max,i}} \frac{dV_c}{dz}dz,
\end{equation}

Where $V_c$ is the co-moving volume and $z_{min,i}$ and $z_{max,i}$ are the minimum and maximum redshift limits 
within which source $i$ would be included in our sample. The full redshift range searched was $z = 0 - 6$, with $\Delta z = 0.01$. To allow for the \wise~ color selection we fitted a second order fit to $\log\nu\,\,{vs.}\,\log F_\nu$  to the four measured \wise~fluxes and used this SED to establish whether the source passed the color selection at each redshift. We did not include the radio source size criterion ($\theta < 45\arcsec$) since our observations indicate that none of our sources would be resolved by NVSS unless they were at a very low redshift ($z < 0.1$) with correspondingly small co-moving volume. In practice, we find that shifting a source to higher redshift usually fails our selection due to becoming too faint in the MIR. Similarly, we find that shifting a source to lower redshift usually fails our selection due to the source becoming too blue in the MIR. Because the color selection is usually affected at lower redshift, where the $1/V_{max}$ factor is small, then the detailed form of the MIR SED does not have a significant impact on the final RLF. 

The luminosity function, $\phi$, is given by:

 \begin{equation}
     \phi = \frac{4\pi}{\Omega}
     \frac{N_{tot}}{N_z}(\Delta \log\,L^{-1})\sum_{i}^{N} \frac{1}{V_{max,i}}, 
 \end{equation}
where $\Omega$ is the solid angle of our survey, which is essentially that of the NVSS since the \wise~ survey is all-sky (a total area of 28,443 sq. deg, \citealt{lonsdale+15}) , $\Delta \log\,L^{-1}$ is the width of each luminosity bin 
(with $L$ measured in units of W~Hz$^{-1}$ here), and $N$ is the number of sources in each luminosity bin. 
Finally, the factor $N_{tot}/N_z$ corrects for the fact that we only measured redshifts for 46\% of the total sample. A simple multiplicative factor is adequate since this subset is itself a significant fraction of the total, and is  relatively unbiased in redshift.  The errors given are simply proportional to $\sqrt{N}$, boosted by $N_{tot}/N_z$. 

Figure~\ref{fig:RLF} shows the RLF of our sample, together with the RLFs of samples of high-excitation (radiative mode) and low-excitation (jet mode) radio-loud AGN from \citet{best+14}.
%and a sample of GPS and HFP radio galaxies from \citet{snellen+00}. 
As expected given the deliberate selection of a rare class in color space, the RLF of our sample falls $\sim 2-3$ dex below that of the radio AGN from \citet{best+14}.
However, this offset is likely to be a lower limit because the radio AGN sample has lower redshift ($0.5 < z < 1$). Given the well known tendency for the co-moving density of radio sources to increase with redshift \citep[e.g.,][]{best+14, pracy+17,  ceraj+18}, a more detailed comparison at matched redshift would likely find an even greater offset.

How should we interpret the lower space density of our sample compared to the other samples of radio AGN? A straightforward explanation that supports our original motivation for selecting this sample is that the sources are in a short-lived phase \citep{lonsdale+15}. Two qualities of the sample point to this: (a) they have compact, high-pressure radio sources, which can plausibly be argued are young, (b) they have high bolometric luminosity but are optically faint, suggesting the sample is dominated by obscured quasars with high columns. Within the fairly well-established theory of this class of object they are thought to be in a very young transient stage following a strong fueling event, probably associated with a merger  \citep[e.g.,][]{hopkins+06}.

Another possible explanation for a low RLF is that the high-column material that yields both the red MIR colors and suppressed optical emission has a low covering factor due to a single cloud that happens to fall along our line of sight. However, we think this is unlikely because another characteristic of our sample is that it has high MIR luminosity. First, a simple optically thick blackbody at T$\sim 60$ K must have a radius of $\sim 1$ kpc to generate such a high MIR luminosity. Second, the high MIR luminosity suggests a large fraction of the AGN output is reprocessed by high-column absorbing material. Thus the covering factor for the high-column material must be reasonably high.

%assuming the MIR luminosity is reprocessed AGN radiation, demands a high, not low,  covering factor for the high-column material.}
%suggesting a large fraction of the AGN output is reprocessed by high-column absorbing material. Thus the covering factor for the high-column material must be reasonably high.}

%\textcolor{red}{\textbf{Other possible explanations for the low RLF include an orientation effect that preferentially selects edge-on sources and  variability of beamed sources.}}
%Thus, the likelihood of the radio source being young is increased by its association with a short-lived phase of the host system. }

\section{Discussion}\label{sec:discussion}
The overall scientific goal of our multi-wavelength program is to identify heavily obscured quasars at the peak epoch of stellar mass assembly and SMBH growth and investigate their connection to galaxy evolution, possibly via the interaction of a powerful jet with the host's ISM. 
Our unique selection criteria of extremely red \wise~ colors, along with compact radio and faint optical emission, promises to identify  galaxies in a key stage of galaxy growth. 
%\textbf{For the remainder of the paper, we will discuss our sources from the viewpoint that they are young, transient sources with expanding radio jets. Consideration of the other explanations for some of the sources will be addressed in the next paper where the SEDs are explored more deeply. Our analysis in that paper suggests that at least --\% of the compact sources are likely to be indeed due to expanding radio lobes of age $<--$ years after a major fueling event (Patil et al. in prep).}
In this section, we discuss the implications of our high-resolution radio imaging survey for the early phases of radio source evolution.  
%Currently, our focus is to explore the overall nature of these sources by conducting multi-wavelength follow-up studies of a representative sample. 
%The MIR SED modelling in \citet{lonsdale+15} showed the central engine of these sources dominate the MIR emission, and possibly some contribution from a young compact straburst. Although similar MIR selection criteria to the other obscured WISE population, the additional radio selection makes our sources different and provides a unique probe into early stages of the radio-loud, and luminous quasars. 

\begin{figure*}[t!]
\centering
\includegraphics[clip=true, trim=3cm 0cm 1.0cm 0cm, width=\linewidth]{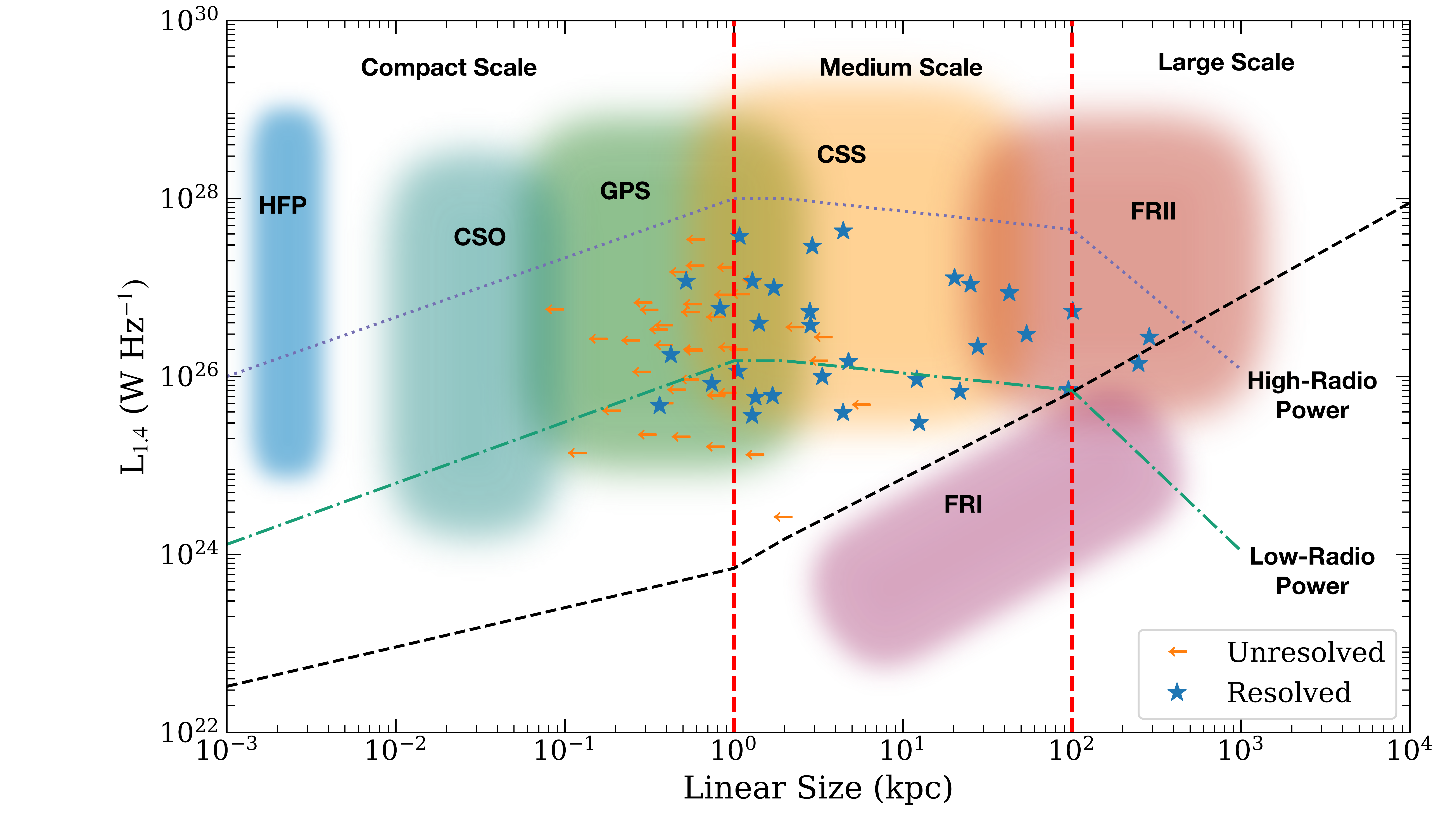}
\caption{1.4~GHz spectral luminosity $vs.$\ largest linear source extent. Blue  stars represent resolved sources and orange arrows indicate unresolved sources from our sample. 
The colored boxes represent the parameter space occupied by different radio populations compiled by \citet{an+12}. 
The purple dotted and green  dash-dotted lines are the evolutionary tracks followed by high-(HRP) and low-radio power (LRP) sources, respectively, based on the model given in \citet{an+12}. The vertical red dotted lines divide the entire plane into three broad size scales.  
%categories based on the literature studies. 
The HFP, CSO, and GPS sources are on the compact scales ($<1$~kpc), CSS and a minority of FRI/FRII sources fall into the medium scales ($\sim 1-100$~kpc), and  FRI/II sources are the large scale populations ($>$100~kpc). The black dashed line is the boundary between the  stable and turbulent jet flows.}
\label{fig:PD}
\end{figure*}

\begin{figure*}[ht]
    \centering
    \includegraphics[width=\linewidth]{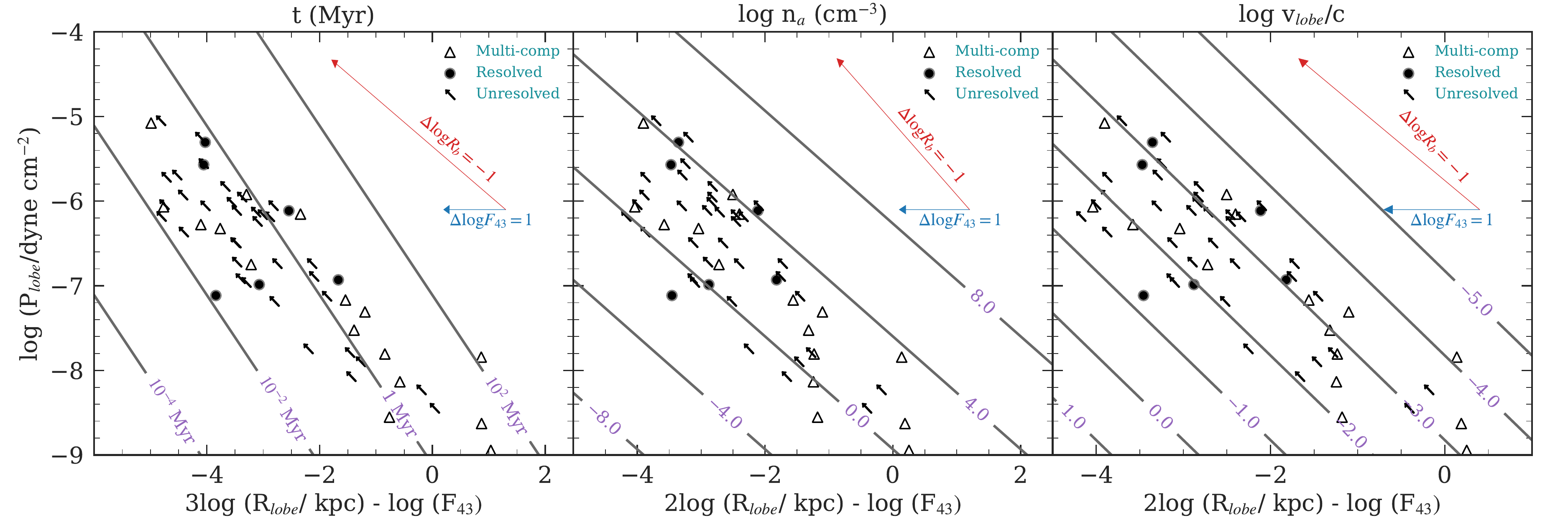}
    \caption{Application of adiabatic lobe expansion model to our sample sources with known redshifts (Equations~\ref{eqn:age_contm}$-$\ref{eqn:vlobem}). These panels isolate  source age ($t_{\scriptsize{\textrm{Myr}}}$), ambient particle density ($n_a$), and lobe expansion speed ($V_l/c$). The observed parameters are $R_{l}$, $p_l$ and $F_{43}$ as described in the text. Open triangles are individual resolved lobe components for double or triple sources; filled circles are partially resolved sources; arrows are unresolved sources. Red and blue vectors illustrate the effect of a decrease in source size by one dex and increase in jet power by one dex. 
    }
    \label{fig:age}
\end{figure*}
\subsection{Radio Source Evolution}\label{sec:pdradio}
%It is now well established that GPS and CSS sources are young counterparts of the large-scale FRI/FRII sources whose jets could be temporarily frustrated by the dense ambient medium  (e.g.,  \citealt{gallimore+06}). 
Several models have been proposed to describe the temporal evolution of the  observed properties of radio sources, such as luminosity and spectral turnover frequency \citep[e.g.,][]{falle+91,fanti+95, readhead+96, kaiser+97, odea+97, snellen+00, kaiser+07, kunert+10, an+12,maciel+14}. Many of the early models assumed %the self-similar
self-similar expansion of radio jets as they move first through dense ISM during their initial growth
%till 
until they emerge into the IGM and ICM to become large-scale, old sources  \citep{kaiser+97}. Early semi-analytic models found that the radio source luminosity  increases as  ram-pressure confined lobes expand within the galaxy. The luminosity reaches a maximum when the jets pass the boundary of the ISM, and then it decreases as the lobes expand into the ICM to become FRI/FRII sources. 

%The radio-power vs.\ linear size (PD) is the most popular way to show the evolutionary tracks different radio-loud AGN as shown in Figure~\ref{fig:PD}. 
We now explore the evolutionary stage of our sample and its connection to the FRI/FRII population by plotting our sources on the radio-power $vs.$\ linear size (PD) diagram in Figure~\ref{fig:PD}. 
The range of linear extents of our sources covers multiple classes of medium- and compact-scaled radio sources, including CSS and GPS populations. 
%We plot our sources with known redshifts in the PD plane in Figure~\ref{fig:PD} to compare with the broader AGN population.  
It is clear from Figure~\ref{fig:PD} and  Section~\ref{sec:rlum} that the radio luminosities of our sample sources lie between those of the classical FRI and FRII  %radio loud AGN. 
populations. 
We also show the two tracks given by \citet{an+12} that follow the high  radio power (dotted) and low  radio power (dash-dotted) sources. 
%^Based on the model given by \citet{an+12}, we show the evolutionary tracks (two dashed lines) followed by two types of AGN; high-jet power corresponding to radiatively efficient accretion and low-jet power corresponding to %the advection dominated, radiatively inefficient accretion rates. 
%radiatively inefficient accretion.  
Our sources, being intermediate in luminosity, lie between 
these two tracks in Figure~\ref{fig:PD}. The dashed line shows the boundary between stable and unstable jets in the model of \citet{an+12}. The fact that all except one (J2318$-$25) of our sources lie above this line is consistent with them having stable jets that yield small-scale edge-brightened double or triple morphologies, as indeed we find in the majority of the resolved sources. 

%The interpretation of Figure~\ref{fig:PD} is limited due to a large number of upper limits on the linear sizes. We would need  higher-angular resolution observations with instruments such as VLBA, e-MERLIN to resolve the morphologies of those sources. 

%However, this simplistic approach does not fully capture the observational trends seen in the different radio classes. \citet{an+12} proposed a modified model to incorporate and explain the effect of the jet power, environment, and the intermittent nature of AGN activity on the observed morphologies. \citet{turner+15} developed a comprehensive framework by combining previous models self-similar FRII expansion and pressure limited expansion seen in FRI sources. 

%\textcolor{magenta}{Concept: We now consider the implications of our source selection to --- set some stage---}

%Based on the evolutionary models given in \citet{an+12} and the location of our sources in the PD diagram, some of them may evolve into classical, large-scale FRI/II radio sources. However, the evolution of radio jet morphologies is dependant on the intrinsic jet power, AGN duty cycles, and ambient environment. Future studies constraining ISM conditions and SMBH accretion rates (e.g. using observations from ALMA and NuSTAR) would be needed to further investigate this issue.  
Based on the evolutionary models given in \citet{an+12} and following similar recent analyses (e.g. \citealt{jarvis+19}), 
it seems the position of our sources on the PD diagram relative to the jet instability criterion supports the possibility that they  might eventually evolve into classical,  FRI/II radio sources.
This possibility is reinforced by the fact that our sources 
are heavily obscured which points to a long term fuel supply  that could sustain the SMBH accretion for the $\sim100$~Myr time span necessary to create larger radio sources. However, a more careful discussion of possible evolutionary links between the \wise-NVSS sources and classical radio galaxies  must consider the source ages. This we now attempt using a simple jet-lobe expansion model.

\subsection{Lobe Expansion Model}\label{sec:ages}

%Our measured parameters $-$ lobe size, pressure, and radio luminosity $-$ only partially describe the physical system.  Ultimately, we are interested in other properties, such as source age, external density, and lobe expansion speed. To find these other properties, we need to frame our measured parameters within a physical model. Ideally, the constraints inherent in the model, such as energy and momentum conservation, then allow us to derive the other properties of the system. Do these properties confirm the overall intent of our sample selection $-$ namely young radio sources expanding within a dense environment?

There has been considerable work on models of radio source evolution in a variety of contexts, both analytic \citep[e.g.,][]{turner+15, hardcastle+18} and numerical \citep[e.g.,][]{mukherjee+16,perucho+19}. Our sources may allow a relatively simple approach because the jets enter a dense, near-nuclear environment and are caught early in their development. While this may seem a potentially complex process, detailed simulations of just this situation \citep[e.g.,][]{mukherjee+16,mukherjee+18} suggest that the radio source develops in a quasi-spherical expansion, and in this case, the analytic model of self-similar expansion is approximately correct.

%Compact radio sources (e.g., GPS, and CSS) are commonly associated with young radio jets, some of which are precursors of large-scale radio sources \citep{fanti+90, alexander+00, snellen+00,perucho+15, orienti+16}.
%Compact radio sources (e.g., GPS, and CSS) are commonly associated with young radio jets, some of which are likely the precursors of large-scale FRI/FRII radio sources \citep{fanti+90, alexander+00, snellen+00,perucho+15, orienti+16}.  To investigate whether our sources are young,  we use analytic expressions of self-similar (i.e. scale-independent) jet expansion to estimate source ages.   

%For %the confined 
%compact jets with sizes of less than a few kpc, as they expand into the dense environment, the interaction with the ambient gas clouds makes expansion nearly spherical \citep{mukherjee+16}. In such cases, the analytic models of self-similar expansion of a bubble are approximately correct. This can be used to derive %upper limit 
%upper limits on the source ages. %Here we attempt to obtain rough estimates on the ages of our sources. 
A simple approach assumes purely adiabatic expansion in which case the dynamics of the early phase of 
%the jet evolution 
jet evolution can be approximated by 
%a 
the presence of a forward shock, a contact discontinuity, and %a 
an inner reverse shock. 
%
%
%The mathematical treatment for the expansion of a spherical lobe driven by continuous energy input is given in \citet{weaver+77}. 
Following the mathematical treatment given in \citet{weaver+77}, a self-similar expansion of a spherical lobe can be expressed in terms of our observed parameters.\footnote{A complete derivation of these relations is given in  \citet{begelman+99} as well as in  Appendix~\ref{sec:app_bubble}. }

 \begin{equation}\label{eqn:age_contm}
 p_l =7.76\times10^{-10}F_{43}t_{\scriptsize{\textrm{Myr}}}R_l^{-3}
 \end{equation}
 \begin{equation}\label{eqn:nam}
 p_{l} =1.17\times10^{-9}\, F_{43}^{2/3} n_a^{1/3}R_l^{-4/3}
 \end{equation}
 \begin{equation}\label{eqn:vlobem}
 p_{l} =1.50\times10^{-12} F_{43}(V_l/c)^{-1} R_l^{-2}
 \end{equation}
 
where $p_l$ is the pressure inside the lobe expressed in dynes/cm$^{2}$, $R_l$ is the radius of a lobe in kpc, $F_{43}$ is the mechanical jet power in units of  $10^{43}$ erg~s$^{-1}$, $n_a$ is the ambient number density in cm$^{-3}$,  $t_{\scriptsize{\textrm{Myr}}}$ is a dynamical age in Myr, $V_l$ is the lobe velocity, and $c$ is the velocity of light.

While Sections~\ref{sec:linsize} and \ref{sec:pressure} describe our  estimates of radio source size, $R_l$, and pressure, $p_l$, estimating the jet power, $F_{43}$, is more uncertain. One approach is to assume that the jet power is related to the radio luminosity. While a number of studies have tried to establish such a link  \citep[e.g.,][]{willott+99, cavagnolo+10}, others have argued that the relation is intrinsically quite scattered and has been amplified by selection bias \citep{godfrey+16}.  Bearing these caveats in mind, we cautiously adopt the relation given by \citet{ineson+17}:
\begin{equation}
    F_{43} = 5 \times 10^{3}
    L_{151}^{0.89 \pm 0.09}
\end{equation}
where $L_{151}$ is the rest-frame 151~MHz radio luminosity in units of 10$^{28}$~W~Hz$^{-1}$~Sr$^{-1}$. For our sample, we estimate $L_{151}$ using the 1.4~GHz luminosity from NVSS and a spectral index $\alpha^{10}_{1.4}$ derived from the NVSS flux and our X-band flux. We exclude sources with flat/inverted indices ($\alpha^{10}_{1.4}>-0.3$) and low S/N sources with very steep indices ($\alpha^{10}_{1.4} < -2.0$) since the uncertainty in the extrapolation to rest-frame $L_{151}$ is  large. 

The left panel in Figure~\ref{fig:age} shows the relation given in Equation~\ref{eqn:age_contm} between source pressure, source size and jet power, by plotting $3\log R_l - \log F_{43}$ against $\log p_l$ so that the source dynamical age, $t_{\scriptsize{Myr}}$, appears as diagonal contours.  

Using estimates for $R_l$ and $p_l$ from Sections~\ref{sec:linsize} and \ref{sec:pressure}, and $L_{151}$ as described above, we find the majority of our sources have dynamical ages in the range  $10^4 - 10^7$ years, with a median around 0.7~Myr. This is consistent with the overall picture that our sample contains young radio sources. 
%possibly triggered by the same event that deposits a significant mass of gas and dust to the nuclear regions to obscure the AGN accretion, which in turn creates the very red  MIR colors. These timescales are also broadly consistent with the RLF analysis in Section~\ref{sec:lumfunc} in which the low space density of our sources relative to the FRI/FRII source population suggested source ages that are younger and/or shorter by factors $\sim10^2 - 10^4$.

The central panel of Figure~\ref{fig:age} plots contours of external density, $n_a$, and the distribution of points reveal relatively high densities, spanning 1$-10^4$ cm$^{-3}$, comparable to the higher-density phases in spiral  disks or near-nuclear ISM. 
Again, this is consistent with our overall picture of a young radio source emerging into a dense medium. 
Most radio sources are within a kpc of a galaxy center where we expect high average gas density, especially given the steep optical-MIR SED colors pointing to high columns. 
Indeed, we can combine the inferred ambient gas density with our measured source size to estimate a column density. The majority span $\log (N_H/\textrm{cm}^{2}) \sim 22-25$ corresponding to A$_V \sim 5-5000$ which is consistent with the red optical-MIR SEDs and the identification of Compton thick columns in the related Hot DOG population \citep[e.g.,][]{stern+14, ricci+17}.

The right panel in Figure~\ref{fig:age} plots contours of lobe expansion speeds. It seems the sources expand with modest, sub-luminal, speeds  $V_l \sim 30-10,000$ km s$^{-1}$ with a median near $450$ km s$^{-1}$. 
%These modest velocities   follow directly from our basic estimates for the source size and age: a source size of 1~kpc with an age of 1~Myr has an associated velocity of $1000$~km~s$^{-1}$. 
We note that our velocities are also similar to those found in much more detailed simulations of a similarly powered jet interacting with a dense clumpy medium \citep[][]{mukherjee+16,mukherjee+18}.   

The discussion of  the growth  of compact radio sources is often framed as two contrasting possibilities: the small sizes result from youth or from ``frustrated'' jets that cannot expand due to a dense surrounding medium \citep[e.g.,][]{van+84, bicknell+18}. Our analysis suggests that both perspectives might be relevant for our sources --- the sources are indeed young, but the ISM is also dense and this  slows the source expansion.   

%We may cautiously use the analysis presented here to go beyond establishing that the sources are young and the surrounding medium is relatively dense, by looking for other correlations using these derived quantities.
%For example, older sources tend to be larger ($R_l$ correlates with $t_{\scriptsize{Myr}}$), consistent with steady source growth. Also, lobes expanding more slowly tend  to have denser environments  ($V_l$ anticorrelates with $n_a$).  At a given ambient density ($n_a$), more powerful jets ($F_{43}$) drive faster lobe expansion speeds ($V_l$). At a given dynamical age ($t_{\scriptsize{Myr}}$) more powerful jets ($F_{43}$) have created larger lobe sizes ($R_l$). All these trends are consistent with the a jet-powered lobe expanding over time via ram pressure balance with an external ISM. 
%However, we do not consider these correlations to provide independent support for  the lobe expansion model because they arise at least in part via the implicit interdependence of the variables within the  model.  Future studies that directly probe the ISM properties near the lobe and independent estimates of the lobe dynamical ages will be needed to further assess the applicability of the lobe expansion model.

\subsection{Prevalence of Gas-rich Mergers}\label{sec:merger}
  
Perhaps the most straightforward indication of youth would be to find a direct association with a short-lived phase in the host galaxy, such as a merger. Unfortunately, by selecting optically faint hosts (to avoid low-redshift sources) a simple inspection of the optical morphology is difficult. In the absence of direct observations, what might we expect?  Despite early numerical simulations suggesting that luminous AGN are associated with gas rich mergers \citep[e.g.,][]{hopkins+08}, the observational evidence has been mixed. For example, \citet{cisternas+10} and \citet{villforth+17} fail to find the AGN-merger connection. However, when the AGN are selected to be dusty and obscured, such as \wise~AGN, the association with mergers is much clearer, particularly at high luminosity \citep[e.g.,][]{satyapal+14, weston+17, goulding+17}. 
%Further evidence for the prevalence of mergers in \wise~ selected objects comes from the similarity in infrared colors between our sample and HotDOGs, which are known to have a high merger fraction \citep[e.g.,][]{fan+16, farrah+17, zappacosta+18}. }

Recent numerical simulations of galaxy mergers also support this association.  \citet{blecha+18} have tracked the evolution of \wise~ colors and luminosities for gas rich mergers, finding the closest match to our sample's very red \wise~ colors during the brief final stage of coalescence. 

Thus, our own sample of \wise~ selected AGN is very likely to contain a significant fraction of recent gas-rich mergers. Such a merger would be consistent with a newly triggered AGN with a radio jet.

\subsection{Are WISE-NVSS Sources Truly Newborn?}\label{sec:newborn}
Another approach that places the \wise-NVSS sources in a wider context is to use the RLF and dynamical age estimates to help establish a link to the other classes of radio source.  First, the RLF analysis in Section~\ref{sec:lumfunc} suggests the \wise-NVSS sources have $\sim300$ times lower space density than classical radio galaxies. Second, comparing the median dynamical age of $\sim 10^{5.8}$ years to a typical age for a classical radio galaxy of $\sim 10^7$ years, suggests a source age ratio of  $\sim7\%$. Combining this age ratio with the ratio in space density of $\sim 0.3\%$, indicates that  $\sim20\%$ of the classical radio galaxies might have been born directly from a \wise-NVSS source.

Finding alternate compact progenitors that might evolve into classical radio galaxies isn't hard.  \citet{odea+98} performs a similar demographic analysis with GPS and CSS sources and shows that they actually over-produce the classical radio galaxies by a factor of $\sim 10$. \citet{odea+98}  interprets this apparent over-production as evidence for recurrent activity in the GPS and CSS populations --- meaning there might be multiple phases of compact emission before the source finally evolves into a large-scale, classical, radio galaxy.

The relation between the \wise-NVSS sources and the GPS and CSS sources is not yet clear. There  seems to be a systematic difference in the MIR properties (Patil et al. in prep.), suggesting that although all these sources may be dynamically young, the \wise-NVSS sources might be truly ``newborn'' --- meaning the radio source has emerged for the first time, into a  dense near-nuclear ISM. In this case, the  \wise-NVSS sources may either evolve directly into the classical radio galaxies, or perhaps join the more common GPS and CSS classes, and from there ultimately evolve into a classical radio phase.

\section{Summary and Conclusions}\label{sec:conclusion}

We have presented a high-resolution 10~GHz VLA imaging study of  a sample of ultra-luminous and heavily obscured quasars in the redshift range $0.4 < z < 3$ with a median $z\sim 1.53$.  Our selection is similar to that of Hot DOGs in MIR colors, but adds a requirement for the presence of compact radio emission that allows us to select objects in which radio emitting jets are present. Of the 155 radio sources in our sample, 86 ($\sim55\%$) remain unresolved even on sub-arcsecond scales.
%the highest angular resolution of our study ($\theta_{\rm FWHM} \sim 0.2^{\prime \prime}$).  
Our main conclusions are as follows:

\begin{enumerate}

    \item The compactness of the majority of the sources on scales $<0.2^{\prime \prime}$ implies typical physical sizes are $\leq 2$ kpc at the median redshift ($z=1.53$) of our sample.
    
    \item We measured in-band spectral indices %for our sample 
    from 8--12~GHz and found a median spectral index of $-1.0$, consistent  with (perhaps slightly steeper than) typical  optically-thin synchrotron emission from radio jets or lobes.

    \item %Radio 
    We estimate equipartition pressures in the radio lobes and find them to be similar to other compact sources such as GPS or CSS, but significantly higher than the lobes of more extended classical radio galaxies.   %This is 
    These high pressures  support the possibility that the \wise-NVSS sources may be powered by recently triggered radio jets emerging into a dense, near-nuclear ISM.

    \item Our radio sources have rest frame $1.4$~GHz luminosities between those of the classical FRI and FRII radio galaxies, in the range $10^{25-27.5}$ W~Hz$^{-1}$. 
    On the well-known Radio Power $vs.$ Linear Size (PD) diagram, our sources fall in the same region as the other compact and medium scale radio sources such as GPS and CSS sources. 
    
    \item
    We perform a standard V/V$_{\rm max}$ analysis to generate a 1.4 GHz radio luminosity function for our sample, and compare it to other samples of radio sources.
    Overall, the \wise-NVSS sources are rare, with space densities roughly $\sim2-3$ dex lower than the population of radio AGN studied by \citet{best+14} and $\sim0.5-1.0$ dex lower than  samples of compact radio AGN (GPS, HFP; \citealt{snellen+00}).

    \item  %The source age estimates are $\lesssim 1$~Myr or less supporting the interpretation that our sources are young. 
    We use a simple adiabatic jet expansion model and an empirical relation between radio luminosity and jet power, to estimate  dynamical ages, ambient densities and expansion velocities for our sample sources. We find source ages in the range $10^{4-7}$~years (median 0.7~Myr), ambient particle densities in the range $1-10^4$~cm$^{-3}$ (median 101~cm$^{-3}$), and lobe expansion speeds in the range $30-10,000$~km s$^{-1}$ (median 450 km s$^{-1}$). Within the framework of this model, these results broadly confirm our expectation that these sources are relatively young and are expanding at modest velocities into a relatively dense ISM, as suggested by their MIR-optical properties.  
    
    % \item
    % Combining the RLF analysis with the dynamical ages, we find that only 10\% of the population of classical radio galaxies could have evolved directly from the \wise-NVSS sources. The fact that there is an over-abundance of the GPS and CSS sources relative to classical radio sources raises the question of the relation between the \wise-NVSS sources and these other compact radio sources. 
    % \textbf{We favour the scenario} \wise-NVSS sources is that their jets have turned on for the very first time, following the merger and dumping of ISM into the nucleus.
    % Following this initial phase, it is possible that the \wise-NVSS sources evolve into GPS or CSS sources, of which some ultimately evolve into the larger classical radio galaxies. 

    \item
    In the absence of unknown selection effects, such as variability \citep{mooley+16}, our RLF and dynamical age analyses suggest that $\sim$10\% of the population of large-scale radio galaxies could have evolved directly from the \wise-NVSS sources. The over-abundance of the GPS and CSS sources relative to classical, large-scale radio sources raises the question of the relation between the \wise-NVSS sources and these other compact radio sources. 
    We favor a scenario in which the \wise-NVSS sources harbor jets that have turned on for the very first time, following the merger and dumping of ISM into the nucleus.
    Following this initial phase, it is possible that the \wise-NVSS sources evolve into GPS or CSS sources, of which some ultimately evolve into the larger classical radio galaxies. 

\end{enumerate}

Overall, we conclude that the radio properties of our sample are consistent with 
emission arising from recently-triggered, young jets.  
%In the next publication in this series, 
In a series of forthcoming studies, we will present an analysis of the broadband radio SEDs of our sources as well as new milliarcsecond-scale-resolution imaging with the VLBA and enhanced Multi Element Remotely Linked Interferometer Network (e-MERLIN). These studies will place %quantitative 
tighter constraints on the source ages and provide deeper insights into their evolutionary stages.  Ultimately, studies of the ISM content and conditions in the vicinity of young, ultra-luminous quasars will be needed to investigate the onset and energetic importance of jet-ISM feedback during the peak epoch of galaxy assembly.  Observations with ALMA and the {\it James Webb Space Telescope}, and eventually the next-generation Very Large Array (e.g., \citealt{nyland+18, patil+18}), will be essential for improving our understanding of feedback driven by young radio AGN at $z\sim2$ and its broader  connection to galaxy evolution.

\section*{Acknowledgments}
We thank the anonymous referee for many helpful suggestions which have significantly improved the paper.
We thank Wiphu Rujopakarn for useful discussions. The National Radio Astronomy Observatory is a facility of the National Science Foundation operated under cooperative agreement by Associated Universities, Inc.  Support for this work was provided by the NSF through the Grote Reber Fellowship Program administered by Associated Universities, Inc./National Radio Astronomy Observatory. 
Basic research in radio astronomy at the U.S. Naval Research Laboratory is supported by 6.1 Base Funding.  This publication makes use of data products from the Wide-field Infrared Survey Explorer, which is a joint project of the University of California, Los Angeles, and the Jet Propulsion Laboratory/California Institute of Technology, funded by the National Aeronautics and Space Administration. 
The authors have made use of {\sc Astropy}, a community-developed core {\sc Python} package for Astronomy \citet{astropy+13}. We also used MONTAGE, which is funded by the National Science Foundation under Grant Number ACI-1440620, and was previously funded by the National Aeronautics and Space Administration's Earth Science Technology Office, Computation Technologies Project, under Cooperative Agreement Number NCC5-626 between NASA and the California Institute of Technology. This research made use of APLpy, an open-source plotting package for Python hosted at \url{http://aplpy.github.com}. This research made use of matplotlib, a Python library for publication quality graphics \citep{hunter+07}.

%This research has made use of the NASA/IPAC Extragalactic Database (NED) which is operated by the Jet Propulsion Laboratory, California Institute of Technology, under contract with the National Aeronautics and Space Administration.

\facilities{VLA, \wise~}

\software{CASA \citep{mcmullin+07},
          astropy \citep{astropy+13},
          matplotlib \citep{hunter+07} 
          AIPS }
          
\bibliography{jvla_snapshot_v1.bib}

\appendix
\section{Radio Lobe Expansion}\label{sec:app_bubble}

The mathematical treatment for the expansion of a spherical lobe driven by continuous energy input is given in \citet{weaver+77}.
The momentum and energy conservation equations are: 
 \begin{equation}
 \label{Eq:momentum}
     \frac{d}{dt}\Bigg(\frac{4}{3}\pi R_l^3 \rho_a V_l\Bigg) = 4\pi R_l^2 p_l
 \end{equation}

\begin{equation}
\label{Eq:energy}
     \frac{d}{dt}\Bigg[\frac{4\pi}{3}\frac{p_l}{\gamma-1}R_l^3\Bigg] + 4\pi R_l^2p_lV_l = F_E,
\end{equation} where $R_l$ is the radius of the lobe's shock, $V_l = dR_l/dt$ is the velocity of the shock, $\rho_a$ is the ambient density of the undisturbed ISM, $p_l$ is the pressure inside the lobe, and $F_E$ is the mechanical power injected by the jet. For a self-similar expansion of the jet lobe, the above equations can be solved to yield:

\begin{equation}\label{eqn:rb}
    R_l = 0.78 F_{43}^{1/5}n_a^{-1/5}t_{\scriptsize{\textrm{Myr}}}^{3/5}\,\,\,\textrm{kpc }
\end{equation}
\begin{equation}\label{eqn:pb}
    p_l = 1.63\times10^{-9} F_{43}^{2/5}n_a^{3/5}t_{\scriptsize{\textrm{Myr}}}^{-4/5} \,\,\,\textrm{dynes  cm}^{-2}
\end{equation} \begin{equation}\label{eqn:vb}
    V_l = 458 F_{43}^{1/5}n_a^{-1/5}t_{\scriptsize{\textrm{Myr}}}^{-2/5}\,\,\,\textrm{km s}^{-1}
    \end{equation} 
 where %$F_{43} = 10^{43}$ erg s$^{-1}$, 
 $F_{43}$ is in units of $10^{43}$ erg~s$^{-1}$, $n_a (= \rho_a/(\mu_m m_p))$ is the ambient number density in cm$^{-3}$, and $t_{\scriptsize{\textrm{Myr}}}$ is a dynamical age in Myr. Here $\mu_{m}$ is the mean molecular weight of the ISM and $m_p$ is the proton mass.
Equations~\ref{eqn:rb}$-$\ref{eqn:vb} can be rearranged to isolate $t_{\scriptsize{\textrm{Myr}}}$, $n_a$, and $V_l$ in terms of our observed parameters:
\begin{equation}\label{eqn:age_cont}
 p_l =7.76\times10^{-10}F_{43}t_{\scriptsize{\textrm{Myr}}}R_l^{-3}
 \end{equation}
 \begin{equation}\label{eqn:na}
 p_{l} =1.17\times10^{-9}\, F_{43}^{2/3} n_a^{1/3}R_l^{-4/3}
 \end{equation}
 \begin{equation}\label{eqn:vlobe}
 p_{l} =1.50\times10^{-12} F_{43}(V_l/c)^{-1} R_l^{-2}
 \end{equation}

\vspace{1cm}
\section{10~GHz Continuum Images }\label{sec:images}
We provide individual 10~GHz Continuum Images of our sample in the online Figure~Set associated with Figure~\ref{fig:cutouts}.
%

%
%%%%%%%%%%%%%%%%%%%%%%%%%%%%%%%%%%%%%%%%%%%%%%

%%%%%%%%%%%%%%%%%%%%%%%%%%%%%%%%%%%%%%%%%%%%%%

%%%%%%%%%%%%%%%%%%%%%%%%%%%%%%%%%%%%%%%%%%%%%%%%%%%%
%%%%%%%%%%%%%%%%%%%%%%%%%%%%%%%%%%%%%%%%%%%%%%%%%%%%
%%%%%%%%%%%%%% Notes on extended sources %%%%%%%%%%%
%%%%%%%%%%%%%%%%%%%%%%%%%%%%%%%%%%%%%%%%%%%%%%%%%%%%

\section{Sources with Extended Emission}\label{sec:app_ext}
We classified radio morphologies by visually inspecting our VLA images as well as several archival radio surveys, namely, TGSS, NVSS, FIRST, and VLASS (see Section~\ref{sec:morph} and ~\ref{sec:extended} for details). We find 25/155 sources have well-extended, complex radio emission on a few arcsecond scales either in our 10 GHz data or other radio surveys. Table~\ref{tab:extended} provides  morphological classes and angular extents for those 25 sources in each of the survey mentioned above except the NVSS.  Figure~\ref{fig:ext_source} compares image cutouts taken from these five radio surveys. 

\begin{deluxetable}{ccccccccchhhhh}
\tablecaption{List of Extended Sources \label{tab:extended}}
\tablehead{
\nocolhead{} & \multicolumn{4}{c}{Source Morphology}  & \multicolumn{4}{c}{Angular Extent}\\
\colhead{Source} & \colhead{VLA-X}  & \colhead{VLASS} & \colhead{FIRST} & \colhead{TGSS}   & \colhead{VLA-X}  & \colhead{VLASS} & \colhead{FIRST} & \colhead{TGSS}  & \nocolhead{} & \nocolhead{} & \nocolhead{} & \nocolhead{} & \nocolhead{}\\
\colhead{} & \colhead{}  & \colhead{} & \colhead{} & \colhead{}   & \colhead{$^{\prime\prime}$}  & \colhead{$^{\prime\prime}$} & \colhead{$^{\prime\prime}$} & \colhead{$^{\prime\prime}$} & \nocolhead{} & \nocolhead{} & \nocolhead{} & \nocolhead{} & \nocolhead{}
} 
\colnumbers
\startdata
J0000+78 & UR & T & \nodata & R & 0.04 & 22.1 & \nodata & 33.4 & PL & NB & UR & \nodata & AX\\
J0010+16 & UR & UR & \nodata & T & 0.07 & 2.5 & \nodata & 153.1 & CV & G & UR & UR & AX\\
J0132+13 & SR & UR & \nodata & D & 0.07 & 2.4 & \nodata & 47.4 & CV & G & SR & UR & AX\\
J0342+37 & UR & UR & \nodata & D & 0.04 & 2.9 & \nodata & 44.3 & PK: & G & \nodata & UR & BX\\
J0543+52 & T & D & \nodata & UR & 5.3 & 4.5 & \nodata & 12.0 & CV: & G & T & T & AX\\
J0602$-$27 & T & D & \nodata & R & 4.4 & 4.3 & \nodata & 26.6 & PL & G & T & T & AX\\
J0737+18 & D & D & UR & R & 9.3 & 8.9 & 5.4 & 29.4 & CX & NA & \nodata & D & BX\\
J1025+61 & T & T & D & D & 46.1 & 46.8 & 47.3 & 47.6 & CX & NA & \nodata & D & BX\\
J1138+20 & UR & D & D & \nodata & 0.02 & 13.8 & 14.3 & \nodata & CV & G & UR & UR & AX\\
J1308$-$34 & T & T & \nodata & R & 8.9 & 9.7 & \nodata & 37.1 & PL & G & T & T & AX\\
J1439$-$37 & D & D & \nodata & R & 11.2 & 15.3 & \nodata & 48.5 & PL & G & UR & D & BX\\
J1525+76 & UR & T & \nodata & D & 0.11 & 46.6 & \nodata & 40.4 & CV & NA & \nodata & UR & BX\\
J1651+34 & M & D & D & R & 12.6 & 12.8 & 11.9 & 15.0 & PL & NA & \nodata & T & BX\\
J1703$-$05 & D & D & \nodata & R & 6.2 & 7.0 & \nodata & 13.2 & PL & G & D & D & AX\\
J1951$-$04 & T & T & \nodata & D & 24.3 & 29.5 & \nodata & 39.9 & CV & G & R & D & BX\\
J2059$-$35 & SR & R & \nodata & D/R & 2.4 & 5.3 & \nodata & 51.3 & CV: & G & \nodata & SR & BX\\
J2124$-$28 & M & T & \nodata & R & 11.4 & 11.2 & \nodata & 30.5 & CV & NB & M & \nodata & AX\\
J2130+20 & T & T & \nodata & D & 39.1 & 37.1 & \nodata & 44.8 & CV: & G & UR & T & BX\\
J2133$-$17 & T & T & \nodata & R & 18.8 & 20.5 & \nodata & 25.2 & PL & G & T & T & BX\\
J2145$-$06 & D & D & D & UR & 3.4 & 10.4 & 10.2 & 25.0 & CV: & G & D & D & AX\\
J2212$-$12 & T & T & \nodata & R & 20.9 & 20.0 & \nodata & 26.8 & PL & G & T & T & BX\\
J2318+25 & T & T & \nodata & R & 34.7 & 36.9 & \nodata & 70.4 & PL & G & \nodata & T & BX\\
J2328$-$02 & SR & D & D & UR & 0.13 & 14.6 & 12.1 & 25.0 & PL & G & SR & SR & AX\\
J2331$-$14 & D & D & \nodata & R & 7.2 & 8.3 & \nodata & 17.6 & CV & G & D & D & AX\\
J2341$-$29 & UR & D & \nodata & R & 0.11 & 5.7 & \nodata & 40.6 & PL & NB & UR & \nodata & AX\\
\enddata
\tablecomments{Column 1: Source name. Column 2-6: Source morphologies in our 10 GHz VLA data, VLASS, FIRST, and TGSS, respectively. The morphological classes are as follows: UR: unresolved; SR: slightly or marginally resolved; D: double; T: triple; M: multicomponent sources. The deatiled description of morphological classes is given in Section~\ref{sec:morph}. Column 6-9: Largest angular extent in arcseconds for the radio emission detected in our 10GHz VLA survey, VLASS, FIRST, and TGSS, respectively. For sources with a single component emission, we provide angular size estimates from their respective source catalogs. For multi-component sources, we provide largest source separation measured manually using CASA task Viewer.  }
\end{deluxetable}

\pagebreak

\begin{figure*}[htpb!]
\includegraphics[clip=true, trim =5cm 14cm 2cm 6cm,  width = \textwidth]{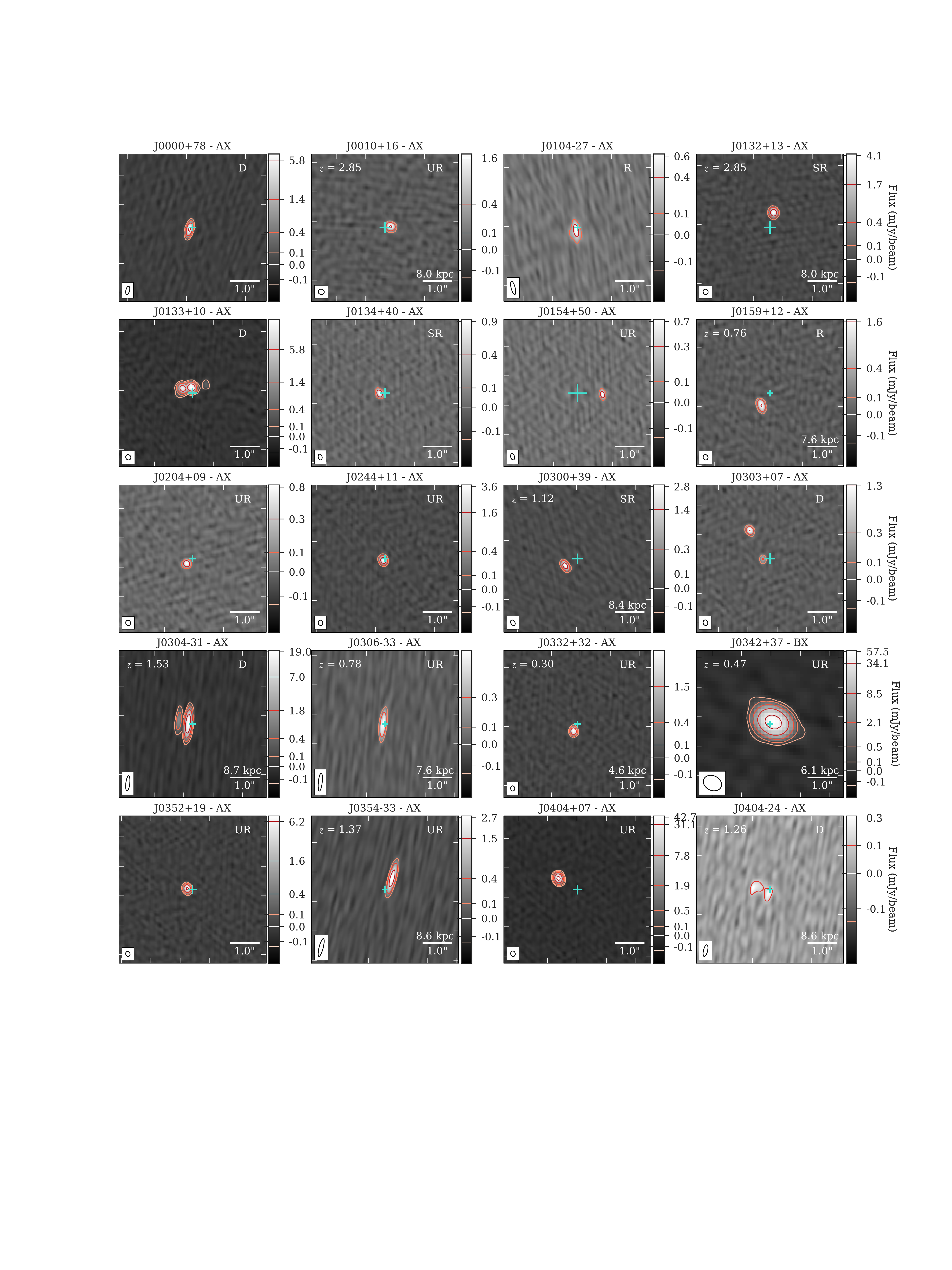}
\caption{10~GHz continuum images for our sample. The source name and VLA array used to produce the image are shown above each image. Contour levels are plotted in units of rms noise which can be found in Table~\ref{tab:obs_info}. The positive contours (solid) increase by a factor of 4 starting from 5$\sigma$ and the negative contours (dashed) are $-5\sigma$.  The contour levels are also marked on the right hand colorbar, including a zero level (which is not plotted on the image as a contour). The cyan plus symbol gives the \wise~ source position with one sigma uncertainty. For clarity, a minimum of  0.2$^{\prime\prime}$ is used. The synthesized beam is shown as a  black ellipse in the lower-left corner. A white solid line on the lower-right gives a scale bar. When available, the redshift is given in the upper-left and the equivalent physical scale is given above the scale bar. The radio morphology code is given in the upper-right. %RA and Dec tick marks are spaced by twice the scale bar length. 
The tick mark spacing is equal to the length of the scale bar.  }\label{fig:cutouts}
\renewcommand{\thefigure}{\arabic{figure} (Cont.)}
\addtocounter{figure}{-1}
\end{figure*}
\begin{figure*}
\centering
\includegraphics[clip=true, trim =5cm 6cm 1cm 6cm, width=\textwidth]{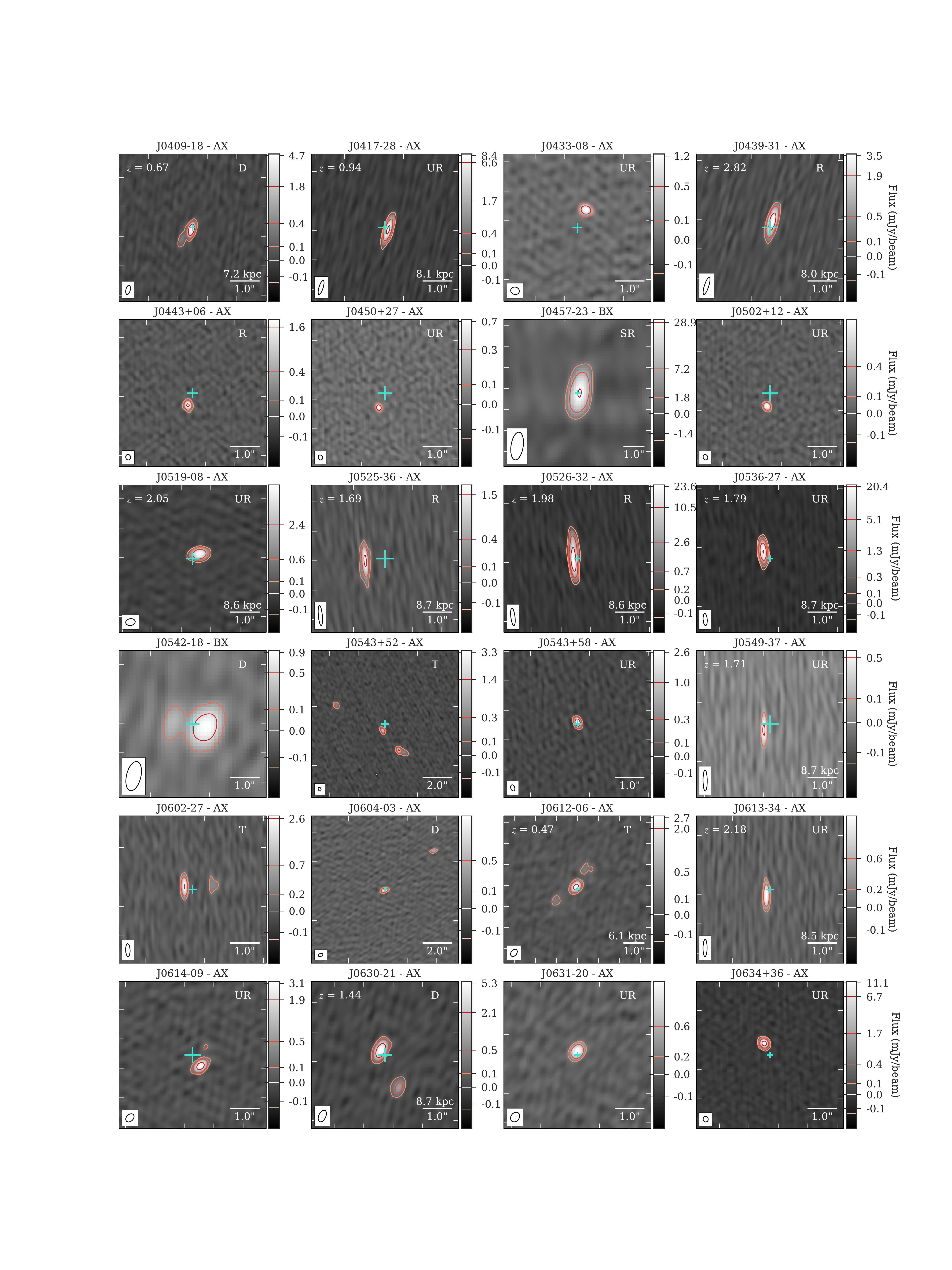}\captcont{\it Continued}
\end{figure*}
\begin{figure*}
\centering
\includegraphics[clip=true, trim =5cm 6cm 1cm 6cm, width=\textwidth]{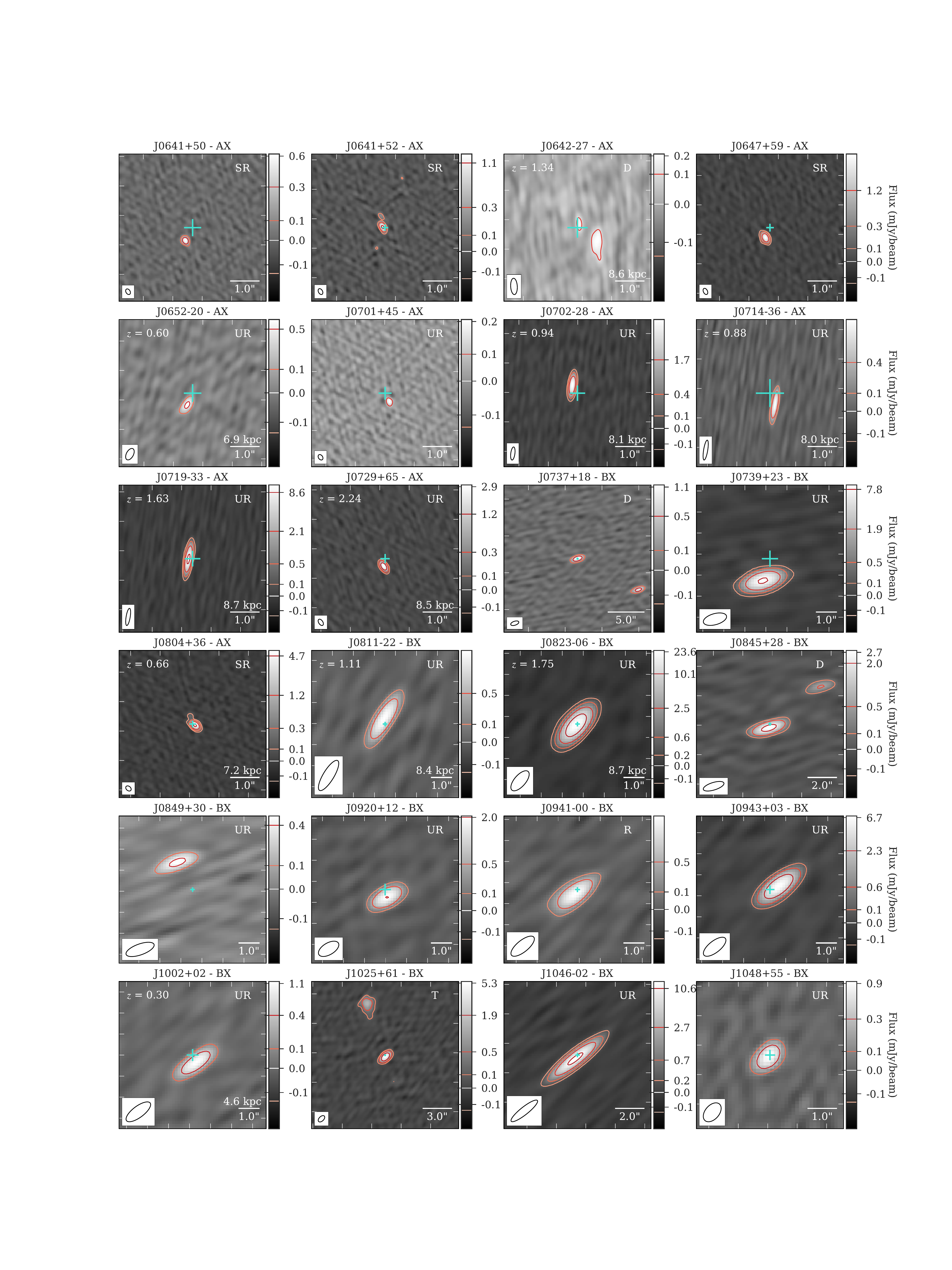}\captcont{\it Continued}
\end{figure*}
\begin{figure*}[htpb!]
\centering
\includegraphics[clip=true, trim =5cm 6cm 1cm 6cm, width=\textwidth]{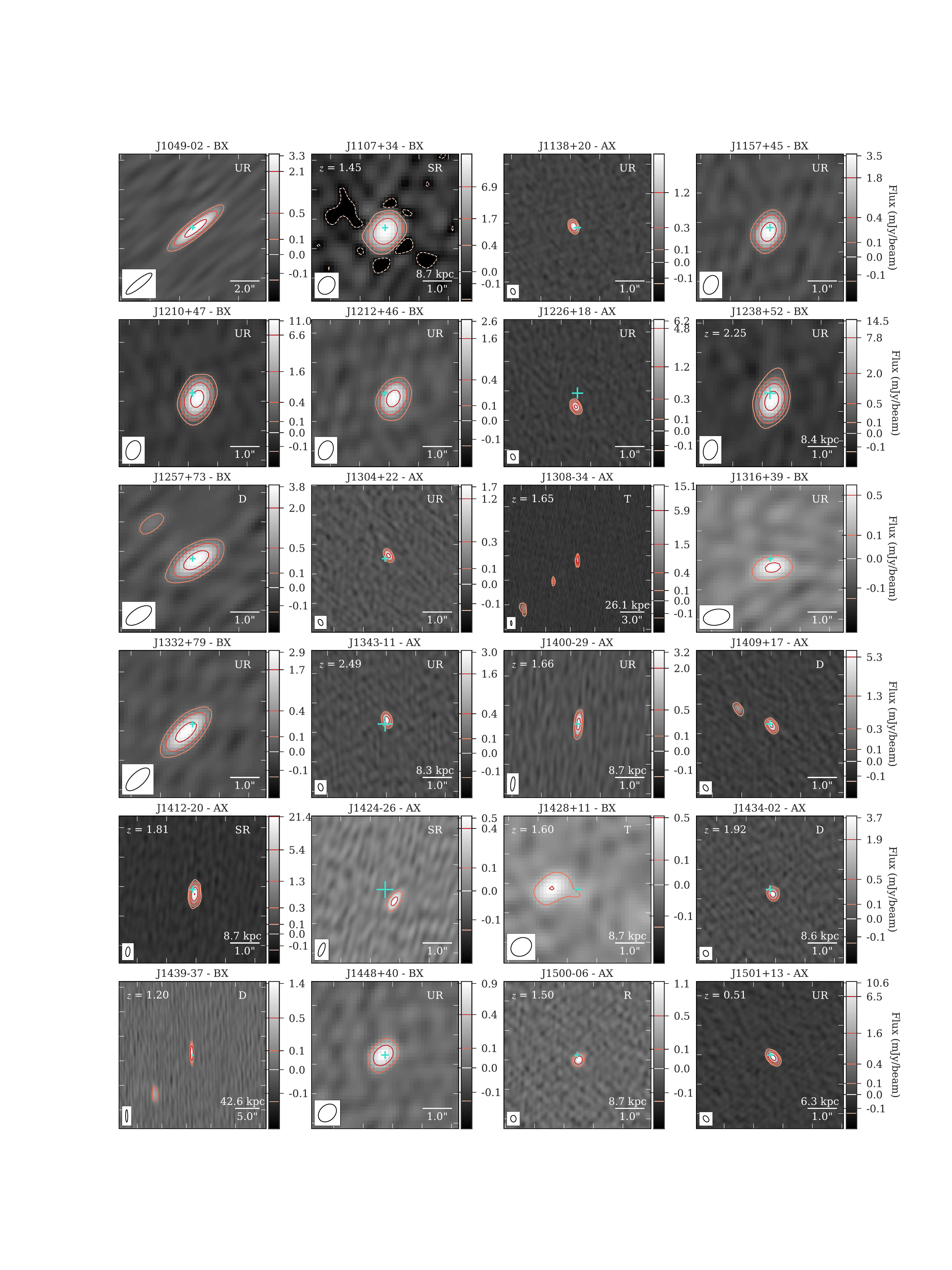}\captcont{\it Continued} 
\end{figure*}
\begin{figure*}[htpb!]
\centering
\includegraphics[clip=true, trim =5cm 6cm 1cm 6cm, width=\textwidth]{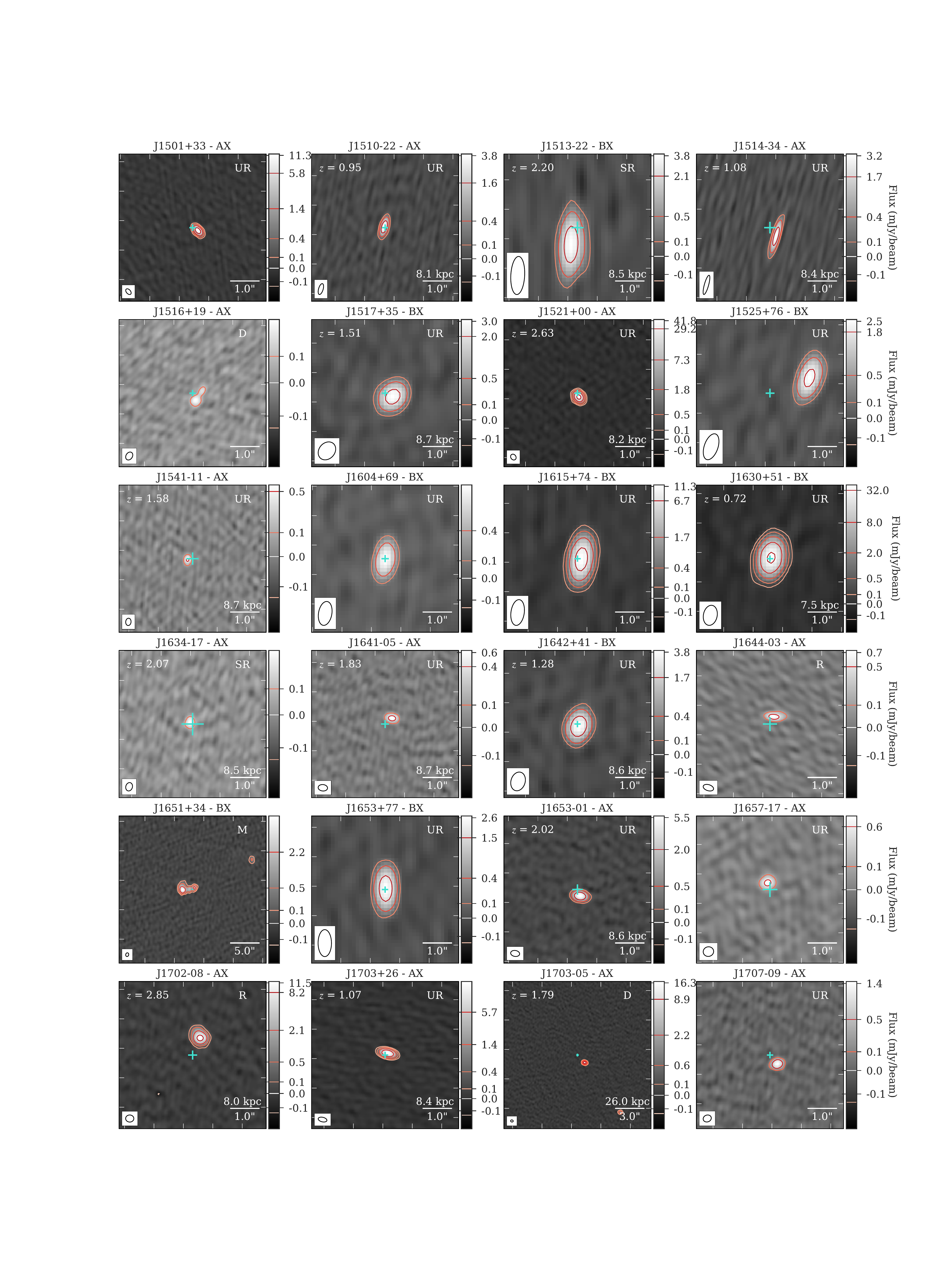}\captcont{\it Continued}
\end{figure*}
\begin{figure*}[htpb!]
\centering
\includegraphics[clip=true, trim =5cm 6cm 1cm 6cm, width=\textwidth]{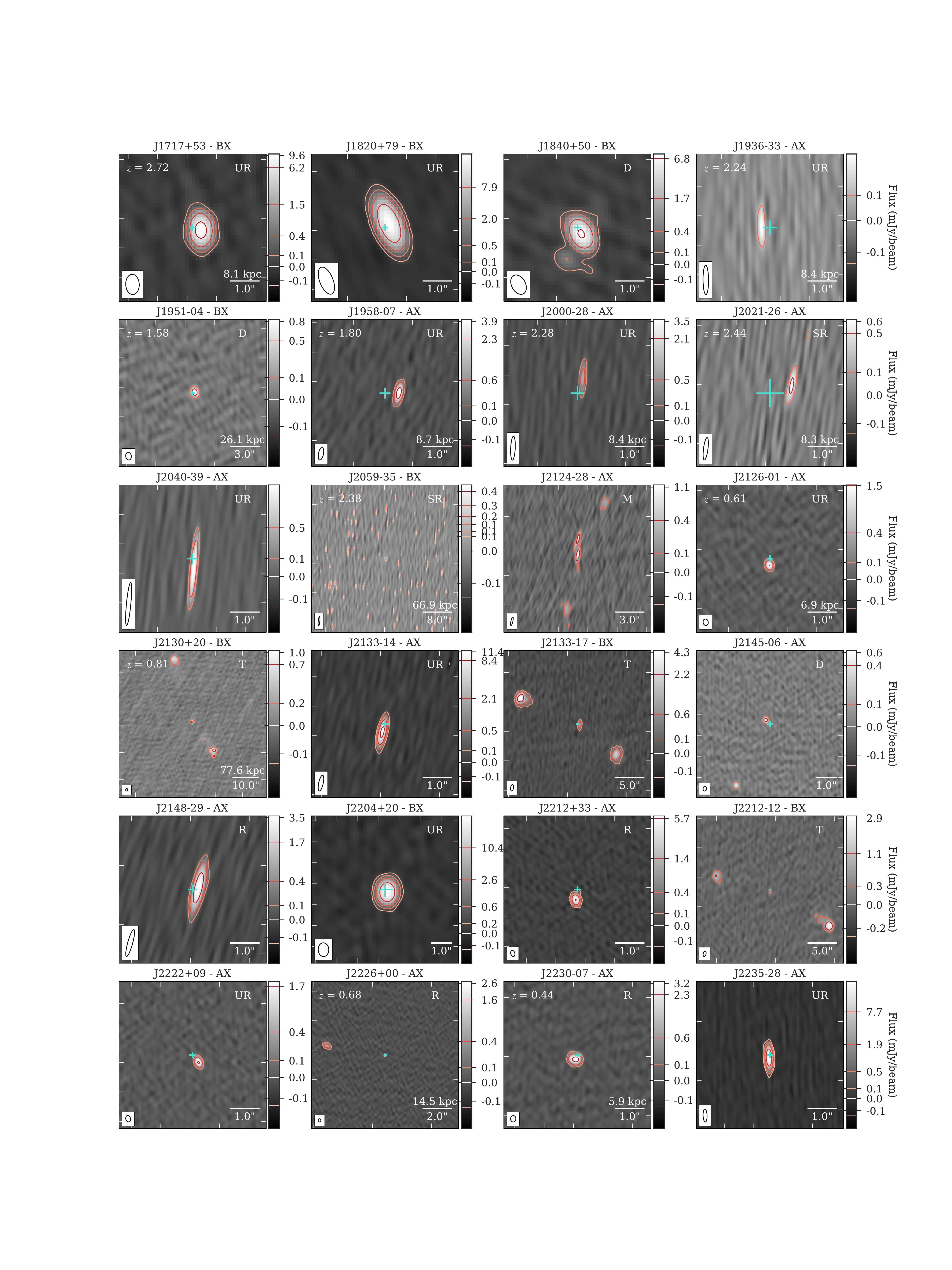}\captcont{\it Continued}
\end{figure*}
\begin{figure*}[htpb!]
\centering
\includegraphics[clip=true, trim =5cm 6cm 1cm 6cm, width=\textwidth]{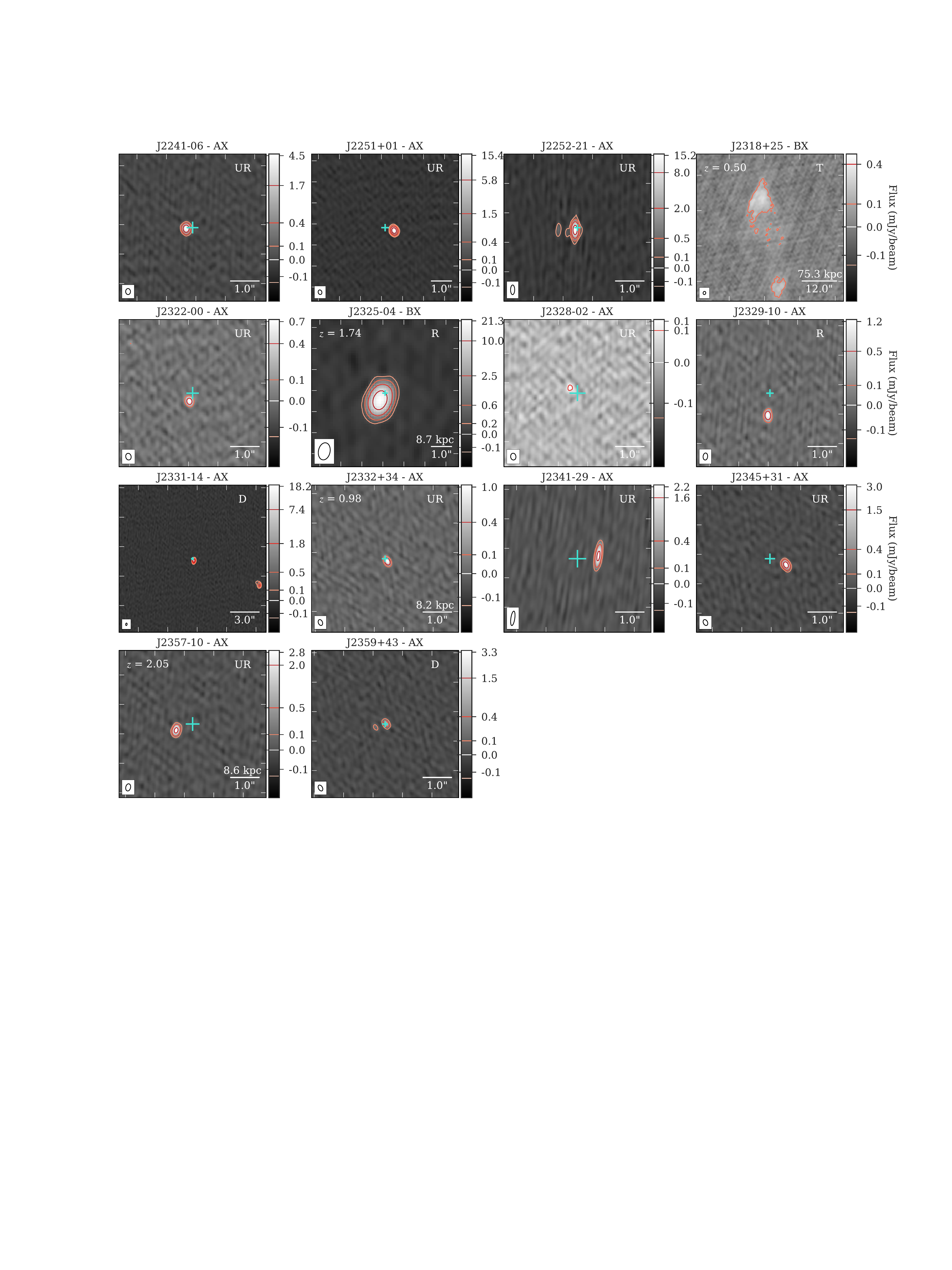}\captcont{\it Continued}
\end{figure*}

\begin{figure*}
\centering
\includegraphics[clip=true, trim =0cm 0cm 0cm 0cm, width=\textwidth]{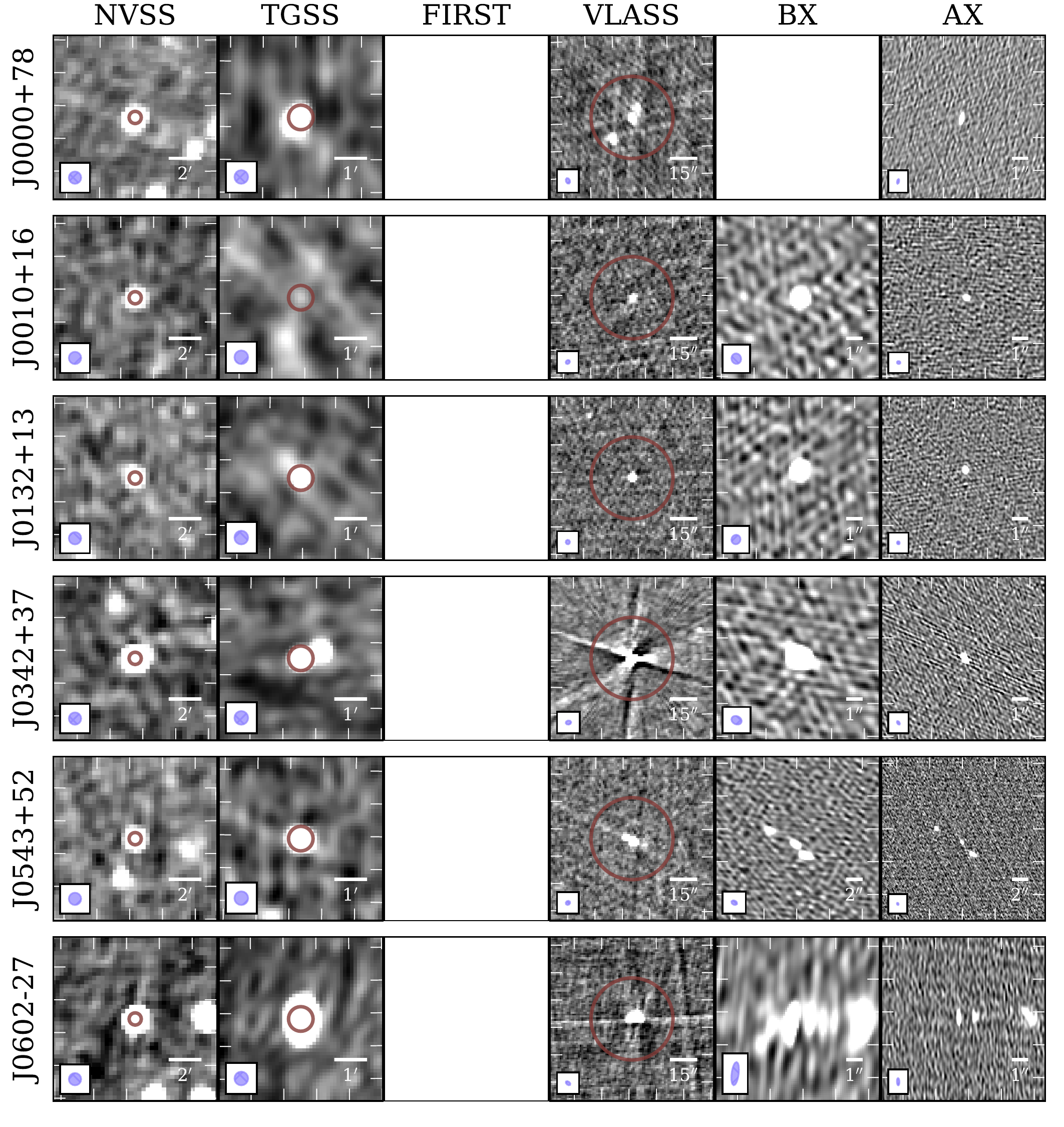}
\addtocounter{figure}{+1}
\captcont{Radio continuum cutouts of our sample sources that have extended emission on angular scales greater than a few arcseconds. The source name is shown to the left of the first column and the name of the radio survey is shown above the first row of cutouts. The red circle corresponds to the typical angular resolution of NVSS ($= 45^{\prime\prime}$). 
%The cyan plus symbol gives the \wise~ source position with one sigma uncertainty. For clarity, a minimum of  0.2$^{\prime\prime}$ is used.
The synthesized beam is shown as a purple ellipse in the lower-left corner. A white solid line on the lower-right denotes the scale bar. %RA and Dec tick marks are spaced by the scale bar length.
The tick mark spacing is equal to the length of the scale bar. 
\label{fig:ext_source}}
\end{figure*}
\begin{figure*}
\centering
\includegraphics[clip=true, trim =0cm 0cm 0cm 0cm, width=\textwidth]{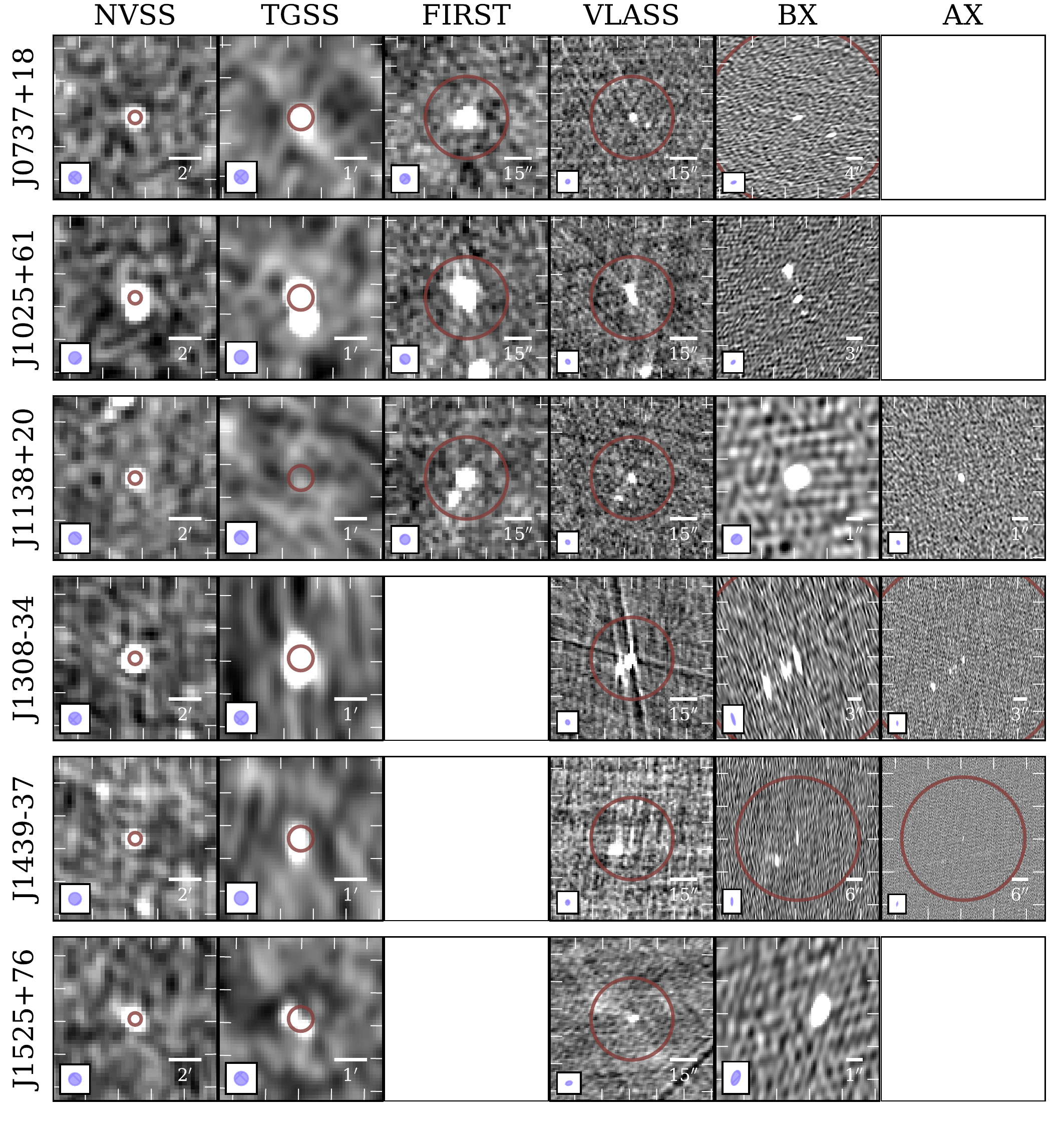}
\captcont{\it Continued}
\end{figure*}
\begin{figure*}
\centering
\includegraphics[clip=true, trim =0cm 0cm 0cm 0cm, width=\textwidth]{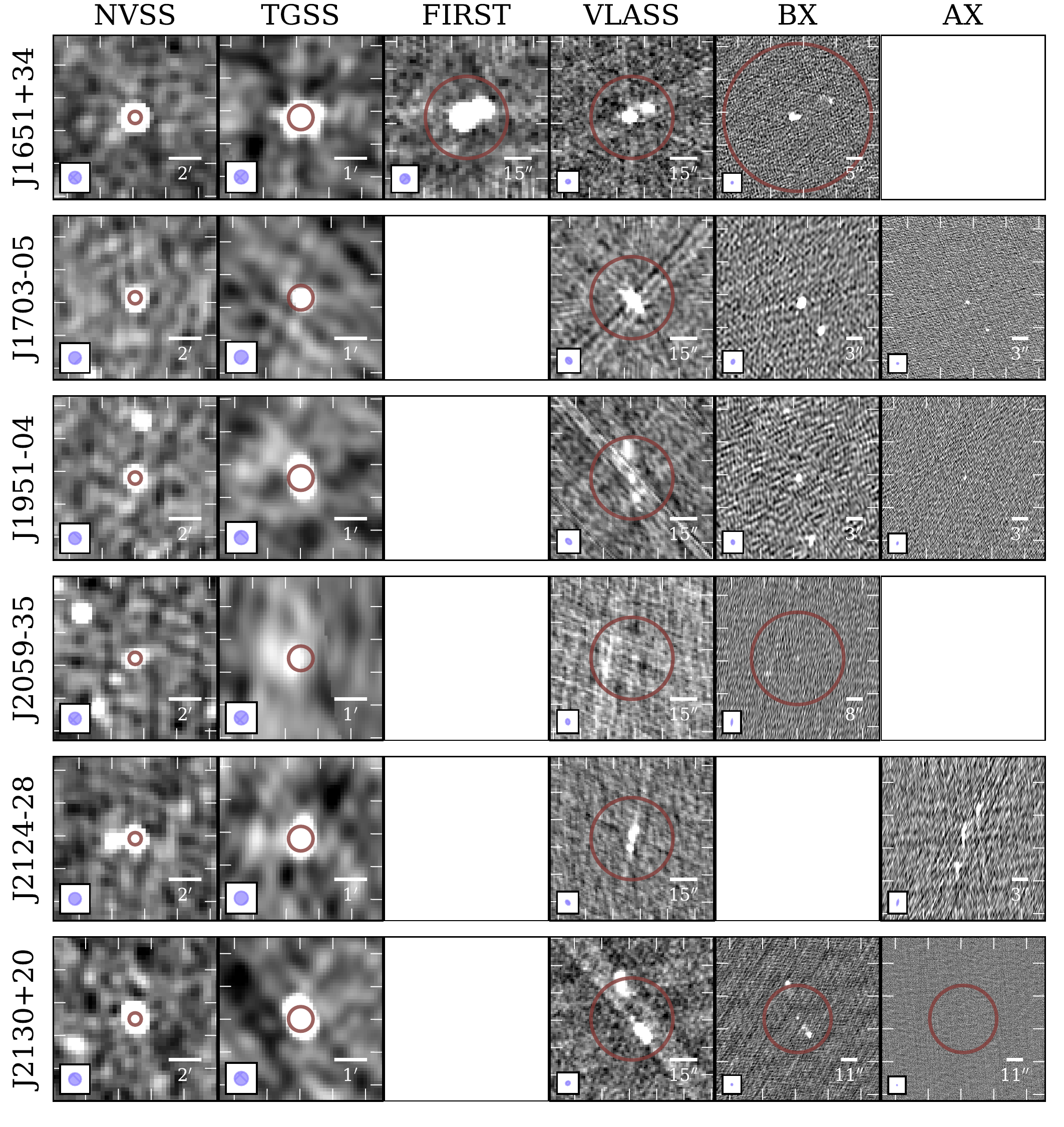}\captcont{\it Continued}
\end{figure*}
\begin{figure*}
\centering
\includegraphics[clip=true, trim =0cm 0cm 0cm 0cm, width=\textwidth]{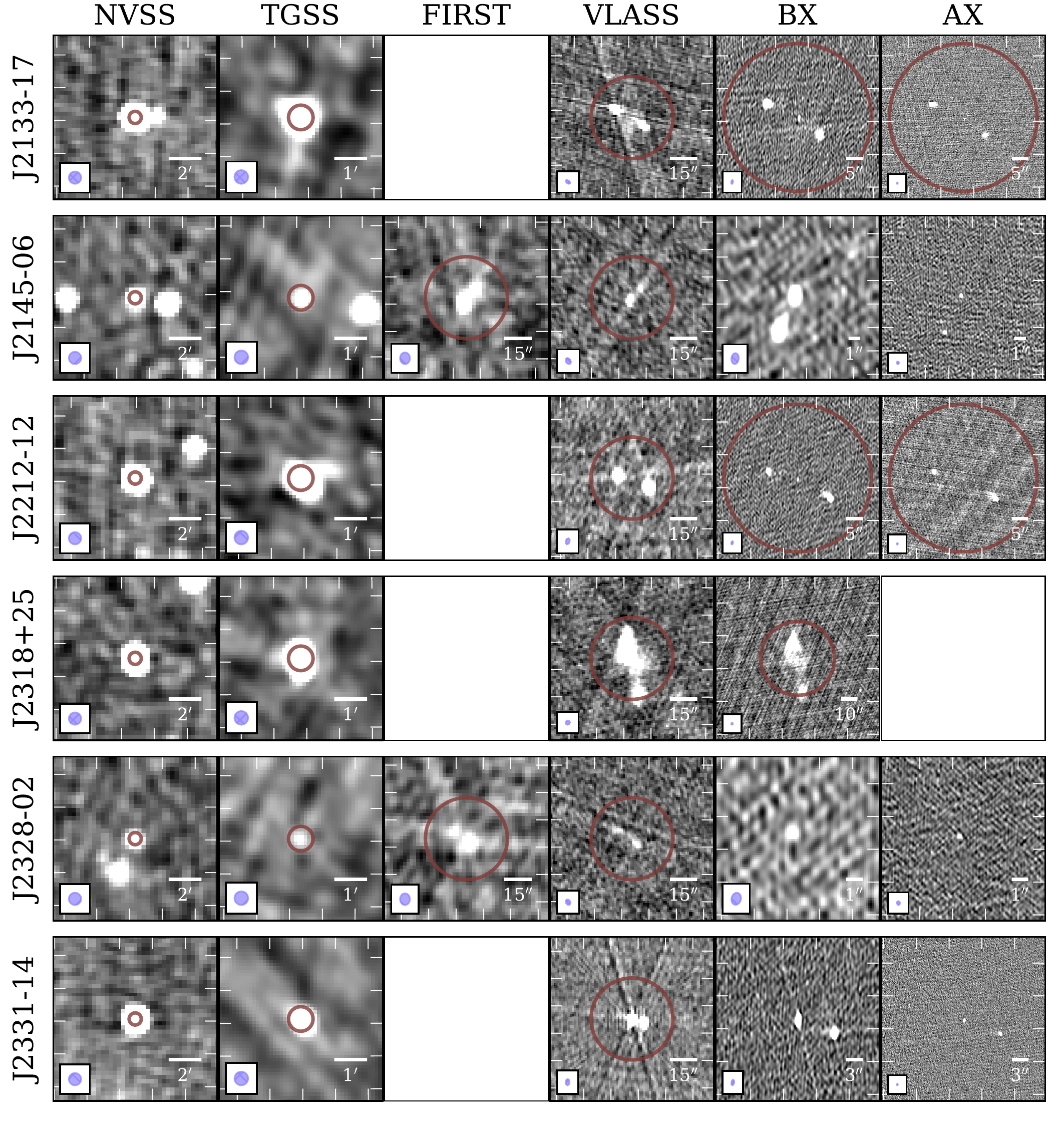}\captcont{\it Continued}
\end{figure*}
\begin{figure*}
\centering
\includegraphics[clip=true, trim =0cm 0cm 0cm 0cm, width=\textwidth]{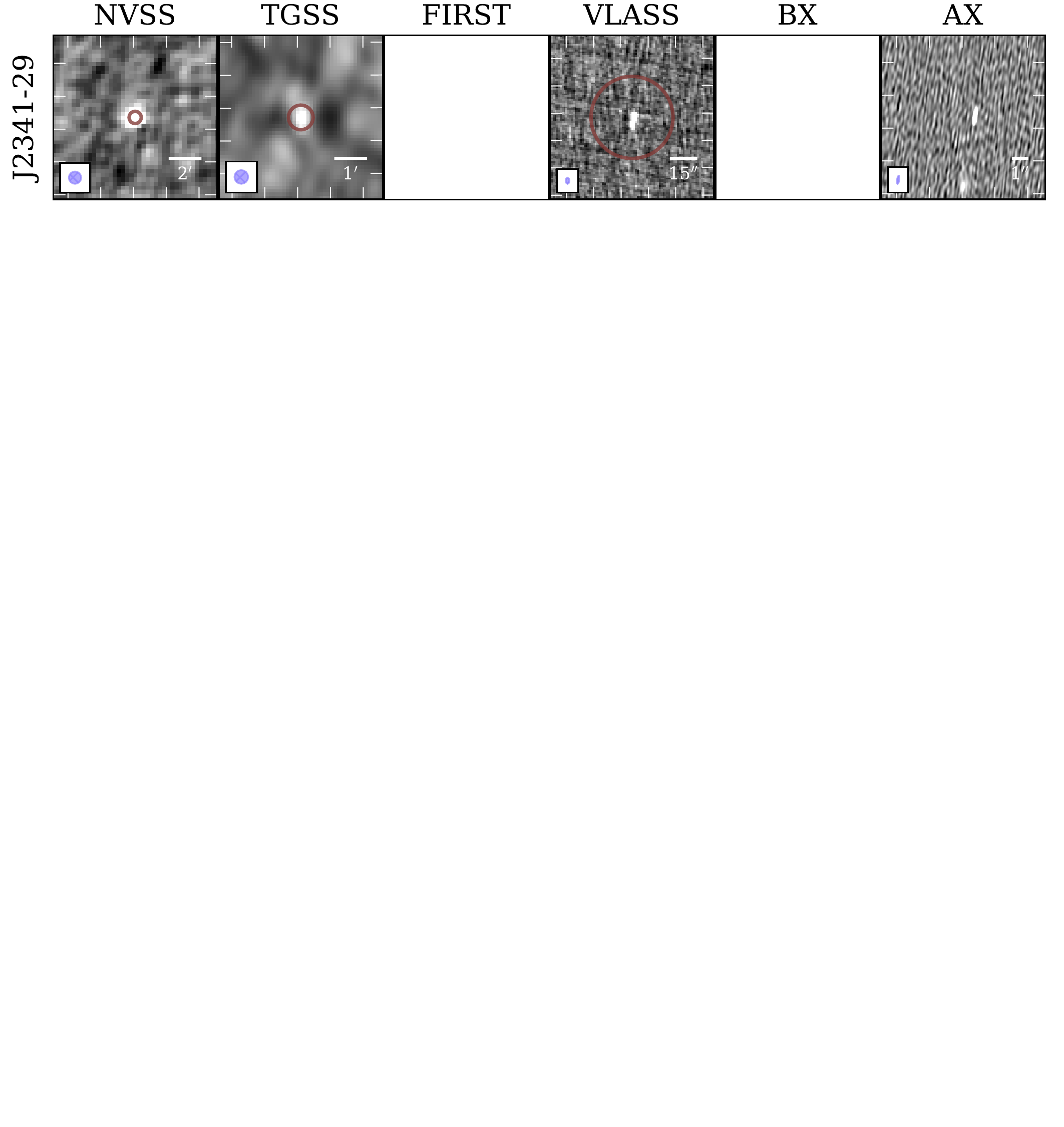}\captcont{\it Continued}
\end{figure*}

\section{Observational Parameters}
Observational details of our sample are provided in Table~\ref{tab:obs_info}.

\section{VLA source measurements}\label{sec:tabs}

Beam sizes and source measurements from the VLA observations are given in Table~\ref{tab:source_full}.
Results from JMFIT for source spatial measurements for the VLA A- and B-array observations are available in Table~\ref{tab:ssize}.
Physical properties for our sample sources with redshift available are given in Table~\ref{tab:redshifts}.

\startlongtable
\begin{deluxetable*}{llcccccccc}
\tablecaption{Observational details of our sample. Column 1; Source name. Column 2: \wise~ ID. Column 3: Date of observation for the A-array data. Column 4: 1$\sigma$ rms noise level in the A-array continuum image. Column 5: Source peak flux S/N. Column 6: A quality flag for the final continuum image. G indicates an image free of any artifacts or calibration issues.  Columns 7-9: Date of observation, 1$\sigma$ rms noise, S/N of the source detection, and an image quality flag for the B-array observations.\label{tab:obs_info}}
\tablewidth{900pt}
\tabletypesize{\footnotesize}
\tablehead{
\colhead{} & \colhead{} & \colhead{A-Array} & \colhead{} & \colhead{} 
& \colhead{}  &   \colhead{B-Array} & \colhead{} & \colhead{}  & \colhead{}\\
\colhead{Source} & \colhead{\wise~ ID} &  \colhead{Obs Date} & \colhead{rms} & \colhead{S/N} & \colhead{Quality} &  \colhead{Obs Date} & \colhead{rms} & \colhead{S/N} & \colhead{Quality}\\
\colhead{} & \colhead{} &  \colhead{yyyy-mm-dd} & \colhead{$\mu$Jy beam$^{-1}$} & \colhead{} & \colhead{} &  \colhead{yyyy-mm-dd} & \colhead{$\mu$Jy beam$^{-1}$} & \colhead{} & \colhead{}
} 
\colnumbers
\startdata
J0000+78 & 000035.88+780717.2 & 2012-10-31 & 19 & 383 & G & \nodata & \nodata & \nodata & \nodata \\
J0010+16 & 001039.54+164328.7 & 2012-12-01 & 20 & 87 & G & 2012-06-13 & 21 & 94 & G \\
J0104$-$27 & 010424.85$-$275029.0 & 2012-11-24 & 19 & 49 & G & 2012-06-13 & 31 & 46 & G \\
J0132+13 & 013211.24+130326.8 & 2012-12-01 & 22 & 202 & G & 2012-08-27 & 27 & 175 & G \\
J0133+10 & 013338.97+101943.9 & 2012-12-01 & 19 & 1064 & G & 2012-08-27 & 35 & 832 & G \\
\enddata
\tablecomments{A complete version of this table is available online. }
\end{deluxetable*}

\startlongtable
\begin{deluxetable*}{lccCRRRCRRR}
\tablecaption{Beam sizes and source measurements. Column 1: Source name. Column 2: The VLA array of the best continuum image. Column 3: Source morphology based on the criteria defined in Section~\ref{sec:morph}. UR=Unresolved, SR=Slightly resolved, R=Fully resolved, D=Double, T=Triple, M=Multi-component Sources. Column 4: Synthesized beam of the A-array data (major axis, $\theta_M$ $\times$ minor axis, $\theta_m$) in arcseconds. Column 5: Position angle of the synthesized beam, measured anti-clockwise from North. Column 6: Peak flux density of the A-array image. Column 7: Integrated flux of source A-array image. In case of multi-component sources, we provide a sum total of fluxes from each component.  Column 8: Synthesized beam of the B-array data (major axis, $\theta_M$ $\times$ minor axis, $\theta_m$) in arcseconds. Column 9: B-array beam position angle, measured anti-clockwise from North. Column 10: Peak flux density of the radio emission in the  B-array image. Column 11: Integrated flux in the  B-array image.  \label{tab:source_full}}
\tablewidth{700pt}
\tabletypesize{\scriptsize}
\tablehead{
\colhead{} & \colhead{} & \colhead{}  & A-Array  & \colhead{} &  \colhead{} &  \colhead{}  & B-Array &\colhead{} &\colhead{} & \colhead{}\\
\colhead{Source} & \colhead{Array} & \colhead{Morph} &\colhead{$\theta_M \times \theta_m$}  & \colhead{PA} & \colhead{S$_{peak}$} & \colhead{S$_{tot}$}  & \colhead{$\theta_M \times \theta_m$} & \colhead{PA} & \colhead{S$_{peak}$ }& \colhead{S$_{tot}$} \\
\colhead{} & \colhead{} & \colhead{} &\colhead{ $\arcsec\times\arcsec$ } & \colhead{deg} & \colhead{mJy~beam$^{-1}$} & \colhead{mJy}  & \colhead{$\arcsec \times\arcsec$}  &\colhead{deg} & \colhead{mJy~beam$^{-1}$} & \colhead{mJy}
} 
\colnumbers
\startdata
J0000+78 & A & D & $0.3\times0.1$ & $-13$ & $7.1\pm0.21$ & $7.96\pm0.22$ & \nodata & \nodata & \nodata & \nodata \\
J0010+16 & A & UR & $0.2\times0.2$ & $74$ & $1.76\pm0.06$ & $1.87\pm0.06$ & $0.6\times0.6$ & $17$ & $1.97\pm0.06$ & $1.95\pm0.07$ \\
J0104$-$27 & A & R & $0.5\times0.1$ & $14$ & $0.68\pm0.02$ & $0.94\pm0.15$ & $2.0\times0.5$ & $-24$ & $1.44\pm0.05$ & $1.42\pm0.07$ \\
J0132+13 & A & SR & $0.2\times0.2$ & $20$ & $4.32\pm0.13$ & $4.64\pm0.14$ & $0.6\times0.5$ & $-26$ & $4.69\pm0.14$ & $4.70\pm0.15$ \\
J0133+10 & A & D & $0.2\times0.2$ & $35$ & $19.9\pm0.6$ & $34.6\pm0.71$ & $0.6\times0.5$ & $-25$ & $28.28\pm0.85$ & $34.59\pm0.85$ \\
J0134+40 & A & SR & $0.2\times0.1$ & $13$ & $0.95\pm0.03$ & $1.13\pm0.05$ & $0.6\times0.5$ & $21$ & $1.29\pm0.05$ & $1.30\pm0.06$ \\
J0154+50 & A & UR & $0.2\times0.1$ & $15$ & $0.69\pm0.03$ & $0.73\pm0.04$ & $0.6\times0.5$ & $23$ & $0.91\pm0.04$ & $0.97\pm0.05$ \\
\enddata
\tablecomments{A complete version of this table is available online. }
\end{deluxetable*}

\begin{longrotatetable}
\begin{deluxetable}{rhcrrhhCRhhrrhhCRhhRRh}
\tabletypesize{\scriptsize}
\tablecaption{Source spatial measurements for the VLA A- and B-array observations: Results from JMFIT \label{tab:ssize}}
\tablehead{
\nocolhead{}&\nocolhead{}&\nocolhead{}&\multicolumn{8}{C}{A-Array}&\multicolumn{8}{C}{B-Array}&\nocolhead{}&\nocolhead{}&\nocolhead{} \\
\colhead{Source} & \nocolhead{WISE~ID} & \colhead{Region} & \colhead{RA-A} & \colhead{Dec-A} & \nocolhead{Beam-A} &\nocolhead{BPA-A}&\colhead{Source Size-A} &\colhead{PA-A}&\nocolhead{$S_{peak}$-A}&\nocolhead{$S_{tot}$-A}&\colhead{RA-B}&\colhead{Dec-B}&\nocolhead{Beam-B}&\nocolhead{BPA-B}&\colhead{Source~Size-B}&\colhead{PA-B}&\nocolhead{$S_{peak}$-B}&\nocolhead{$S_{tot}$-B}&\colhead{$\alpha_{IB}$}&\colhead{S/N}&\nocolhead{nregs} \\
\colhead{}&\nocolhead{}&\colhead{}&\colhead{(hh:mm:ss.s)}&\colhead{(dd:mm:ss.s)}&\nocolhead{($mas\times mas$)}&\nocolhead{(deg)}&\colhead{($mas\times mas$)} &\colhead{(deg)}&\nocolhead{(mJy/beam)}&\nocolhead{(mJy)}&\colhead{(hh:mm:ss.s)}&\colhead{(dd:mm:ss.s)}&\nocolhead{($mas\times mas$)}&\nocolhead{(deg)}&\colhead{($mas\times mas$)}&\colhead{(deg)}&\nocolhead{(mJy/beam)}&\nocolhead{(mJy)} &\colhead{} &\colhead{}& \nocolhead{} \\
}
\colnumbers
\startdata
J0000+78 & W000035.88+780717.2 & Reg 1 & 00:00:35.918 & 78:07:17.15 & 0.29 $\times$ 0.12 & -13.2 & < 32 & 158 $\pm$ 0 &7.10 \pm 0.21&7.15 \pm 0.22& \nodata & \nodata & \nodata & \nodata & \nodata & \nodata & \nodata & \nodata &$-0.61 \pm 0.02$& 383 & 2.0 \\
 &  & Reg 2 & -15.72 & 5.82 & 0.29 $\times$ 0.12 & -13.2 & 163 $\pm$  32 $\times$ 100 $\pm$  10 & 54 $\pm$ 3 &0.47 \pm 0.02&0.81 \pm 0.05& \nodata & \nodata & \nodata & \nodata & \nodata & \nodata & \nodata & \nodata &$-1.48 \pm 0.38$& 27 & 2.0 \\
J0010+16 & W001039.54+164328.7 & Reg 1 & 00:10:39.529 & 16:43:28.81 & 0.21 $\times$ 0.18 & 74.3 & < 73 & 12 $\pm$ 4 &1.76 \pm 0.06&1.87 \pm 0.06& 00:10:39.531 & 16:43:28.82 & 0.64 $\times$ 0.59 & 17.0 & <126 & 0 $\pm$ 5 &1.97 \pm 0.06&1.95 \pm 0.07&$-1.71 \pm 0.09$& 87 & 1.0 \\
J0104$-$27 & W010424.85-275029.0 & Reg 1 & 01:04:24.862 & -27:50:29.16 & 0.46 $\times$ 0.13 & 14.3 & 976  $\times$ 472  & \nodata &0.68 \pm 0.02&0.94 \pm 0.15& 01:04:24.867 & -27:50:28.98 & 2.02 $\times$ 0.49 & -24.1 & <262 & 145 $\pm$ 0 &1.44 \pm 0.05&1.42 \pm 0.07&$-1.71 \pm 0.33$& 49 & 1.0 \\
J0132+13 & W013211.24+130326.8 & Reg 1 & 01:32:11.240 & 13:03:27.39 & 0.19 $\times$ 0.17 & 20.2 & 65 $\pm$   4 $\times$  23 $\pm$   3 & 153 $\pm$ 2 &4.32 \pm 0.13&4.64 \pm 0.14& 01:32:11.239 & 13:03:27.39 & 0.58 $\times$ 0.54 & -25.5 & <127 & 168 $\pm$ 2 &4.69 \pm 0.14&4.70 \pm 0.15&$-1.26 \pm 0.12$& 202 & 1.0 \\
J0133+10 & W013338.97+101943.9 & Reg 1 & 01:33:38.973 & 10:19:44.09 & 0.19 $\times$ 0.17 & 34.6 & < 40 & 105 $\pm$ 0 &12.34 \pm 0.37&13.73 \pm 0.37&  01:33:38.978 & 10:19:44.01 & 0.62 \times 0.54 & -25.2 & 388 \pm   2 \times  60 \pm   2 & 98 \pm 0 &28.28 \pm 0.85&34.59 \pm 0.85&$-1.46 \pm 0.10$& 1064 & 3.0 \\
 &  & Reg 2 & 0.30 & -0.03 & 0.19 $\times$ 0.17 & 34.6 & 64 $\pm$   1 $\times$  56 $\pm$   1 & 94 $\pm$ 1 &19.90 \pm 0.60&20.44 \pm 0.60& \nodata & \nodata & \nodata & \nodata & \nodata & \nodata & \nodata & \nodata &$-1.01 \pm 0.01$& 664 & 3.0 \\
 &  & Reg 3 & -0.50 & 0.10 & 0.19 $\times$ 0.17 & 34.6 & 164 $\pm$  24 $\times$  62 $\pm$  24 & 155 $\pm$ 12 &0.29 \pm 0.02&0.42 \pm 0.04& \nodata & \nodata & \nodata & \nodata & \nodata & \nodata & \nodata & \nodata &$-1.69 \pm 0.86$& 16 & 3.0 \\
J0134+40 & W013419.27+403049.3 & Reg 1 & 01:34:19.290 & 40:30:49.34 & 0.20 $\times$ 0.13 & 13.3 & 93 $\pm$  15 $\times$   7 $\pm$   6 & 71 $\pm$ 2 &0.95 \pm 0.03&1.13 \pm 0.05& 01:34:19.281 & 40:30:49.41 & 0.57 $\times$ 0.47 & 20.6 & <182 & 45 $\pm$ 4 &1.29 \pm 0.05&1.30 \pm 0.06&$-1.66 \pm 0.25$& 51 & 1.0 \\
J0154+50 & W015442.57+504600.4 & Reg 1 & 01:54:42.481 & 50:46:00.36 & 0.22 $\times$ 0.13 & 14.9 & < 89 & 157 $\pm$ 2 &0.69 \pm 0.03&0.73 \pm 0.04& 01:54:42.480 & 50:46:00.40 & 0.61 $\times$ 0.46 & 23.0 & <238 & 52 $\pm$ 4 &0.91 \pm 0.04&0.97 \pm 0.05&$-1.52 \pm 0.18$& 39 & 1.0 \\
J0159+12 & W015919.56+120137.4 & Reg 1 & 01:59:19.581 & 12:01:37.04 & 0.19 $\times$ 0.16 & 20.7 & 223 $\pm$   4 $\times$  28 $\pm$  11 & 18 $\pm$ 1 &1.73 \pm 0.06&2.73 \pm 0.07& 01:59:19.576 & 12:01:37.04 & 0.58 $\times$ 0.57 & -20.7 & 179 $\pm$  21 $\times$ 105 $\pm$  31 & 1 $\pm$ 9 &2.47 \pm 0.08&2.63 \pm 0.09&$-1.20 \pm 0.15$& 87 & 1.0 \\
J0204+09 & W020411.96+092030.2 & Reg 1 & 02:04:11.979 & 09:20:30.11 & 0.19 $\times$ 0.16 & 25.0 & < 80 & 23 $\pm$ 78 &0.81 \pm 0.03&0.90 \pm 0.04& 02:04:11.975 & 09:20:30.18 & 0.61 $\times$ 0.56 & -44.0 & <212 & 179 $\pm$ 11 &1.04 \pm 0.04&1.04 \pm 0.06&$-1.24 \pm 0.16$& 49 & 1.0 \\
J0244+11 & W024423.99+112354.4 & Reg 1 & 02:44:24.000 & 11:23:54.36 & 0.19 $\times$ 0.16 & 15.9 & < 50 & 39 $\pm$ 1 &3.81 \pm 0.12&3.91 \pm 0.12& 02:44:24.006 & 11:23:54.40 & 0.62 $\times$ 0.56 & -62.4 & <123 & 121 $\pm$ 2 &4.05 \pm 0.12&4.04 \pm 0.13&$-1.24 \pm 0.04$& 182 & 1.0 \\
J0300+39 & W030037.53+390125.3 & Reg 1 & 03:00:37.571 & 39:01:25.14 & 0.20 $\times$ 0.14 & 30.7 & 88 $\pm$   3 $\times$  19 $\pm$   9 & 41 $\pm$ 1 &2.89 \pm 0.09&3.16 \pm 0.09& 03:00:37.571 & 39:01:25.06 & 0.60 $\times$ 0.48 & 45.2 & 157 $\pm$  32 $\times$ <   0 & 23 $\pm$ 1 &3.29 \pm 0.10&3.35 \pm 0.11&$-0.16 \pm 0.12$& 163 & 1.0 \\
J0303+07 & W030333.57+073648.1 & Reg 1 & 03:03:33.621 & 07:36:49.14 & 0.19 $\times$ 0.16 & 18.4 & < 51 & 65 $\pm$ 3 &1.36 \pm 0.04&1.41 \pm 0.05& 03:03:33.617 & 07:36:49.17 & 0.71 $\times$ 0.57 & -57.5 & <197 & 125 $\pm$ 5 &0.40 \pm 0.03&0.39 \pm 0.05&$-0.97 \pm 0.10$& 79 & 2.0 \\
 &  & Reg 2 & -0.45 & -0.98 & 0.19 $\times$ 0.16 & 18.4 & < 72 & 164 $\pm$ 7 &0.37 \pm 0.02&0.38 \pm 0.03& -0.46 & -0.94 & 0.71 $\times$ 0.57 & -57.5 & <287 & 58 $\pm$ 13 &1.53 \pm 0.05&1.58 \pm 0.06&$-1.52 \pm 0.37$& 22 & 2.0 \\
J0304$-$31 & W030427.53-310838.2 & Reg 1 & 03:04:27.549 & -31:08:38.27 & 0.52 $\times$ 0.13 & -6.3 & < 19 & 21 $\pm$ 0 &20.08 \pm 0.60&20.20 \pm 0.60& 03:04:27.557 & -31:08:38.43 & 1.61 $\times$ 0.44 & -171.7 & 357 $\pm$  16 $\times$ 133 $\pm$   3 & 179 $\pm$ 0 &18.25 \pm 0.55&19.65 \pm 0.55&$-0.89 \pm 0.01$& 890 & 2.0 \\
 &  & Reg 2 & 0.36 & 0.10 & 0.52 $\times$ 0.13 & -6.3 & < 74 & 11 $\pm$ 0 &1.15 \pm 0.04&1.21 \pm 0.05& \nodata & \nodata & \nodata & \nodata & \nodata & \nodata & \nodata & \nodata &$-1.12 \pm 0.14$& 51 & 2.0 \\
\enddata
\tablecomments{Column 1: Source name. Column 2: Region. For a single component source, the entire component is named Reg 1. For multi-component sources, brightest radio emission component is named Reg 1. Column 3 and 4: J2000 Right ascension and declination of the fitted source in the A-array image. In case of sources with more than one component, a source separation (in arcseconds) from the Reg 1 is provided. Column 5: Deconvolved source sizes for the A-array data. If source is resolved only along the major axis, the deconvolved minor axis is specified as 0. In case of unresolved source, we provide an upper limit on the major axis. A detailed description is provided in Section~\ref{sec:SAS}. For extended sources with non-gaussian like emission, we provide the size of 3$\sigma$ contour as the angular size of the respective region. Column 6: Position angle of the fitted gaussian, measured anti-colockwise from North. Column 7 and 8: J2000 right ascension and declination for the B-array image. Column 9: B-array deconvolved source sizes from the JMFIT source fitting. Column 10: Position angle for the fitted source. Column 11: In-band spectral index for the best image available. We used A-array data when a good quality image is available. (see Section~\ref{sec:ibalph}). Column 12: Source detection S/N avearaged from the 8.6 and 11.4~GHz images used for calculating $\alpha_{IB}$. \\
A complete version of this table is available online. }
\end{deluxetable}
\end{longrotatetable}

\startlongtable
\begin{deluxetable*}{cChhcChhCRC}
\tablecaption{Physical properties for our sample sources with  redshift available. Column 1: Source name. Column 2: Redshift. Column 3: Region name. Column 4: Linear dimensions of the radio emission in each region. For an unresolved source, we use an upper limit on the angular major axis to estimate the limit on the source linear size.  Column 5: Rest-frame 1.4~GHz luminosity. We use NVSS flux and the spectral index between NVSS and 10~GHz continuum observations to calculate the luminosity. Column 6: Spectral index between NVSS and 10~GHz observations. Fluxes from all of the regions are added up to estimate the spectral indices. Column 7: Equipartition lobe pressures as described in Section~\ref{sec:pressure}\\
A complete version of this table is available online.\label{tab:redshifts}}
\tabletypesize{\footnotesize}
\tablehead{
\colhead{Source} & \colhead{$z$} & \nocolhead{Config} & \nocolhead{Morphology} & \colhead{Region} & \colhead{Linear Size} & \nocolhead{$\alpha_{IB}$} & \nocolhead{S/N} & \colhead{log$_{10}$ $L_{1.4\,GHz}$}& \colhead{$\alpha^{10}_{1.4}$}& \colhead{$\log P_l$}  \\
\colhead{} & \colhead{} & \nocolhead{} & \nocolhead{} & \colhead{} & \colhead{(kpc$\times$kpc)} & \nocolhead{} & \nocolhead{} & \colhead{(W~Hz$^{-1}$)}& \colhead{}& \colhead{dyne cm$^{-2}$} 
}
\colnumbers
\startdata
J0010+16 & 2.85 & A & UR & Reg 1 & $<0.6$ & $-1.71 \pm 0.09$ & 87 & 27.2 & $-1.17 \pm 0.03$ & $-6.4$  \\
J0132+13 & 2.85 & A & SR & Reg 1 & $0.5\times 0.2$ & $-1.26 \pm 0.12$ & 202 & 27.1 & $-0.79 \pm 0.02$ & $-5.6$  \\
J0159+12 & 0.76 & A & R & Reg 1 & $1.7\times 0.2$ & $-1.20 \pm 0.15$ & 87 & 25.8 & $-1.04 \pm 0.02$ & $-6.5$  \\
J0300+39 & 1.12 & A & SR & Reg 1 & $0.7\times 0.2$ & $-0.16 \pm 0.12$ & 163 & 25.9 & $-0.75 \pm 0.03$ & $-6.1$  \\
J0304$-$31 & 1.53 & A & D & Reg 1 & $<0.2$ & $-0.89 \pm 0.01$ & 890 & 26.7 & $-0.48 \pm 0.02$ & $-5.1$  \\
 &  &  &  & Reg 2 & $<0.6$ & $-1.12 \pm 0.14$ & 51 & \nodata & \nodata & $-6.7$  \\
J0306$-$33 & 0.78 & A & UR & Reg 1 & $<0.8$ & $-0.95 \pm 0.09$ & 77 & 25.2 & $-0.77 \pm 0.05$ & $-7.1$  \\
J0332+32 & 0.30 & A & UR & Reg 1 & $<0.1$ & $-1.28 \pm 0.03$ & 255 & 25.1 & $-1.08 \pm 0.02$ & $-5.7$  \\
J0342+37 & 0.47 & B & UR & Reg 1 & $<0.3$ & $-1.34 \pm 0.01$ & $999$ & 26.0 & $-0.48 \pm 0.02$ & $-5.8$  \\
J0354$-$33 & 1.37 & A & UR & Reg 1 & $<0.8$ & $-0.96 \pm 0.05$ & 152 & 25.8 & $-0.50 \pm 0.04$ & $-6.7$  \\
J0404$-$24 & 1.26 & A & D & Reg 1 & $5.2\times 3.8$ & $-3.08 \pm nan$ & 20 & 26.2 & $-1.38 \pm 0.09$ & $-8.6$  \\
 &  &  &  & Reg 2 & $4.9\times 2.5$ & $-0.74 \pm nan$ & 15 & \nodata & \nodata & $-7.8$  \\
J0409$-$18 & 0.67 & A & D & Reg 1 & $<0.4$ & $-1.02 \pm 0.03$ & 219 & 26.0 & $-1.08 \pm 0.02$ & $-6.3$  \\
 &  &  &  & Reg 2 & $2.1\times 0.7$ & $-2.42 \pm 0.71$ & 15 & \nodata & \nodata & $-7.2$  \\
J0417$-$28 & 0.94 & A & UR & Reg 1 & $<0.4$ & $-1.00 \pm 0.02$ & 417 & 25.7 & $-0.31 \pm 0.03$ & $-6.1$  \\
J0439$-$31 & 2.82 & A & R & Reg 1 & $1.7\times<0.1$ & $-0.98 \pm 0.03$ & 153 & 27.0 & $-0.71 \pm 0.02$ & $-7.0$  \\
J0519$-$08 & 2.05 & A & UR & Reg 1 & $<0.5$ & $-0.57 \pm 0.03$ & 298 & 26.7 & $-0.55 \pm 0.02$ & $-6.1$  \\
J0525$-$36 & 1.69 & A & R & Reg 1 & $1.3\times 0.4$ & $-1.14 \pm 0.16$ & 99 & 25.8 & $-0.43 \pm 0.06$ & $-6.5$  \\
J0526$-$32 & 1.98 & A & R & Reg 1 & $4.4\times 0.2$ & $-0.41 \pm 0.09$ & 713 & 27.6 & $-0.84 \pm 0.02$ & $-5.9$  \\
J0536$-$27 & 1.79 & A & UR & Reg 1 & $<0.2$ & $0.94 \pm 0.01$ & 999 & 25.6 & $0.50 \pm 0.04$ & $-4.4$  \\
J0549$-$37 & 1.71 & A & SR & Reg 1 & $<2.3$ & $-1.49 \pm 0.33$ & 23 & 26.5 & $-1.43 \pm 0.05$ & $-7.7$  \\
J0612$-$06 & 0.47 & A & T & Reg 1 & $<0.5$ & $-0.38 \pm nan$ & 111 & 25.5 & $-1.12 \pm 0.03$ & $-6.9$  \\
 &  &  &  & Reg 2 & $3.6\times 3.1$ & $nan \pm nan$ & 12 & \nodata & \nodata & $-7.2$  \\
 &  &  &  & Reg 3 & $4.7\times 3.8$ & $-0.86 \pm nan$ & 15 & \nodata & \nodata & $-8.9$  \\
J0613$-$34 & 2.18 & A & UR & Reg 1 & $<1.3$ & $-1.75 \pm 0.12$ & 65 & 27.1 & $-1.23 \pm 0.03$ & $-7.0$  \\
J0630$-$21 & 1.44 & A & D & Reg 1 & $<0.3$ & $0.23 \pm 0.03$ & 209 & 26.0 & $-0.32 \pm 0.03$ & $-5.9$  \\
 &  &  &  & Reg 2 & $1.6\times 1.0$ & $-0.78 \pm 0.37$ & 31 & \nodata & \nodata & $-7.3$  \\
J0642$-$27 & 1.34 & A & D & Reg 1 & $12.2$ & $0.24 \pm nan$ & 18 & 25.8 & $-0.93 \pm 0.07$ & $-7.8$  \\
 &  &  &  & Reg 2 & $10.7\times 4.1$ & $-4.21 \pm nan$ & 17 & \nodata & \nodata & $-8.9$  \\
J0652$-$20 & 0.60 & A & UR & Reg 1 & $<1.3$ & $-0.99 \pm 0.30$ & 25 & 25.1 & $-1.23 \pm 0.05$ & $-7.8$  \\
J0702$-$28 & 0.94 & A & UR & Reg 1 & $<0.5$ & $-0.68 \pm 0.02$ & 305 & 25.3 & $-0.08 \pm 0.04$ & $-6.4$  \\
J0714$-$36 & 0.88 & A & UR & Reg 1 & $<0.5$ & $-1.31 \pm 0.12$ & \nodata & 25.7 & $-1.04 \pm 0.03$ & $-8.4$  \\
 &  &  &  & Reg 2 & $<0.5$ & $-0.38 \pm 0.24$ & \nodata & \nodata & \nodata & $-9.1$  \\
J0719$-$33 & 1.63 & A & UR & Reg 1 & $<0.2$ & $-2.11 \pm 0.02$ & 405 & 26.4 & $-0.40 \pm 0.02$ & $-5.2$  \\
\enddata
\end{deluxetable*}

\end{document}